\documentclass[12pt,a4paper]{article}
\usepackage[utf8]{inputenc}
\usepackage{amsmath}
\usepackage{amsfonts} 
\usepackage{mathtools} 
\usepackage{amssymb} 
\usepackage{subfigure} 
\usepackage{graphicx}
\usepackage{xcolor}
\usepackage{cite} 
\usepackage{mathrsfs}
\usepackage{hyperref}
\usepackage{slashbox}
\usepackage[normalem]{ulem}

\graphicspath{{Figs/}}

\newcommand{\abs}[1]{|#1|} 
\newcommand{\re}[1]{\text{Re}\left(#1\right)}
\newcommand{\im}[1]{\text{Im}\left(#1\right)}
\newcommand{\refEQ}[1]{eq.\,\eqref{#1}} 
\newcommand{\refEQS}[1]{eqs.\,\eqref{#1}} 

\newcommand{\UqX}[2]{\mathcal U_{#2}^{#1}}\newcommand{\UqXd}[2]{\mathcal U_{#2}^{#1\dagger}}
\newcommand{\UuL}{\UqX{u}{L}}\newcommand{\UuLd}{\UqXd{u}{L}}
\newcommand{\UdL}{\UqX{d}{L}}\newcommand{\UdLd}{\UqXd{d}{L}}
\newcommand{\UuR}{\UqX{u}{R}}
\newcommand{\UdR}{\UqX{d}{R}}
\newcommand{\OqX}[2]{\mathcal O_{#2}^{#1}}\newcommand{\OqXt}[2]{\mathcal O_{#2}^{#1\,T}}
\newcommand{\OuL}{\OqX{u}{L}}\newcommand{\OuLt}{\OqXt{u}{L}}
\newcommand{\OdL}{\OqX{d}{L}}\newcommand{\OdLt}{\OqXt{d}{L}}
\newcommand{\OuR}{\OqX{u}{R}}\newcommand{\OuRt}{\OqXt{u}{R}}
\newcommand{\OdR}{\OqX{d}{R}}\newcommand{\OdRt}{\OqXt{d}{R}}
\newcommand{\OROTmat}{R}
\newcommand{\OROT}[1]{R_{#1}}
\newcommand{\SROT}[1]{S_{#1}}
\newcommand{\Yd}[1]{\Gamma_{#1}}
\newcommand{\Yu}[1]{\Delta_{#1}}
\newcommand{\Ydc}[1]{\Gamma_{#1}^\ast}
\newcommand{\Yuc}[1]{\Delta_{#1}^\ast}

\newcommand{\PR}[1]{{\rm P}_{\!#1}}

\newcommand{\id}{\mathbf{1}}

\newcommand{\un}[2]{\hat n_{{\rm [#1]}#2}^{\phantom{\ast}}}
\newcommand{\unC}[2]{\hat n_{{\rm [#1]}#2}^{\ast}}
\newcommand{\und}[1]{\un{d}{#1}}
\newcommand{\unu}[1]{\un{u}{#1}}
\newcommand{\undC}[1]{\unC{d}{#1}}
\newcommand{\unuC}[1]{\unC{u}{#1}}
\newcommand{\unvec}[1]{\hat n_{{\rm [#1]}}^{\phantom{\ast}}}

\newcommand{\undvec}{\unvec{d}}
\newcommand{\unuvec}{\unvec{u}}

\newcommand{\rn}[2]{\hat r_{{\rm [#1]}#2}}
\newcommand{\rnd}[1]{\rn{d}{#1}}
\newcommand{\rnu}[1]{\rn{u}{#1}}
\newcommand{\rnvec}[1]{\hat r_{{\rm [#1]}}}
\newcommand{\rndvec}{\rnvec{d}}
\newcommand{\rnuvec}{\rnvec{u}}
\newcommand{\Hd}[1]{\Phi_{#1}^{\phantom{\dagger}}}
\newcommand{\Hdd}[1]{\Phi_{#1}^\dagger}
\newcommand{\HdC}[1]{\tilde\Phi_{#1}^{\phantom{\dagger}}}
\newcommand{\Hdc}[1]{\Phi_{#1}^{\phantom{\dagger}\!\!\!\ast}}

\newcommand{\nHH}{{H}^0}
\newcommand{\nHR}{{R}^0}
\newcommand{\nHI}{{I}^0}
\newcommand{\nh}{{\rm h}}
\newcommand{\nH}{{\rm H}}
\newcommand{\nA}{{\rm A}}
\newcommand{\cH}{{\rm H}^\pm}
\newcommand{\cHm}{{\rm H}^-}
\newcommand{\cHp}{{\rm H}^+}
\newcommand{\mh}{m_{\nh}}
\newcommand{\mH}{m_{\nH}}
\newcommand{\mA}{m_{\nA}}
\newcommand{\mcH}{m_{\cH}}
\newcommand{\mNSc}{\mathcal M_0^2}
\newcommand{\mNScT}{{\mathcal M_0^{2}}^T}

\newcommand{\cb}{c_\beta}
\renewcommand{\sb}{s_\beta}
\newcommand{\cbb}{c_{2\beta}}
\newcommand{\sbb}{s_{2\beta}}

\newcommand{\cbCP}{c_{\beta}}
\newcommand{\sbCP}{s_{\beta}}

\newcommand{\tb}{t_\beta}
\newcommand{\tbinv}{\tb^{-1}}
\newcommand{\tti}{\tb+\tbinv}

\newcommand{\vev}[1]{v_{#1}}

\newcommand{\tCP}{\theta}
\newcommand{\stCP}{s_{\theta}}
\newcommand{\sttCP}{s_{2\theta}}
\newcommand{\ctCP}{c_{\theta}}
\newcommand{\ROTmat}{\mathcal R}\newcommand{\ROTmatT}{\ROTmat^T}\newcommand{\ROTmatinv}{\ROTmat^{-1}}
\newcommand{\ROT}[1]{\ROTmat_{#1}}

\newcommand{\HbROT}{\mathcal R_{\beta}^{\phantom{T}}}\newcommand{\HbROTt}{\mathcal R_{\beta}^T}\newcommand{\HbROTinv}{\mathcal R_{\beta}^{-1}}
\newcommand{\CKM}{V}\newcommand{\CKMd}{\CKM^\dagger}
\newcommand{\V}[1]{{\CKM_{#1}^{\phantom{\ast}}}}
\newcommand{\Vc}[1]{{\CKM_{#1}^\ast}}

\newcommand{\mMQ}[1]{M_{#1}^{\phantom{\dagger}}}
\newcommand{\mMU}{\mMQ{u}}
\newcommand{\mMD}{\mMQ{d}}
\newcommand{\wMQ}[1]{M_{#1}^{0}}\newcommand{\wMQd}[1]{M_{#1}^{0\dagger}}
\newcommand{\wMU}{\wMQ{u}}\newcommand{\wMUd}{\wMQd{u}}
\newcommand{\wMD}{\wMQ{d}}\newcommand{\wMDd}{\wMQd{d}}
\newcommand{\hwMQ}[1]{\hat M_{#1}^{0}}\newcommand{\hwMQt}[1]{\hat M_{#1}^{0\,T}}
\newcommand{\hwMU}{\hwMQ{u}}
\newcommand{\hwMD}{\hwMQ{d}}\newcommand{\hwMDt}{\hwMQt{d}}

\newcommand{\mNQ}[1]{N_{#1}^{\phantom{\dagger}}}\newcommand{\mNQd}[1]{N_{#1}^{\dagger}}
\newcommand{\mNU}{\mNQ{u}}\newcommand{\mNUd}{\mNQd{u}}
\newcommand{\mND}{\mNQ{d}}\newcommand{\mNDd}{\mNQd{d}}
\newcommand{\wNQ}[1]{N_{#1}^{0}}
\newcommand{\wNU}{\wNQ{u}}
\newcommand{\wND}{\wNQ{d}}
\newcommand{\mHQ}[1]{H_{#1}}\newcommand{\mAQ}[1]{A_{#1}}
\newcommand{\mHD}{\mHQ{d}}\newcommand{\mAD}{\mAQ{d}}
\newcommand{\mHU}{\mHQ{u}}\newcommand{\mAU}{\mAQ{u}}
\newcommand{\Zn}[1]{\mathbb{Z}_{#1}}
\newcommand{\ZZ}{\Zn{2}}
\newcounter{notas}
\graphicspath{{Figs/}}

\begin{document}

\hfill\begin{minipage}[r]{0.3\textwidth}\begin{flushright}  CFTP/18-012\\    IFIC/18-nnn \end{flushright} \end{minipage}

\begin{center}

\vspace{0.50cm}

{\large \bf {Vacuum Induced CP Violation Generating a Complex CKM Matrix with Controlled Scalar FCNC}}

\vspace{0.50cm}

Miguel Nebot $^{a,}$\footnote{\texttt{miguel.r.nebot.gomez@tecnico.ulisboa.pt}}
Francisco J. Botella $^{b,}$\footnote{\texttt{Francisco.J.Botella@uv.es}},
Gustavo C. Branco $^{a,}$\footnote{\texttt{gbranco@tecnico.ulisboa.pt}},

\end{center}
\vspace{0.50cm}
\begin{flushleft}
\emph{$^a$ Departamento de F\'\i sica and Centro de F\' \i sica Te\' orica de Part\' \i culas (CFTP),\\
\quad Instituto Superior T\' ecnico (IST), U. de Lisboa (UL),\\ 
\quad Av. Rovisco Pais 1, P-1049-001 Lisboa, Portugal.} \\
\emph{$^b$ Departament de F\`\i sica Te\`orica and Instituto de F\' \i sica Corpuscular (IFIC),\\
\quad Universitat de Val\`encia -- CSIC, E-46100 Valencia, Spain.}
\end{flushleft}
\vspace{0.5cm}

\begin{abstract}
We propose a viable minimal model with spontaneous CP violation in the framework of a Two Higgs Doublet Model. The model is based on a generalised Branco-Grimus-Lavoura model with a flavoured $\mathbb{Z}_2$ symmetry, under which two of the quark families are even and the third one is odd. The lagrangian respects CP invariance, but the vacuum has a CP violating phase, which is able to generate a complex CKM matrix, with the rephasing invariant strength of CP violation compatible with experiment. The question of scalar mediated flavour changing neutral couplings is carefully studied. In particular we point out a deep connection between the generation of a complex CKM matrix from a vacuum phase and the appearance of scalar FCNC. The scalar sector is presented in detail, showing that the new scalars are necessarily lighter than 1 TeV. A complete analysis of the model including the most relevant constraints is performed, showing that it is viable and that it has definite implications for the observation of New Physics signals in, for example, flavour changing Higgs decays or the discovery of the new scalars at the LHC. We give special emphasis to processes like $t\to \nh c,\nh u$, as well as $\nh\to bs, bd$, which are relevant for the LHC and the ILC.
\end{abstract}



\clearpage
\section{Introduction\label{SEC:intro}}
The first model of spontaneous T and CP violation was proposed \cite{Lee:1973iz} by T.D. Lee in 1973 at a time when only two incomplete quark generations were known. The main motivation for Lee's seminal work was to put the breaking of CP and T on the same footing as the breaking of gauge symmetry. In Lee's model, the Lagrangian is CP and T invariant, but the vacuum violates these discrete symmetries. This was achieved through the introduction of two Higgs doublets, with vacuum expectation values with a relative phase which violates T and CP invariance.
In Lee's model, CP violation would arise solely from Higgs exchange, since at the time only two generations were known and therefore the CKM matrix was real. 
The general two Higgs Doublet Model (2HDM) \cite{Branco:2011iw,Ivanov:2017dad} has Scalar Flavour Changing Neutral Couplings (SFCNC) at tree level which need to be controlled in order to conform to the stringent experimental constraints. This can be achieved by imposing Natural Flavour Conservation (NFC) in the scalar sector, as suggested by Glashow and Weinberg (GW) \cite{Glashow:1976nt}. Alternatively, it was suggested by Branco, Grimus and Lavoura (BGL) \cite{Branco:1996bq} that one may have 2HDM with tree level SFCNC but with their flavour structure only dependent on the CKM matrix $\CKM$.

BGL models have been extensively analysed in the literature \cite{Botella:2009pq,Botella:2011ne,Bhattacharyya:2014nja,Botella:2014ska,Celis:2014iua,Botella:2015hoa}, and their phenomenological consequences have been studied, in particular in the context of LHC.
Recently BGL models have been generalised \cite{Alves:2017xmk} in the framework of 2HDM. 
Both the GW and the BGL schemes can be implemented through the introduction of extra symmetries in the 2HDM. On the other hand, it has been shown \cite{Branco:1980sz} that the introduction of these symmetries in the 2HDM prevents the generation of either spontaneous or explicit CP violation in the scalar sector, unless they are softly broken \cite{Branco:1985aq}. It was recently discussed \cite{Branco:2015bfb} that for a scalar potential with an extra symmetry beyond gauge symmetry, there is an intriguing correlation between the capability of the potential to generate explicit and spontaneous CP violation.

In this paper we propose a realistic model of spontaneous CP violation in the framework of 2HDM. At this stage it is worth recalling the obstacles which have to be surmounted by any model of spontaneous CP violation:
\begin{itemize}
\item[(i)] The scalar potential should be able to generate spontaneous CP breaking by a phase of the vacuum, denoted $\theta$.
\item[(ii)] The phase $\theta$ should be able to generate a complex CKM matrix, with the strength of CP violation compatible with experiment. Recall that the 
CKM matrix has to be complex even in the presence of New Physics \cite{Botella:2005fc}.
\item[(iii)] SFCNC effects should be under control so that they do not violate experimental bounds.
\end{itemize}
The paper is organised as follows. In the next section we present the structure of the model and specify the flavoured symmetry introduced. In the third section we show how a complex CKM matrix is generated from the vacuum phase. Section \ref{SEC:scalar} contains a detailed analysis of the scalar potential with real couplings. In section \ref{SEC:Yukawa} we derive the physical Yukawa couplings and the phenomenological analysis of the model is presented in section \ref{SEC:Pheno}. Finally we present our conclusions in the last section.

\section{The Structure of the Model and the Flavoured symmetry\label{SEC:model}}
The Yukawa couplings in the 2HDM read
\begin{equation}\label{eq:Yukawa:00}
\mathscr L_{\rm Y}=
-\bar Q_L^0(\Yd{1}\Hd{1}+\Yd{2}\Hd{2}) d_R^0
-\bar Q_L^0(\Yu{1}\HdC{1}+\Yu{2}\HdC{2}) u_R^0+\text{H.c.}\,,
\end{equation}
with summation over generation indices understood and $\HdC{j}=i\sigma_2\Hdc{j}$.
We consider the following $\ZZ$ transformations to define the model:
\begin{align}
&\Hd{1}\mapsto \Hd{1},\quad \Hd{2}\mapsto -\Hd{2},\quad Q_{L3}^0\mapsto -Q_{L3}^0,\quad Q_{Lj}^0\mapsto Q_{Lj}^0,\quad j=1,2,\nonumber\\
&d_{Rk}^0\mapsto d_{Rk}^0,\quad u_{Rk}^0\mapsto u_{Rk}^0,\quad k=1,2,3.
\label{eq:Symmetry:00}
\end{align}
Invariance under \refEQ{eq:Symmetry:00} gives the following form of the Yukawa coupling matrices:
\begin{alignat}{3}
\nonumber
&\Yd{1}=\begin{pmatrix}\times&\times&\times\\ \times&\times&\times\\ 0&0&0\end{pmatrix},&\quad
&\Yd{2}=\begin{pmatrix}0&0&0\\ 0&0&0\\ \times&\times&\times\end{pmatrix},\\
&\Yu{1}=\begin{pmatrix}\times&\times&\times\\ \times&\times&\times\\ 0&0&0\end{pmatrix},&\quad
&\Yu{2}=\begin{pmatrix}0&0&0\\ 0&0&0\\ \times&\times&\times\end{pmatrix}.
\label{eq:Yukawa:Texture:00}
\end{alignat}
The symmetry assignment in \refEQ{eq:Symmetry:00} and the Yukawa matrices in \refEQ{eq:Yukawa:Texture:00} correspond to the generalised BGL models introduced in \cite{Alves:2017xmk}. We impose CP invariance at the Lagrangian level, so we require the Yukawa couplings to be real:
\begin{equation}\label{eq:Yukawas:real}
\Ydc{j}=\Yd{j},\qquad \Yuc{j}=\Yu{j}\,.
\end{equation}
We write the scalar doublets $\Hd{j}$ in the ``Higgs basis'' $\{H_{1},H_{2}\}$ \cite{Georgi:1978ri,Donoghue:1978cj,Botella:1994cs} (see section \ref{SEC:scalar} and appendix \ref{APP:scalar} for further details on the scalar sector)
\begin{equation}\label{eq:HiggsBasis:00}
\begin{pmatrix}H_{1}\\ H_{2}\end{pmatrix}=\HbROT\,
\begin{pmatrix}e^{-i\theta_1}\Hd{1}\\ e^{-i\theta_2}\Hd{2}\end{pmatrix},\quad \text{with}\quad 
\HbROT=\begin{pmatrix}\phantom{-}\cb & \sb\\ -\sb & \cb \end{pmatrix},\ \HbROTt=\HbROTinv.
\end{equation}
In this basis, only $H_1$ acquires a vacuum expectation value
\begin{equation}\label{eq:HiggsBasis:01}
\langle H_{1}\rangle=\frac{v}{\sqrt 2}\begin{pmatrix} 0\\ 1\end{pmatrix},\quad \langle H_{2}\rangle=\begin{pmatrix} 0\\ 0\end{pmatrix}.
\end{equation}
Equation \eqref{eq:Yukawa:00} can then be rewritten as
\begin{equation}\label{eq:Yukawa:01}
\mathscr L_{\rm Y}=
-\frac{\sqrt 2}{v}\bar Q_L^0(\wMD H_{1}+\wND H_{2}) d_R^0
-\frac{\sqrt 2}{v}\bar Q_L^0(\wMU \tilde H_{1}+\wNU \tilde H_{2}) u_R^0+\text{H.c.}\,,
\end{equation}
where the quark mass matrices $\wMD$, $\wMU$ and the $\wND$, $\wNU$ matrices read
\begin{alignat}{3}
\wMD&=\frac{ve^{i\theta_1}}{\sqrt 2}(\cb\Yd{1}+e^{i\theta}\sb\Yd{2})\, ,&\quad \wND&=\frac{ve^{i\theta_1}}{\sqrt 2}(-\sb\Yd{1}+e^{i\theta}\cb\Yd{2})\, ,\label{eq:MN:matrices:00:1}\\
\wMU&=\frac{ve^{-i\theta_1}}{\sqrt 2}(\cb\Yu{1}+e^{-i\theta}\sb\Yu{2})\, ,&\quad \wNU&=\frac{ve^{-i\theta_1}}{\sqrt 2}(-\sb\Yu{1}+e^{-i\theta}\cb\Yu{2})\,,\label{eq:MN:matrices:00:2}
\end{alignat}
where $\theta=\theta_2-\theta_1$ is the relative phase among $\langle\Hd{2}\rangle$ and $\langle\Hd{1}\rangle$. For simplicity, we remove the irrelevant global phases $e^{\pm i\theta_1}$ setting $\theta_1=0$.\\ %
Notice that the matrices $\wND$, $\wNU$ can be written:
\begin{align}
\wND&=\tb\wMD+e^{i\theta}\frac{v}{\sqrt 2}(\tti)\sb\Yd{2}=\tb\wMD-(\tti)\PR{3}\wMD\,,\label{eq:MN:matrices:01:1}\\
\wNU&=\tb\wMU+e^{-i\theta}\frac{v}{\sqrt 2}(\tti)\sb\Yu{2}=\tb\wMU-(\tti)\PR{3}\wMU\,,\label{eq:MN:matrices:01:2}
\end{align}
where $\PR{3}$ is the projector
\begin{equation}\label{eq:P3:00}
\PR{3}=\begin{pmatrix}0&0&0\\ 0&0&0\\ 0&0&1\end{pmatrix}\,.
\end{equation}

\section{Generation of a complex CKM matrix from the vacuum phase\label{SEC:CKM}}
In this section, we show how the vacuum phase $\theta$ is capable of generating a complex CKM matrix. As previously emphasized, this is a necessary requirement for the model to be consistent with experiment.
Following \refEQS{eq:Yukawa:Texture:00}-\eqref{eq:Yukawas:real} and \eqref{eq:MN:matrices:00:1}-\eqref{eq:MN:matrices:00:2}, we write:
\begin{equation}\label{eq:RedMass:real:00}
\wMD=\begin{pmatrix}1&0&0\\ 0&1&0\\ 0&0& e^{i\theta}\end{pmatrix}\hwMD,\qquad \wMU=\begin{pmatrix}1&0&0\\ 0&1&0\\ 0&0& e^{-i\theta}\end{pmatrix}\hwMU,
\end{equation}
with $\hwMD$ and $\hwMU$ \emph{real}. 
Then,
\begin{equation}\label{eq:RedMass:fact}
\wMD\wMDd=\begin{pmatrix}1&0&0\\ 0&1&0\\ 0&0& e^{i\theta}\end{pmatrix}\hwMD\hwMDt\begin{pmatrix}1&0&0\\ 0&1&0\\ 0&0& e^{-i\theta}\end{pmatrix}
\end{equation}
with $\hwMD\hwMDt$ \emph{real} and \emph{symmetric}, which is diagonalised with a real orthogonal transformation:
\begin{equation}\label{eq:RedMass:diag}
\OdLt\,\hwMD\hwMDt\,\OdL=\text{diag}(m_{d_i}^2)\,.
\end{equation}
Consequently, \refEQ{eq:RedMass:fact} gives
\begin{equation}\label{eq:Mass:diag:01}
\OdLt\begin{pmatrix}1&0&0\\ 0&1&0\\ 0&0& e^{-i\theta}\end{pmatrix}\wMD\wMDd\begin{pmatrix}1&0&0\\ 0&1&0\\ 0&0& e^{i\theta}\end{pmatrix}\OdL=
\text{diag}(m_{d_i}^2)\,.
\end{equation}
That is, the diagonalisation of $\wMD\wMDd$ is accomplished with
\begin{equation}\label{eq:Mass:diag:03}
\UdLd\wMD\wMDd\UdL=\text{diag}(m_{d_i}^2)\,,\quad\text{where}\quad \UdL=\begin{pmatrix}1&0&0\\ 0&1&0\\ 0&0& e^{i\theta}\end{pmatrix}\OdL\,.
\end{equation}
Similarly,
\begin{equation}\label{eq:Mass:diag:04}
\UuLd\wMU\wMUd\UuL=\text{diag}(m_{u_i}^2)\,,\quad\text{with}\quad\UuL=\begin{pmatrix}1&0&0\\ 0&1&0\\ 0&0& e^{-i\theta}\end{pmatrix}\OuL\,.
\end{equation}
Notice the important sign difference in $\theta$ between \refEQS{eq:Mass:diag:03} and \eqref{eq:Mass:diag:04}, which give the following CKM matrix $\CKM\equiv\UuLd\UdL$,
\begin{equation}\label{eq:CKM}
\CKM=\OuLt\begin{pmatrix}1&0&0\\ 0&1&0\\ 0&0& e^{i2\theta}\end{pmatrix}\OdL\,.
\end{equation}
Notice also that, if $e^{i2\theta}= \pm 1$, $\CKM$ is real, i.e. it does not generate CP violation. This can be understood through a careful analysis of the potential, which will be presented in section \ref{SEC:scalar}. The model we present here has spontaneous CP violation and thus a physical phase in the CKM matrix can only arise from $\theta$. In section \ref{sSEC:min} we show that for $\theta=\pi/2$ the vacuum is CP invariant and no CP violation can be generated in this model. In particular CKM is necessarily real for this value of $\theta$, as noticed in \refEQ{eq:CKM}.


It is also straightforward to observe that $\wMDd\wMD$ and $\wMUd\wMU$ are real and symmetric, and are thus diagonalised with real orthogonal matrices $\OdR$ and $\OuR$,
\begin{equation}\label{eq:Mass:diag:05}
\OdRt\wMDd\wMD\OdR=\text{diag}(m_{d_i}^2),\qquad \OuRt\wMUd\wMU\OuR=\text{diag}(m_{u_i}^2)\,,
\end{equation}
such that the bi-diagonalisation of $\wMD$ and $\wMU$ reads
\begin{equation}
\mMD=\text{diag}(m_{d_i})=\UdLd\wMD\OdR,\quad \mMU=\text{diag}(m_{u_i})=\UuLd\wMU\OuR\,.
\end{equation}
Following \refEQ{eq:MN:matrices:01:1},
\begin{multline}\label{eq:N:diag:00}
\mND\equiv\UdLd\wND\OdR=\tb\UdLd\wMD\OdR-(\tti)\UdLd\PR{3}\wMD\OdR=\\
\tb\mMD-(\tti)\UdLd\PR{3}\UdL\mMD\,,
\end{multline}
with $\PR{3}$ the projector in \refEQ{eq:P3:00} and $\UdL$ in \refEQ{eq:Mass:diag:03}.
In the last term of \refEQ{eq:N:diag:00},
\begin{equation}\label{eq:PR3:01}
\UdLd\PR{3}\UdL=\OdLt\PR{3}\OdL\,,
\end{equation}
that is, $\mND$ in \refEQ{eq:N:diag:00} is \emph{real}. 
 Introducing a real unit vector $\rndvec$ and a complex unit vector $\undvec$ with components
\begin{equation}\label{eq:ndvec:00}
\rnd{j}\equiv [\OdL]_{3j}\,,\quad \und{j}\equiv[\UdL]_{3j}=e^{i\theta}\rnd{j}\,,
\end{equation}
one has, for $\UdLd\PR{3}\UdL$ in \refEQ{eq:PR3:01},
\begin{equation}\label{eq:ndnd:00}
[\UdLd\PR{3}\UdL]_{ij}=\undC{i}\und{j}=\rnd{i}\rnd{j}\,.
\end{equation}
Similarly, for $\mNU$ we have
\begin{equation}\label{eq:N:diag:01}
\mNU\equiv\UuLd\wNU\OuR=\tb\mMU-(\tti)\UuLd\PR{3}\UuL\mMU\,,
\end{equation}
with 
\begin{equation}\label{eq:PR3:02}
\UuLd\PR{3}\UuL=\OuLt\PR{3}\OuL\,,
\end{equation}
and
\begin{equation}\label{eq:nuvec:00}
\rnu{j}\equiv[\OuL]_{3j}\,\quad \unu{j}\equiv[\UuL]_{3j}=e^{-i\theta}\rnu{j}\,,\quad [\UuLd\PR{3}\UuL]_{ij}=\unuC{i}\unu{j}=\rnu{i}\rnu{j}\,.
\end{equation}
Like $\mND$, $\mNU$ is \emph{real}; $\mND$ and $\mNU$ have the form:
\begin{align}
[\mND]_{ij}&=\tb\delta_{ij}m_{d_i}-(\tti)\undC{i}\und{j}m_{d_j}\, ,\label{eq:Nq:gen:00:1}\\
[\mNU]_{ij}&=\tb\delta_{ij}m_{u_i}-(\tti)\unuC{i}\unu{j}m_{u_j}\, .\label{eq:Nq:gen:00:2}
\end{align}
Since $\CKM=\UuLd\UdL$, the complex unitary vectors $\undvec$ and $\unuvec$ are not independent:
\begin{equation}\label{eq:ndnu:00}
\und{i}=\unu{j}\V{ji}\,,\quad \unu{i}=\Vc{ij}\und{j}\,.
\end{equation}
%
It is interesting to notice that the 2HDM scenario studied in \cite{Ferreira:2011xc}, where the soft breaking of a $\Zn{3}$ symmetry is the source of CP violation, shares some interesting properties with the present one: there, the CKM matrix can also be factorised in terms of real orthogonal rotations and a diagonal matrix containing the CP violating depence; the tree level SFCNC are also real in that phase convention. Other aspects of the model like the structure of the Yukawa couplings as well as the scalar sector to be discussed in section \ref{SEC:scalar} are, however, completely different.\\
In the rest of this section, we analyse in detail the generation of a complex CKM matrix from the vacuum phase $\theta$. The couplings of the physical scalars to the fermions are discussed in section \ref{SEC:Yukawa}, after the discussion of the scalar sector in \ref{SEC:scalar}.

It is clear that $e^{i2\theta}\neq \pm 1$ is necessary in order to have an irreducibly complex CKM matrix. However, one has to verify that one can indeed obtain a realistic CKM matrix, one that it is in agreement with the experimental constraints on the moduli $\abs{\V{ij}}$ (in particular of the moduli of the first and second rows), and on the CP violating phase $\gamma\equiv\arg(-\V{ud}\Vc{ub}\V{cb}\Vc{cd})$ (the only one accessible through tree level processes alone). Concerning CP violation, one can alternatively analyse that the unique (up to a sign) imaginary part of a rephasing invariant quartet $\im{\V{i_1j_1}\Vc{i_1j_2}\V{i_2j_2}\Vc{i_2j_1}}$ ($i_1\neq i_2$, $j_1\neq j_2$) has the correct size $\sim 3\times 10^{-5}$. Starting with \refEQ{eq:CKM}, one can compute that imaginary part. For the task, it is convenient to trade $\OdL$ and $\OuL$ for the real unit vector $\rndvec$ in \refEQ{eq:ndvec:00} and the real orthogonal matrix $\OROTmat$:
\begin{equation}
\rnd{j}=[\OdL]_{3j},\quad \OROTmat\equiv\OuLt\OdL\,.
\end{equation}
Then, one can rewrite
\begin{equation}\label{EQ:CKM:ROT:rd:ru}
\CKM=\OuLt[\id+(e^{i2\theta}-1)\PR{3}]\OdL\,\Rightarrow
 \V{ij}=\OROT{ij}+(e^{i2\theta}-1)\SROT{ij}\,,
\end{equation}
and we introduce $\SROT{ij}$ to allow for compact expressions:
\begin{equation}\label{EQ:CKM:Sij:00}
\SROT{ij}\equiv[\OuLt\PR{3}\OdL]_{ij}=\rnu{i}\rnd{j}=\sum_{k=1}^3\OROT{ik}\rnd{k}\rnd{j}=\sum_{k=1}^3\rnu{i}\rnu{k}\OROT{kj}\,.
\end{equation}
The real and imaginary parts of $\V{ij}$ are\footnote{Here and in the following $c_x\equiv \cos x$, $s_x\equiv \sin x$.}
\begin{equation}\label{eq:ReIm:Vij:00}
\re{\V{ij}}=\OROT{ij}-2\stCP^2\SROT{ij},\quad \im{\V{ij}}=\sttCP\SROT{ij}\,.
\end{equation}
Notice that, although \refEQ{eq:ReIm:Vij:00} is not rephasing invariant, this poses no problem when considering rephasing invariant quartets. With \refEQ{eq:ReIm:Vij:00}, one can obtain:
\begin{multline}\label{eq:ImQuartet:00}
\im{\V{i_1j_1}\Vc{i_1j_2}\V{i_2j_2}\Vc{i_2j_1}}=\sin 2\theta\Big\{
4\stCP^2(\OROT{i_1j_1}\SROT{i_2j_1}\OROT{i_2j_2}\SROT{i_1j_2}-\SROT{i_1j_1}\OROT{i_2j_1}\SROT{i_2j_2}\OROT{i_1j_2})+\\
4\stCP^2(\SROT{i_1j_1}\SROT{i_2j_1}\SROT{i_2j_2}\OROT{i_1j_2}-\SROT{i_1j_1}\SROT{i_2j_1}\OROT{i_2j_2}\SROT{i_1j_2}+\SROT{i_1j_1}\OROT{i_2j_1}\SROT{i_2j_2}\SROT{i_1j_2}-\OROT{i_1j_1}\SROT{i_2j_1}\SROT{i_2j_2}\SROT{i_1j_2})\\
+\SROT{i_1j_1}\OROT{i_2j_1}\OROT{i_2j_2}\OROT{i_1j_2}-\OROT{i_1j_1}\SROT{i_2j_1}\OROT{i_2j_2}\OROT{i_1j_2}+\OROT{i_1j_1}\OROT{i_2j_1}\SROT{i_2j_2}\OROT{i_1j_2}-\OROT{i_1j_1}\OROT{i_2j_1}\OROT{i_2j_2}\SROT{i_1j_2}\Big\}\,.
\end{multline}
Although \refEQ{eq:ImQuartet:00} is not very illuminating, one can nevertheless illustrate that realistic values of $\im{\V{i_1j_1}\Vc{i_1j_2}\V{i_2j_2}\Vc{i_2j_1}}$ can be obtained even in cases with less parametric freedom, as done in subsection \ref{sSEC:CKMexample} below. The general case is analysed in subsection \ref{sSEC:CKMnumeric}. Before addressing those questions, we discuss two important aspects that deserve attention in the next two subsections: (i) the number of independent parameters and the most convenient choice for them, (ii) the fact that in this model, if tree level SFCNC were completely absent in one quark sector, then the CKM matrix would not be CP violating. One encounters again a deep connection \cite{Branco:1979pv} between the complexity of CKM and SFCNC, in the context of models with spontaneous CP violation.

\subsection{Parameters\label{sSEC:Parameters}}
\begin{figure}[h!tb]
\begin{center}
\includegraphics[width=0.3\textwidth]{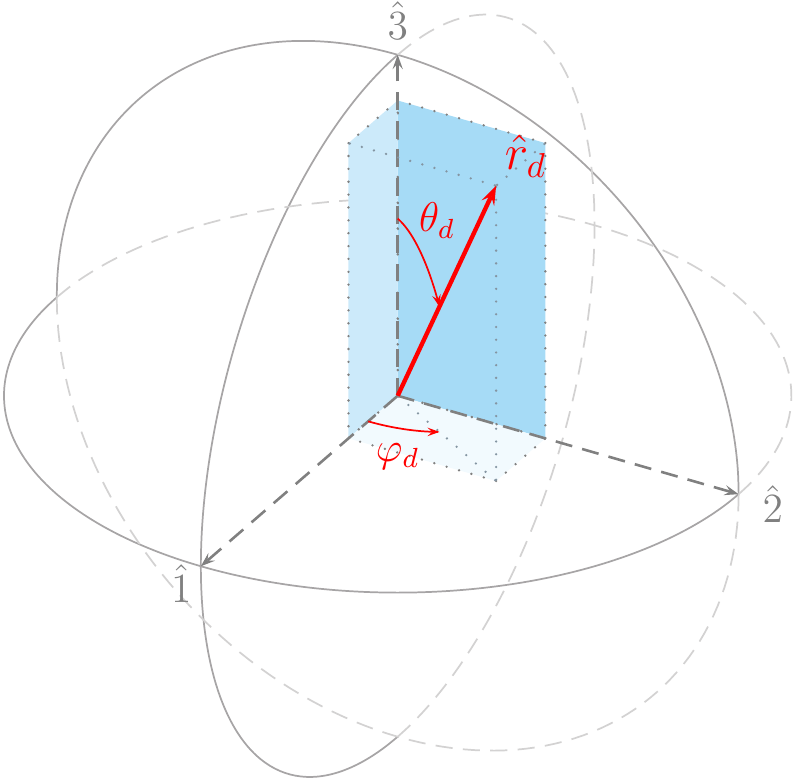}
\caption{$\rndvec$.\label{FIG:par:rd}}
\end{center}
\end{figure}

The CKM matrix $\CKM$ requires 4 physical parameters, while the tree level SFCNC require 2, since $\rndvec$ is a unit real vector. One can parametrise $\rndvec$ in terms of two angles $\theta_d$, $\varphi_d$:
\begin{equation}
\rndvec = (\sin\theta_{d}\cos\varphi_{d},\,\sin\theta_{d}\sin\varphi_{d},\,\cos\theta_{d})\,,
\end{equation}
as shown in Figure \ref{FIG:par:rd}\footnote{The different products $\rnd{i}\rnd{j}$ controlling SFCNC are, simply, the areas of the shaded rectangular projections in the $(\hat i,\hat j)$ planes in Figure \ref{FIG:par:rd}.}. 
The orthogonal matrix $\OROTmat$ requires 3 real parameters; together with $\theta$ and $\rndvec$, these 6 parameters match the parameters necessary to describe $\CKM$ (4 parameters) and the products $\rnd{j}\rnd{k}$, $\rnu{j}\rnu{k}$ (2 parameters). However, in terms of $\OuL$ and $\OdL$, there are a priori 3+3 real parameters; together with $\theta$, 7 parameters in all. This apparent mismatch can be readily understood: a common redefinition
\begin{equation}\label{eq:OdL:OuL:Inv:00}
\OdL\mapsto \left(\begin{smallmatrix}\cos\alpha & \sin\alpha & 0\\ -\sin\alpha & \cos\alpha & 0\\ 0& 0 & 1\end{smallmatrix}\right)\,\OdL,\qquad
\OuL\mapsto \left(\begin{smallmatrix}\cos\alpha & \sin\alpha & 0\\ -\sin\alpha & \cos\alpha & 0\\ 0& 0 & 1\end{smallmatrix}\right)\,\OuL,
\end{equation}
leaves $\OROTmat$, $\rndvec$, $\rnuvec$ and $\CKM$ unchanged, effectively removing one parameter from the $\OuL$, $\OdL$, parameter count. Consequently, it is convenient to adopt a parametrisation of $\OROTmat$ of the form
\begin{multline}\label{eq:Orth:Rot:00}
\OROTmat=
\begin{pmatrix}1&0&0\\ 0&c_{\alpha_3}&s_{\alpha_3}\\ 0&-s_{\alpha_3}&c_{\alpha_3}\end{pmatrix}
\begin{pmatrix}c_{\alpha_2}&0&s_{\alpha_2}\\0&1&0\\-s_{\alpha_2}&0&c_{\alpha_2}\end{pmatrix}
\begin{pmatrix}c_{\alpha_1}&s_{\alpha_1}&0\\-s_{\alpha_1}&c_{\alpha_1}&0\\ 0&0&1\end{pmatrix}\\=
\begin{pmatrix}c_{\alpha_1} c_{\alpha_2} & s_{\alpha_1}c_{\alpha_2} & s_{\alpha_2}\\ -s_{\alpha_1}c_{\alpha_3}-c_{\alpha_1}s_{\alpha_2}s_{\alpha_3} & c_{\alpha_1}c_{\alpha_3}-s_{\alpha_1}s_{\alpha_2}s_{\alpha_3} & c_{\alpha_2}s_{\alpha_3}\\ s_{\alpha_1}s_{\alpha_3}-c_{\alpha_1}s_{\alpha_2}c_{\alpha_3} & -c_{\alpha_1}s_{\alpha_3}-s_{\alpha_1}s_{\alpha_2}c_{\alpha_3} & c_{\alpha_2}c_{\alpha_3}\end{pmatrix}\,,
\end{multline}
and a parametrisation of $\OdL$ of the form\footnote{The rows in \refEQ{eq:OdL:Par:00} are simply unit vectors in spherical coordinates at $\vec r=\rndvec$, while $\alpha$, in \refEQ{eq:OdL:Par:00} as in \refEQ{eq:OdL:OuL:Inv:00}, produces an irrelevant rotation around $\rndvec$.}
\begin{equation}\label{eq:OdL:Par:00}
\OdL=
\begin{pmatrix}c_\alpha & s_\alpha & 0\\ -s_\alpha & c_\alpha & 0\\ 0& 0 & 1\end{pmatrix}
\begin{pmatrix}
c_{\theta_{d}}c_{\varphi_{d}} & c_{\theta_{d}}s_{\varphi_{d}} & -s_{\theta_{d}}\\
-s_{\varphi_{d}} & c_{\varphi_{d}} & 0\\
s_{\theta_{d}}c_{\varphi_{d}} & s_{\theta_{d}}s_{\varphi_{d}} & c_{\theta_{d}}
\end{pmatrix},
\end{equation}
where $\rndvec$ is readily identified in the third row and the redundant $\alpha$, as in \refEQ{eq:OdL:OuL:Inv:00}, can be set to $\alpha=0$. One can then concentrate on $\{\alpha_1,\alpha_2,\alpha_3,\theta_d,\varphi_d,\theta\}$ in order to reproduce a realistic CKM matrix.

\subsection{SFCNC and CP Violation in CKM\label{sSEC:FCNC:CKM}}
In \refEQS{eq:Nq:gen:00:1}--\eqref{eq:Nq:gen:00:2}, tree level SFCNC are a priori present in both the up and the down quark sectors and controlled by $\unC{q}{i}\un{q}{j}=\rn{q}{i}\rn{q}{j}$. Therefore, if $\rnvec{q}$ has a vanishing component, SFCNC in that sector ($q=u$ or $d$) do only appear in one type of transition (the one not involving that component). If $\rnvec{q}$ had two vanishing components (then the remaining one equals $\pm 1$), there would not be SFCNC in that sector: interstingly, in this model, having no tree level SFCNC in one quark sector is \emph{incompatible} with a CP violating CKM matrix. This can be readily checked by noticing that, in that case, in \refEQ{EQ:CKM:Sij:00}, the matrix with entries $S_{ij}$ has only a non vanishing row (column), corresponding to the absence of tree level SFCNC in the up (down) sector, for which $S_{ij}=\OROT{ij}$. Then, with $i_1\neq i_2$ and $j_1\neq j_2$, in \refEQ{eq:ImQuartet:00} all terms except two out of the last four automatically vanish, and those two terms appear with opposite sign, giving $\im{\V{i_1j_1}\Vc{i_1j_2}\V{i_2j_2}\Vc{i_2j_1}}=0$. As illustrated below, in subsection \ref{sSEC:CKMnumeric}, this implies that a lower bound on the size of the second largest component in $\rnvec{q}$ should exist, that is a lower bound on the intensity of some SFCNC in both quark sectors. Appendix \ref{APP:FCNC:CKM} completes the discussion of the interplay  in this model among flavour non-conservation and CP violation in the CKM matrix.

\subsection{A simple example\label{sSEC:CKMexample}}
As a simplified example of how a realistic CKM matrix can be obtained, consider a scenario with 
\begin{equation}\label{eq:nd:example:00}
\rndvec=(\cos\varphi_{d},\ \sin\varphi_{d},\ 0)\,.
\end{equation}
Then, for $i_1=j_1=1$, $i_2=j_2=2$, \refEQ{eq:ImQuartet:00} reduces to
\begin{equation}\label{eq:ImQuartet:01}
\im{\V{11}\Vc{12}\V{22}\Vc{21}}=\frac{1}{2}(\OROT{11}\OROT{21}+\OROT{12}\OROT{22})(\OROT{12}\OROT{21}-\OROT{11}\OROT{22})\sin 2\varphi_d\,\sin 2\tCP\,.
\end{equation}
Since the rows of $\OROTmat$ form a complete orthonormal set of 3-vectors,
\begin{equation}
\OROT{11}\OROT{21}+\OROT{12}\OROT{22}=-\OROT{13}\OROT{23},\qquad \OROT{12}\OROT{21}-\OROT{11}\OROT{22}=-\OROT{33},
\end{equation}
and \refEQ{eq:ImQuartet:02} is further reduced to
\begin{equation}\label{eq:ImQuartet:01b}
\im{\V{11}\Vc{12}\V{22}\Vc{21}}=\frac{1}{2}\OROT{13}\OROT{23}\OROT{33}\sin 2\varphi_d\,\sin 2\tCP\,.
\end{equation}
With $\OROTmat$ in \refEQ{eq:Orth:Rot:00}, \refEQ{eq:ImQuartet:01} gives
\begin{equation}\label{eq:ImQuartet:02}
\im{\V{11}\Vc{12}\V{22}\Vc{21}}=\frac{1}{8}\cos\alpha_2\,\sin 2\alpha_2\,\sin 2\alpha_3\,\sin 2\varphi_d\,\sin 2\tCP\,.
\end{equation}
A complete example of this type which reproduces correctly the CKM matrix, is given by: 
\begin{equation}
\theta=\pi/8,
\end{equation}
\begin{equation}\label{eq:R:example:00}
\OROTmat=
\begin{pmatrix}
1-7\times 10^{-6} & 0 & -3.746\times 10^{-3}\\ -1.536\times 10^{-4} & -1+8.41\times 10^{-4} & -0.041\\ -3.743\times 10^{-3} & 0.041 & -1+8.48\times 10^{-4}
\end{pmatrix}\,,
\end{equation}
and
\begin{equation}\label{eq:OdL:example:00}
\OdL=
\begin{pmatrix}
0 & 0 & -1\\ 0.9509 & 0.3096 & 0\\ 0.3096 & -0.9509 & 0
\end{pmatrix}\,.
\end{equation}
The parameters underlying \refEQS{eq:R:example:00}--\eqref{eq:OdL:example:00} have the values
\begin{align}
\OROTmat:&\quad \alpha_1=0,\ \alpha_2=-3.746\times 10^{-3},\ \alpha_3=0.041-\pi,\nonumber\\
\OdL:&\quad \theta_d=\pi/2,\ \varphi_d=-1.2561,\ \alpha=0\,.
\end{align}
For $\OROTmat$, $\alpha_1=0$ has been chosen for simplicity, since in this scenario it does not enter \refEQ{eq:ImQuartet:02}. With the previous values, one can easily check that
\begin{equation}
\abs{\V{us}}=0.2253,\ \abs{\V{ub}}=3.75\times 10^{-3},\ \abs{\V{cb}}=0.041,\ \im{\V{11}\Vc{12}\V{22}\Vc{21}}=3.195\times 10^{-5}\,.
\end{equation}
Concerning the discussion in subsection \ref{sSEC:FCNC:CKM}, this example shows that, although the complete absence of tree level FCNC in one sector is incompatible with a CP violating CKM matrix, this incompatibility does not extend to the case of tree level SFCNC circumscribed to only one type of transition (in this example, $d\leftrightarrow s$ transitions).

\subsection{General case\label{sSEC:CKMnumeric}}
\begin{figure}[h!tb]
\begin{center}
\subfigure[$\sin 2\theta$ vs. $\theta_d$\label{FIG:logs2tthd}]{\includegraphics[width=0.29\textwidth]{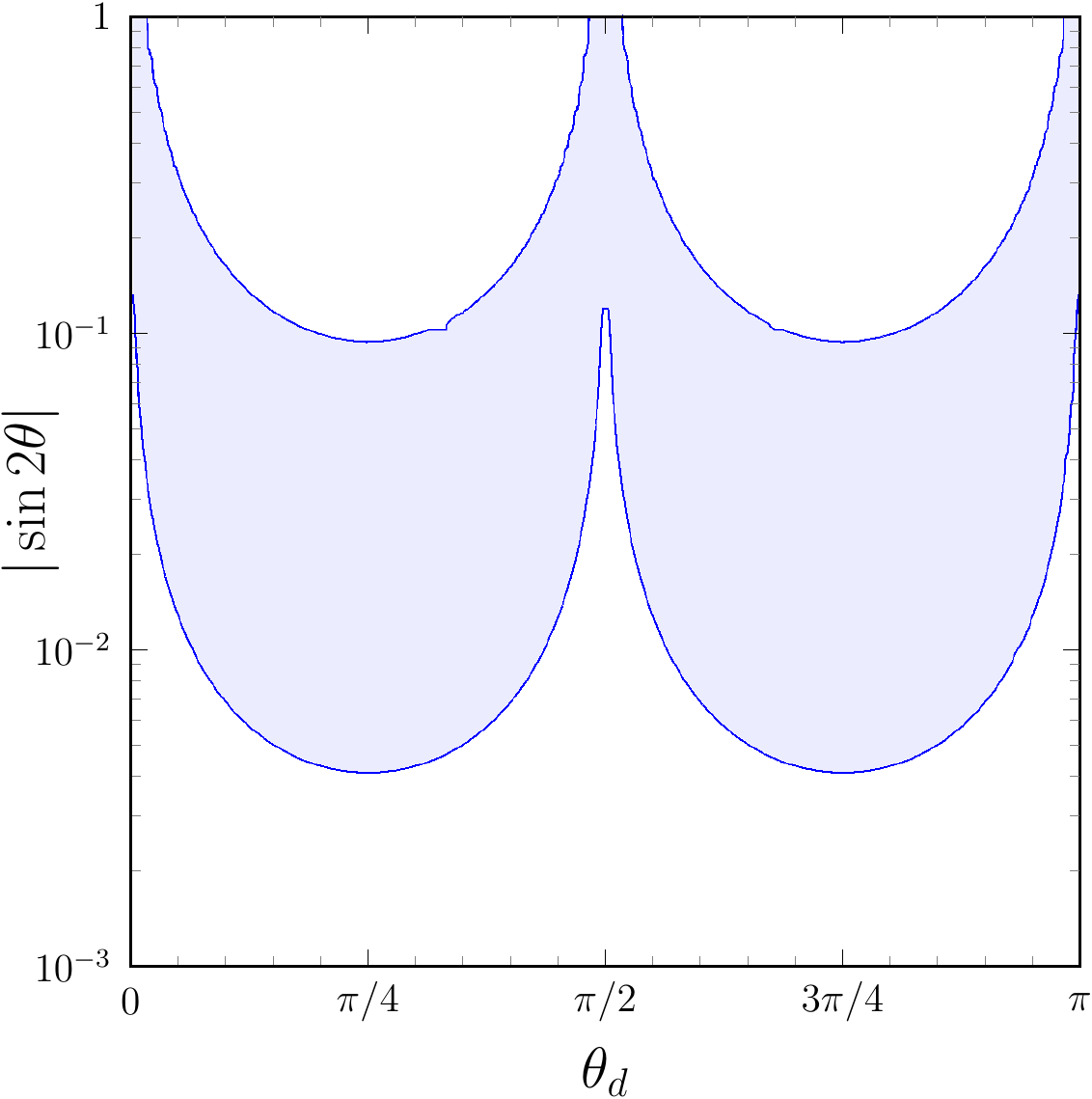}}\quad
\subfigure[$|\rnd{Min}|$ vs. $|\rnd{Mid}|$.\label{FIG:rd:Min}]{\includegraphics[width=0.3\textwidth]{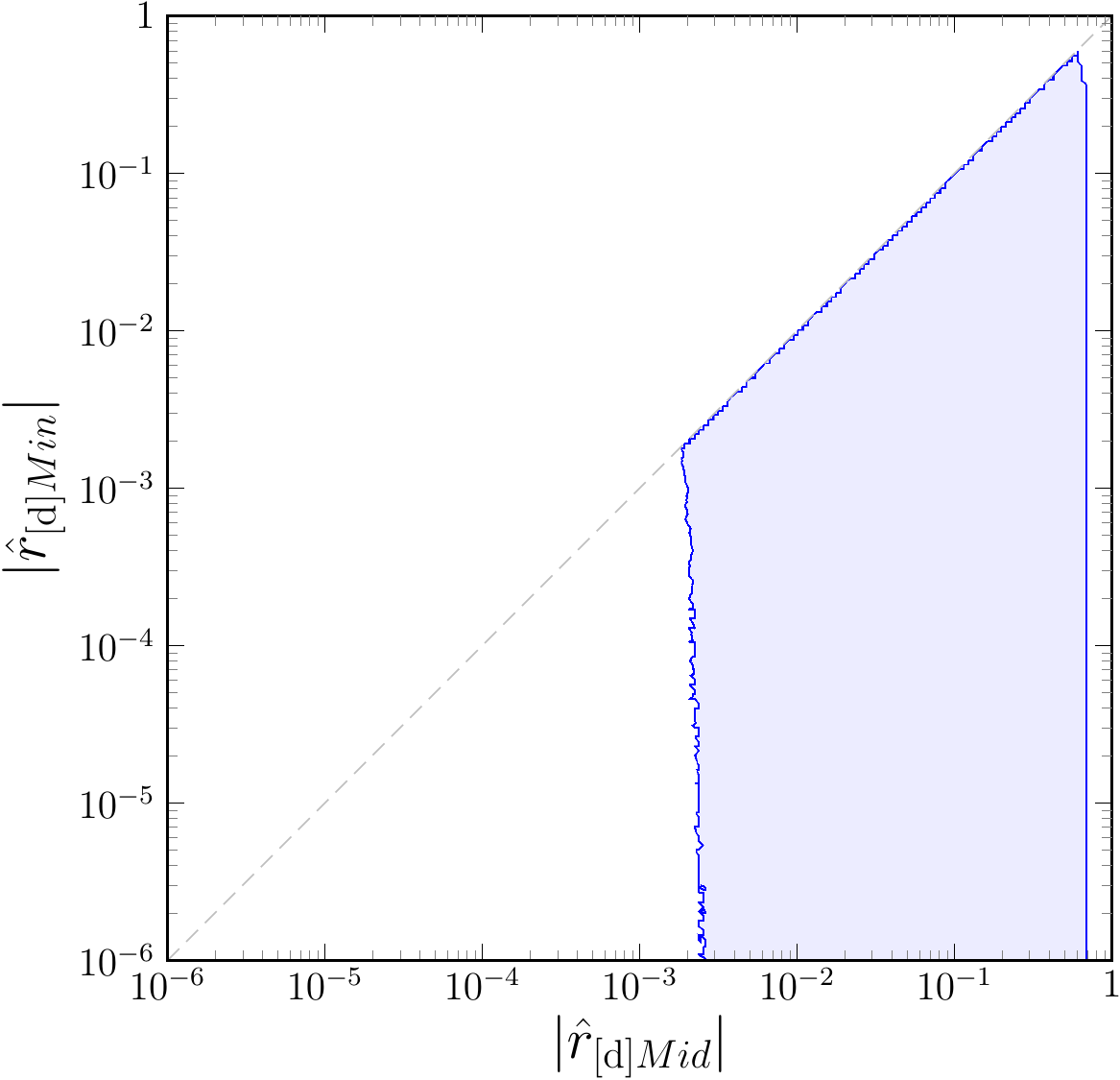}}\quad
\subfigure[$|\rnu{Min}|$ vs. $|\rnu{Mid}|$.\label{FIG:ru:Min}]{\includegraphics[width=0.3\textwidth]{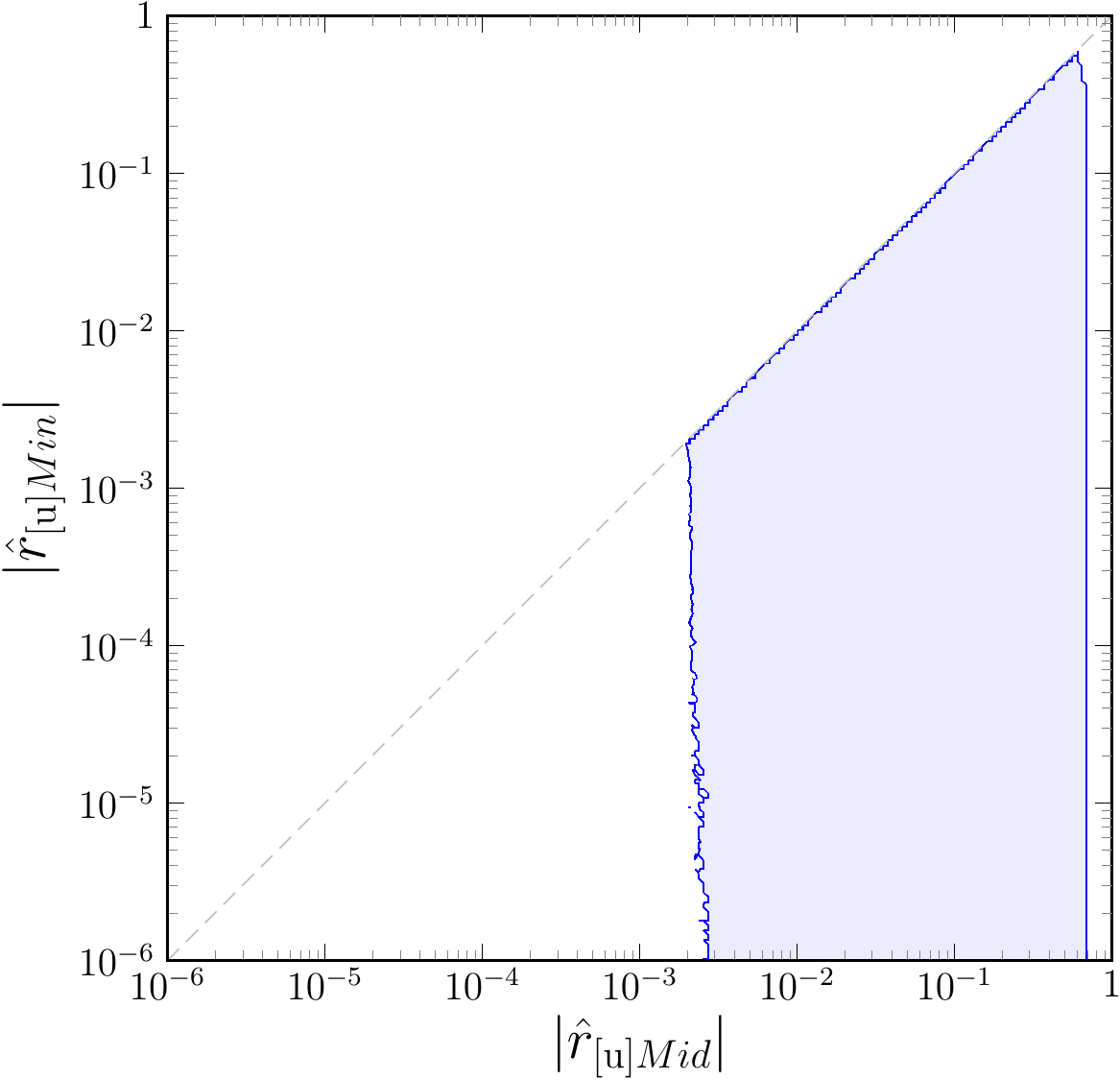}}\\
\subfigure[\label{FIG:goodCKM:1}]{\includegraphics[width=0.3\textwidth]{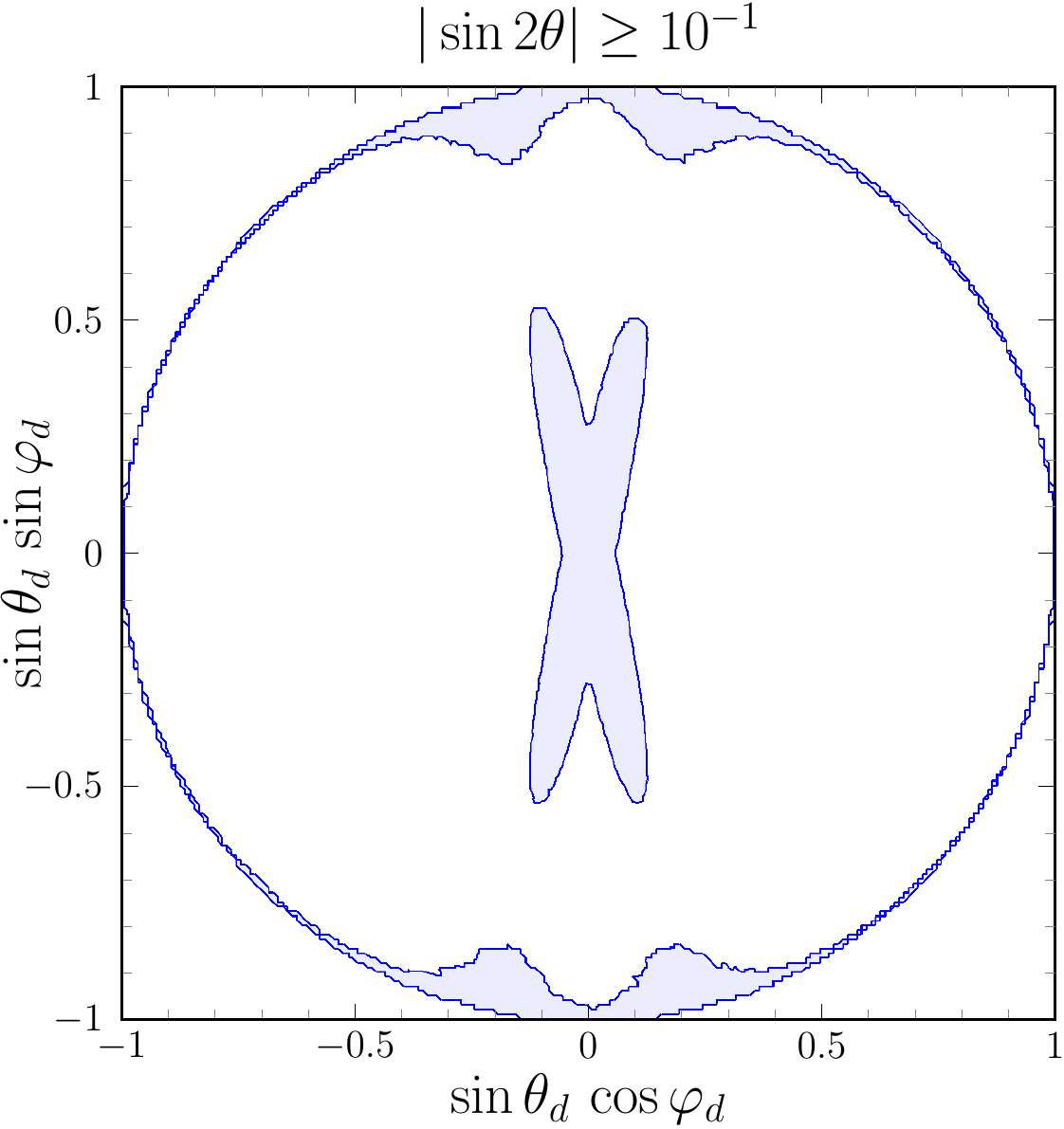}}\quad
\subfigure[\label{FIG:goodCKM:2}]{\includegraphics[width=0.3\textwidth]{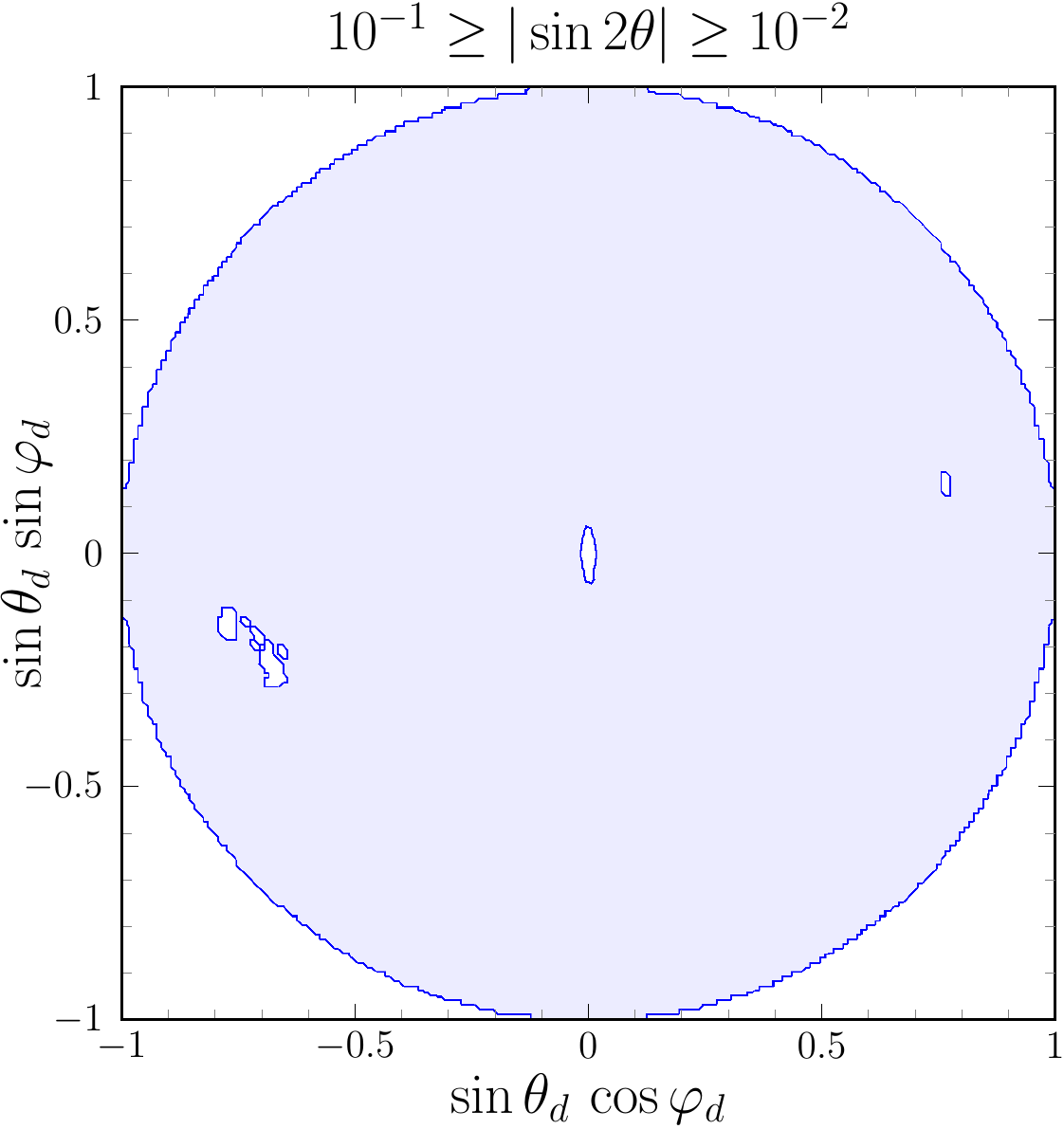}}\quad 
\subfigure[\label{FIG:goodCKM:3}]{\includegraphics[width=0.3\textwidth]{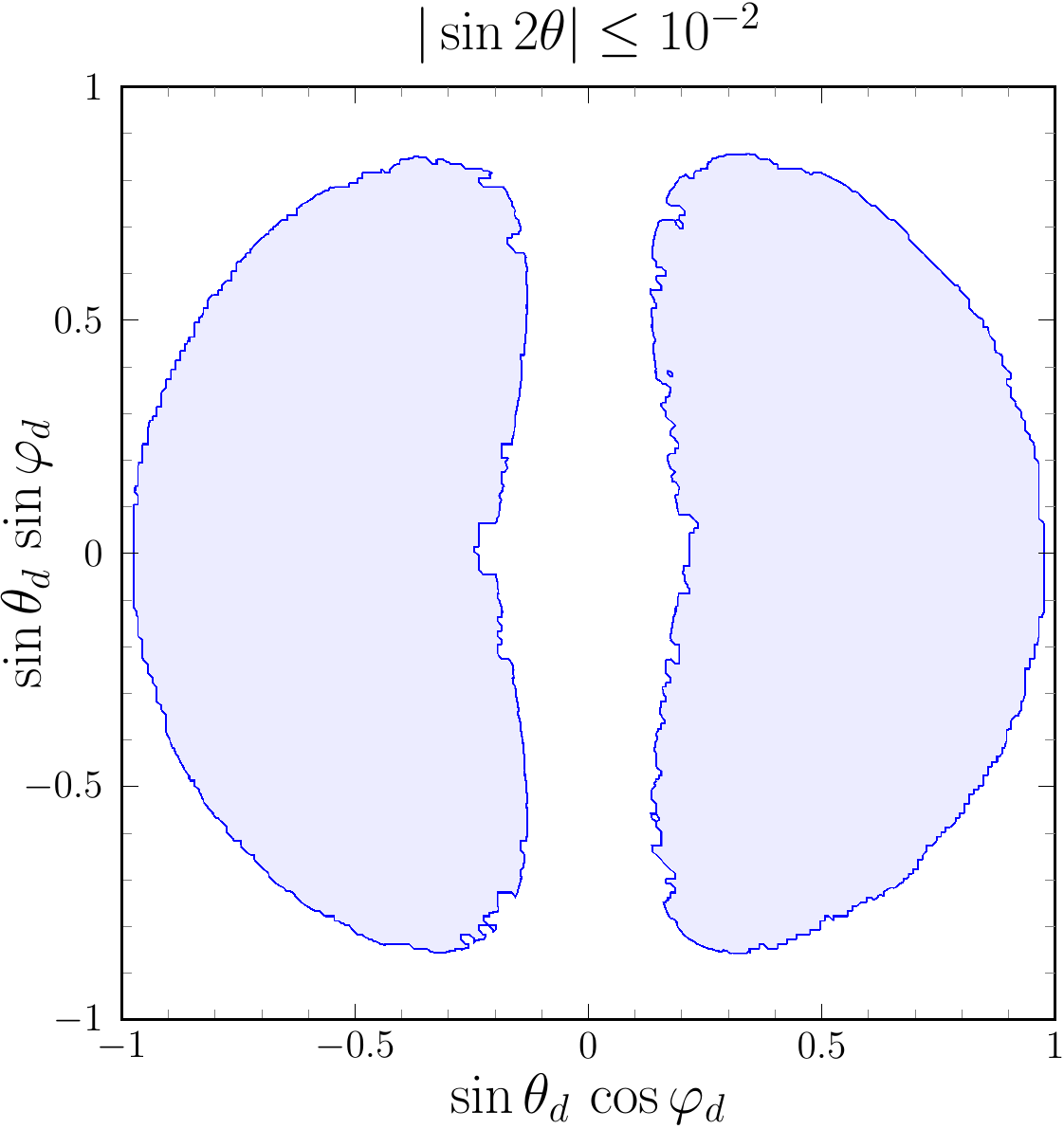}}
\caption{Regions (at 99\% C.L.) which reproduce a realistic CKM matrix are shown.\label{FIG:goodCKM}}
\end{center}
\end{figure}

The previous example illustrates that the CKM matrix can be adequately reproduced even in a restricted scenario where one has less number of free parameters. For the general case one can explore with a simple numerical analysis the regions of parameter space where the CKM matrix is in agreement with data, that is moduli $\abs{\V{ij}}$ in the first two rows and the phase $\gamma$ agree with experimental results \cite{Patrignani:2016xqp}.\\ 
Figure \ref{FIG:logs2tthd} shows the region of the plane $\abs{\sin 2\theta}$ vs. $\theta_d$ which can yield a good CKM matrix. It is to be noticed that (i) regions rather close to $\theta=0,\pi/2,\pi$, with $\abs{\sin 2\theta}<10^{-2}$, are allowed and require $\theta_d\sim \pi/4,3\pi/4$, while (ii) for $\abs{\sin 2\theta}\sim 1$, allowed regions require $\theta_d\sim 0,\pi/2,\pi$. In any case, $\sin 2\theta\neq 0$ is a necessary requirement, as expected, since there is no CP violation in that limit.\\
Figures \ref{FIG:goodCKM:1}--\ref{FIG:goodCKM:3} also show the allowed regions in the $\sin\theta_d\sin\varphi_d$ vs. $\sin\theta_d\cos\varphi_d$ plane\footnote{That is, the projection on the $(\hat 1,\hat 2)$ plane of the allowed regions in the surface of the sphere of Figure \ref{FIG:par:rd}.}, separating $\abs{\sin 2\theta}\geq 10^{-1}$ in \ref{FIG:goodCKM:1}, $10^{-1}\geq\abs{\sin 2\theta}\geq 10^{-2}$ in \ref{FIG:goodCKM:2}, and  $10^{-2}\geq\abs{\sin 2\theta}$ in \ref{FIG:goodCKM:3}.\\
Following the discussion in subsection \ref{sSEC:FCNC:CKM}, one can sort the components of $\rndvec$ according to their size:
\begin{equation}\label{eq:rd:size:00}
|\rnd{Min}|\leq |\rnd{Mid}|\leq |\rnd{Max}|\,.
\end{equation}
Incompatibility with a CP violating CKM matrix implies that $|\rnd{Mid}|\neq 0$, while the simple example in subsection \ref{sSEC:CKMexample} shows that $|\rnd{Min}|=0$ is allowed. Figure \ref{FIG:rd:Min} shows the allowed region of $|\rnd{Min}|$ vs. $|\rnd{Mid}|$ confirming that it is necessary that $|\rnd{Mid}|\geq 2\times 10^{-3}$, while figure \ref{FIG:ru:Min} corresponds similarly to $\rnuvec$: both figures illustrate that, in this model, necessarily, there is at least some minimal presence of tree SFCNC\footnote{Notice that, consequently, $\theta_d=0,\pi$ are excluded: the resolution in Figures \ref{FIG:logs2tthd} and \ref{FIG:goodCKM:1} is too coarse to observe that, while Figures \ref{FIG:rd:Min} and \ref{FIG:ru:Min} clearly illustrate the point.}.
%

%

In conclusion, it is clear that the first requirement on the model, i.e. that it can reproduce the observed CKM matrix, can be fulfilled.

\clearpage
\section{The scalar potential with real couplings\label{SEC:scalar}}
We consider the 2HDM with CP invariance and impose the $\ZZ$ symmetry of \refEQ{eq:Symmetry:00} which is only softly broken by a $\mu_{12}$ term. All couplings are real, so that CP holds at the Lagrangian level. The scalar potential can be written:
\begin{multline}\label{eq:ScalarPotential:00}
\mathscr V(\Hd{1},\Hd{2})=
\mu_{11}^2\Hdd{1}\Hd{1}+\mu_{22}^2\Hdd{2}\Hd{2}+\mu_{12}^2(\Hdd{1}\Hd{2}+\Hdd{2}\Hd{1})+\lambda_1(\Hdd{1}\Hd{1})^2+\lambda_2(\Hdd{2}\Hd{2})^2\\
+2\lambda_3(\Hdd{1}\Hd{1})(\Hdd{2}\Hd{2})+2\lambda_4(\Hdd{1}\Hd{2})(\Hdd{2}\Hd{1})+\lambda_5[(\Hdd{1}\Hd{2})^2+(\Hdd{2}\Hd{1})^2]\,.
\end{multline}
The vacuum expectation values are
\begin{equation}\label{eq:EWSSB:00}
\langle\Hd{1}\rangle=\begin{pmatrix} 0\\ e^{i\theta_1}v_1/\sqrt{2}\end{pmatrix},\quad \langle\Hd{2}\rangle=\begin{pmatrix} 0\\ e^{i\theta_2}v_2/\sqrt{2}\end{pmatrix},
\end{equation}
and break electroweak symmetry spontaneously. As anticipated in section \ref{SEC:model}, we use $\theta=\theta_2-\theta_1$, $v^2=v_1^2+v_2^2$, $\cb=\cos\beta\equiv v_1/v$, $\sb=\sin\beta\equiv v_2/v$ and $\tb\equiv\tan\beta$, with $\vev{1}\geq 0$, $\vev{2}\geq 0$.
\subsection{Minimization\label{sSEC:min}}
The minimization conditions for $V(v_1,v_2,\theta)\equiv \mathscr V(\langle\Hd{1}\rangle,\langle\Hd{2}\rangle)$ are
\begin{align}
&\frac{\partial V}{\partial \theta}=-\vev{1}\vev{2}\sin\theta(\mu_{12}^2+2\lambda_5 \vev{1}\vev{2}\cos\theta)=0,\label{eq:V:min:theta}\\
&\frac{\partial V}{\partial \vev{1}}=\mu_{11}^2\vev{1}+\lambda_1\vev{1}^3+(\lambda_3+\lambda_4)\vev{1}\vev{2}^2+\vev{2}(\mu_{12}^2\cos\theta+\lambda_5\vev{1}\vev{2}\cos 2\theta)=0,\label{eq:V:min:v1}\\
&\frac{\partial V}{\partial \vev{2}}=\mu_{22}^2\vev{2}+\lambda_2\vev{2}^3+(\lambda_3+\lambda_4)\vev{1}^2\vev{2}+\vev{1}(\mu_{12}^2\cos\theta+\lambda_5\vev{1}\vev{2}\cos 2\theta)=0.\label{eq:V:min:v2}
\end{align}
In order to have spontaneous CP violation, we consider a solution $\{\vev{1},\vev{2},\theta\}$ of \refEQS{eq:V:min:theta}--\eqref{eq:V:min:v2} with $\theta\neq 0,\pm\pi/2,\pm\pi$. From \refEQ{eq:V:min:theta} one obtains
\begin{equation}\label{eq:thetaCP}
\cos\theta=\frac{-\mu_{12}^2}{2\lambda_5\vev{1}\vev{2}}\,.
\end{equation}
Notice that, in addition to $\theta$, $-\theta$ is also a solution. It is obvious that for $\theta=0,\pi$ the vacuum is CP invariant. It has also been shown \cite{Branco:1980sz} that for $\theta=\pi/2$ the vacuum is also CP invariant. Note from \refEQ{eq:thetaCP} that $\theta=\pm\pi/2$ is obtained when $\mu_{12}^2=0$. In this case, the scalar potential is invariant under the $\ZZ$ symmetry of \refEQ{eq:Symmetry:00}. This symmetry allows the two scalar fields $\Hd{1}$, $\Hd{2}$ to have either equal or opposite CP parities. It is this freedom that is used to construct a simple proof \cite{Branco:1980sz} that for $\theta=\pm\pi/2$, the vacuum is CP invariant.\\ 
One can trade $\mu_{11}^2$, $\mu_{22}^2$ and $\mu_{12}^2$ for other parameters  using \refEQS{eq:V:min:theta}--\eqref{eq:V:min:v2}:
\begin{align}
&\mu_{12}^2=-2\lambda_5\vev{1}\vev{2}\cos\theta,\label{eq:V:min:mu12}\\
&\mu_{11}^2=-(\lambda_1\vev{1}^2+(\lambda_3+\lambda_4-\lambda_5)\vev{2}^2),\label{eq:V:min:mu11}\\
&\mu_{22}^2=-(\lambda_2\vev{2}^2+(\lambda_3+\lambda_4-\lambda_5)\vev{1}^2).\label{eq:V:min:mu22}
\end{align}
That is, imposing \refEQS{eq:V:min:mu12}--\eqref{eq:V:min:mu22} on $\mathscr V(\Hd{1},\Hd{2})$ in \refEQ{eq:ScalarPotential:00}, one is selecting a scalar potential where, at least, the necessary minimization conditions in \refEQS{eq:V:min:theta}--\eqref{eq:V:min:v2} are satisfied for generic $\{\vev{1},\vev{2},\theta\}$. One can in addition choose $\vev{1}^2+\vev{2}^2=v^2=(246\text{ GeV})^2$ for appropriate electroweak symmetry breaking without loss of generality (this is enforced, for example, by a simple rescaling of the parameters in the potential). Fixing $v^2$ in that manner, one is left with a candidate minimum characterised by the values of $\theta$ and $\tan\beta=\vev{2}/\vev{1}$, which remain free parameters that we can choose at will, up to the different constraints on the scalar potential to be discussed later:
\begin{enumerate}
\item the potential is bounded from below and $V(\vev{1},\vev{2},\theta)$ is the lowest lying minimum;
\item perturbative unitarity bounds on scattering processes in the scalar sector are respected.
\end{enumerate}
Expanding $\Hd{j}$ around the candidate vacuum in \refEQ{eq:EWSSB:00}
\begin{equation}
\Hd{j}=e^{i\theta_j}\begin{pmatrix}\varphi^+_j\\ \frac{1}{\sqrt 2}(\vev{j}+\rho_j+i\eta_j)\end{pmatrix}\,,
\end{equation}
we can now explore the different mass terms for the charged and neutral scalars. Requiring that the mass parameters of all the physical scalars are positive ensures, at least, that the candidate minimum is a local minimum of the potential. In the Higgs basis of \refEQ{eq:HiggsBasis:00}, the expansion of the fields reads
\begin{equation}\label{eq:HiggsBasis:10}
H_{1}=\begin{pmatrix} G^+\\ (v+\nHH+iG^0)/\sqrt{2}\end{pmatrix},\quad H_{2}=\begin{pmatrix} \cHp\\ (\nHR+i\nHI)/\sqrt{2}\end{pmatrix},
\end{equation}
\begin{equation}\label{eq:HiggsBasis:11}
\begin{pmatrix}G^+\\ \cHp\end{pmatrix}=\HbROT\begin{pmatrix}\varphi_1^+\\ \varphi_2^+\end{pmatrix},\ 
\begin{pmatrix}G^0\\ \nHI\end{pmatrix}=\HbROT\begin{pmatrix}\eta_1\\ \eta_2\end{pmatrix},\ 
\begin{pmatrix}\nHH\\ \nHR\end{pmatrix}=\HbROT\begin{pmatrix}\rho_1\\ \rho_2\end{pmatrix},
\end{equation}
with the would-be Goldstone bosons $G^\pm$ and $G^0$ readily identified
\begin{equation}
G^\pm=\cb\varphi_1^\pm-\sb\varphi_2^\pm,\quad G^0=\cb\eta_1-\sb\eta_2\,.
\end{equation}

\subsection{Scalar masses and mixings\label{sSEC:M2}}
\subsubsection{Charged scalar\label{sSEC:M2:Charged}}
The transformation into the Higgs basis also gives the mass term of the charged scalar $\cH=\sbCP\varphi_1^\pm-\cbCP\varphi_2^\pm$,
\begin{equation}\label{eq:ChargedMass:00}
\mathscr V(\Hd{1},\Hd{2})\ \supset\ v^2(\lambda_5-\lambda_4)\cHp\cHm\,\Rightarrow \mcH^2=v^2(\lambda_5-\lambda_4)\,.
\end{equation}
Notice that, in order to choose a set of independent parameters, \refEQ{eq:ChargedMass:00} will allow us to trade $\lambda_4$ for $\mcH^2$ and $\lambda_5$. Furthermore, since $\lambda_5$ and $\lambda_4$ are subject to the constraints on the scalar potential discussed in appendix \ref{APP:scalar}, $\mcH$ has a limited allowed range: for example, if $\lambda_5-\lambda_4<20$, then it follows that $\mcH<9\mh$.
\subsubsection{Neutral scalars\label{sSEC:M2:Neutral}}
For the neutral scalar sector, the mass terms are
\begin{equation}
\mathscr V(\Hd{1},\Hd{2})\ \supset\ \frac{1}{2}\begin{pmatrix}\nHH& \nHR& \nHI\end{pmatrix}\ \mNSc\ \begin{pmatrix}\nHH\\ \nHR\\ \nHI\end{pmatrix}\,,
\end{equation}
with $\mNSc=\mNScT$, and
\begin{align}
& [\mNSc]_{11}=2v^2\left\{\lambda_1\cb^4+\lambda_2\sb^4+2\cb^2\sb^2[\lambda_{345}+2\lambda_5\ctCP^2]\right\},\nonumber\\
& [\mNSc]_{22}=2v^2\left\{\cb^2\sb^2(\lambda_1+\lambda_2-2\lambda_{345})+\lambda_5(\cb^2-\sb^2)^2\ctCP^2\right\},\nonumber\\
& [\mNSc]_{12}=2v^2\sb\cb\left\{-\lambda_1\cb^2+\lambda_2\sb^2+(\cb^2-\sb^2)[\lambda_{345}+2\lambda_5\ctCP^2])\right\},\nonumber\\
& [\mNSc]_{13}=-v^2\lambda_5\sbb \sttCP,\nonumber\\
& [\mNSc]_{23}=-v^2\lambda_5\cbb \sttCP,\nonumber\\
& [\mNSc]_{33}=2v^2\lambda_5 \stCP^2,\label{eq:M2:33}
\end{align}
with, we recall, the shorthand notation $c_x=\cos x$, $s_x=\sin x$, and $\lambda_{345}\equiv\lambda_3+\lambda_4-\lambda_5$.\\ 
For $\lambda_5\sttCP\neq 0$, attending to $[\mNSc]_{13}\neq 0$ and $[\mNSc]_{23}\neq 0$ above, there is scalar-pseudoscalar mixing, as it is expected from spontaneous breaking of CP in the scalar sector. $\mNSc$ is diagonalised through a real orthogonal transformation $\ROTmat$ 
\begin{equation}
\ROTmatT\,\mNSc\,\ROTmat=\text{diag}(\mh^2,\mH^2,\mA^2)\,,\quad \ROTmatinv=\ROTmatT.
\end{equation}
The physical neutral scalars are
\begin{equation}\label{eq:ScalarROT:00}
\begin{pmatrix}\nh\\ \nH\\ \nA\end{pmatrix}=\ROTmatT\begin{pmatrix}\nHH\\ \nHR\\ \nHI\end{pmatrix}\,,
\end{equation}
and we assume $\nh$ to be the lightest one, the Higgs-like neutral scalar with $\mh=125$ GeV. With $\mNSc$ in \refEQS{eq:M2:33}, $\ROTmat$ ``mixes'', a priori, all three neutral scalars.
It is interesting to notice that
\begin{equation}\label{eq:M2:tr}
\text{Tr}[\mNSc]=\mh^2+\mH^2+\mA^2=2v^2[\lambda_1\cb^2+\lambda_2\sb^2+\lambda_5]\,,
\end{equation}
and 
\begin{equation}\label{eq:M2:det}
\det[\mNSc]=\mh^2\,\mH^2\,\mA^2=2v^6\lambda_5(\lambda_1\lambda_2-\lambda_{345}^2)\sin^2 2\beta\,\sin^2\theta\,.
\end{equation}
As explained in appendix \ref{APP:scalar}, since it is necessary that $\lambda_5>0$, it is also required that $\lambda_1\lambda_2>\lambda_{345}^2$ for $V(\vev{1},\vev{2},\theta)$ to be, at least, a local minimum (for which, necessarily, $\det[\mNSc]>0$).\\ 
Equations \eqref{eq:M2:tr} and \eqref{eq:M2:det} encode in a transparent manner several interesting properties of the model. First, since the different $\lambda_j$ are bounded by the requirements discussed in appendix \ref{APP:scalar} (in particular by perturbative unitarity), and $\sin^2 2\beta\leq 1$ and $\sin^2\tCP\leq 1$, the masses of the new scalars $\nH$, $\nA$, $\cH$, have a limited allowed range. For a very crude estimate, consider for example $\lambda_1\cb^2+\lambda_2\sb^2+\lambda_5\sim 10$ in \refEQ{eq:M2:tr}: with $v\simeq 2\mh$, $\mH^2+\mA^2\sim 80\mh^2$ and it is clear that the smaller among $\mH$ and $\mA$ cannot be larger than $\sim 6\mh$, while the larger among $\mH$ and $\mA$ cannot be larger than $\sim 9\mh$. \\ 
On the other hand, from \refEQ{eq:M2:det}, for $\sin 2\beta\ll 1$, at least one neutral scalar should be light and either $\tan\beta\gg 1$ or $\tan^{-1}\beta\gg 1$, which enhance SFCNC couplings. One can than expect that $\sin 2\beta\ll 1$ will be disfavoured by the constraints discussed in section \ref{SEC:Pheno}, while $\tan\beta\sim\tan^{-1}\beta\sim 1$ are easier to accommodate. Finally, it is to be noticed that for $\sin\theta=1$, there is no mixing among $\{\nh,\nH\}$ and $\nA$ (and $\mA^2=2\lambda_5 v^2$), and, as discussed, no spontaneous CP violation and a real CKM matrix. For $\sin\theta=0$, $\mA=0$ and thus for $\abs{\sin\tCP}\ll 1$, one could expect again the presence of at least one light scalar.\\
From the previous comments, it emerges that in this model there is limited room to have a scalar sector where (i) $\nh$ is a Higgs boson with quite SM-like properties and (ii) $\mcH$, $\mH$, $\mA\gg \mh$. In this model, there is no decoupling regime for the new scalars. It is also clear, with these values, that the new scalars should be produced at the LHC. Nevertheless, the most relevant production and decay modes for their discovery will vary significantly between different regions of parameter space, including the Yukawa couplings discussed in section \ref{SEC:Yukawa}, and also the details of the lepton sector, and are thus beyond the scope of this work.

\subsection{A simple analysis of the scalar sector\label{sSEC:ScalarAnalysis}}
\begin{figure}[h!tb]
\begin{center}
\subfigure[$M_{Min}$ vs. $\tan\beta$.\label{FIG:Mminlogtanbeta:goodScalar}]{\includegraphics[width=0.3\textwidth]{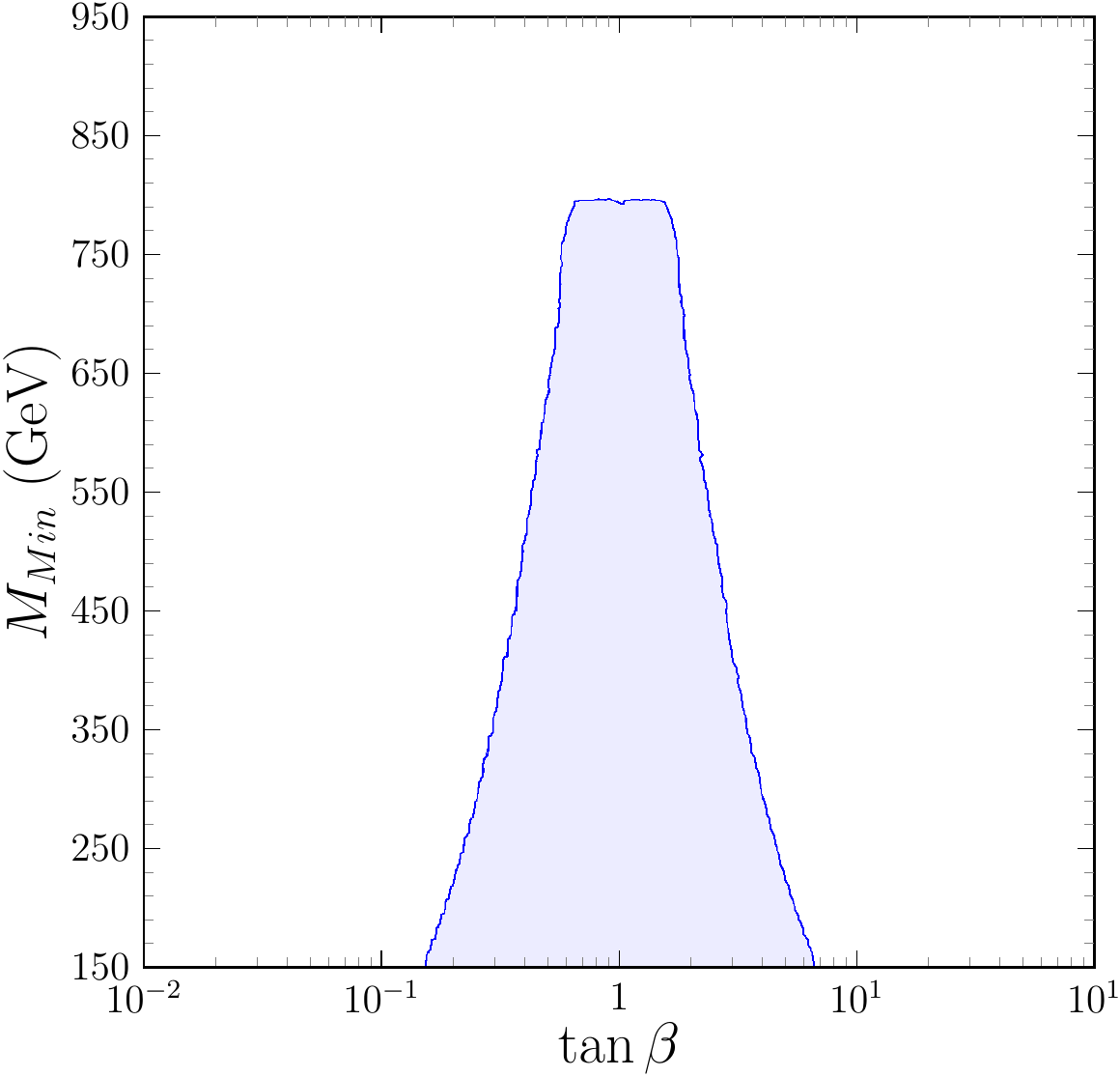}}\qquad
\subfigure[$M_{Max}$ vs. $\tan\beta$.\label{FIG:Mmaxlogtanbeta:goodScalar}]{\includegraphics[width=0.3\textwidth]{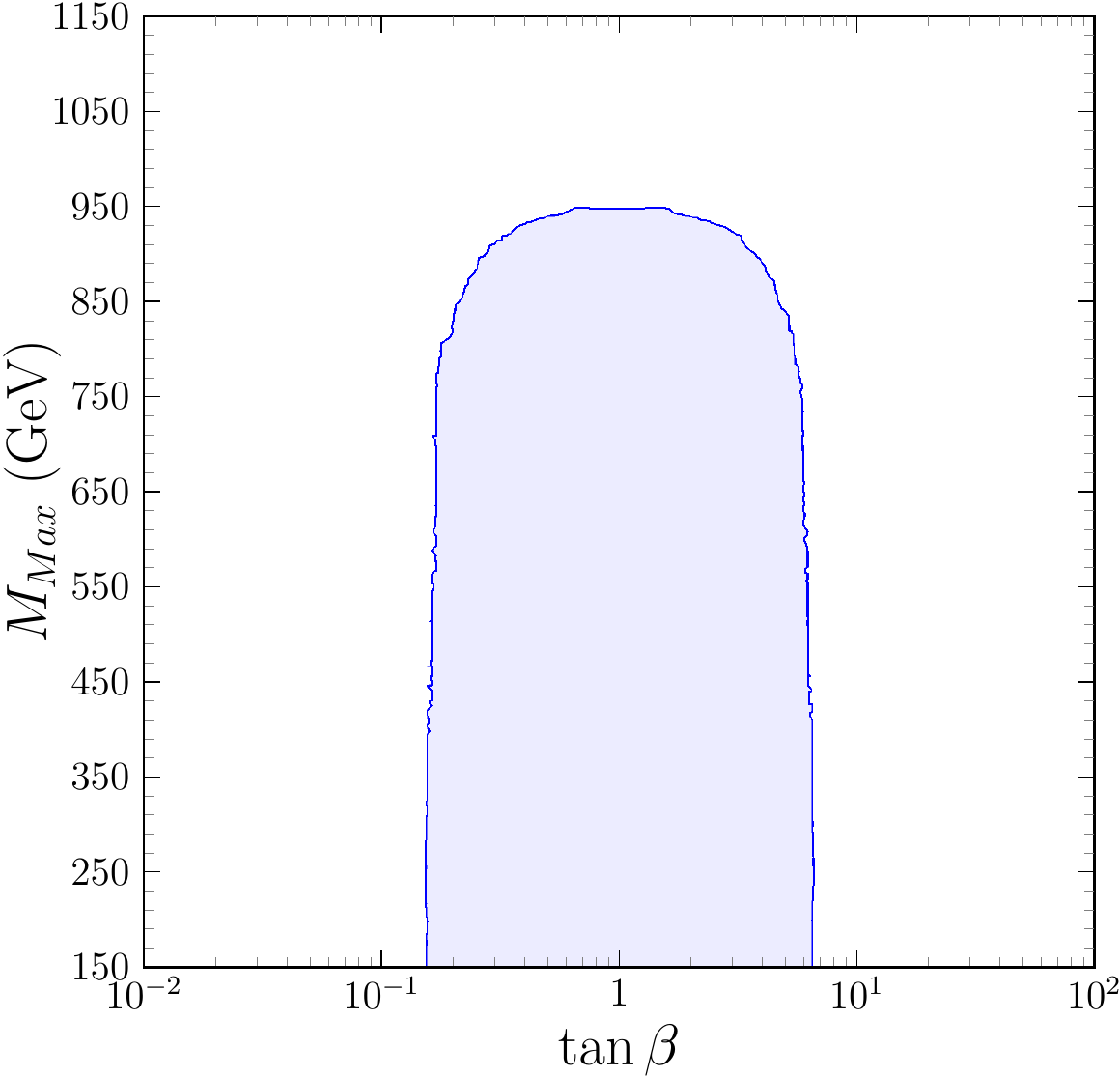}}\\
\subfigure[$M_{Max}$ vs. $M_{Min}$.\label{FIG:MmaxMmin:goodScalar}]{\includegraphics[width=0.3\textwidth]{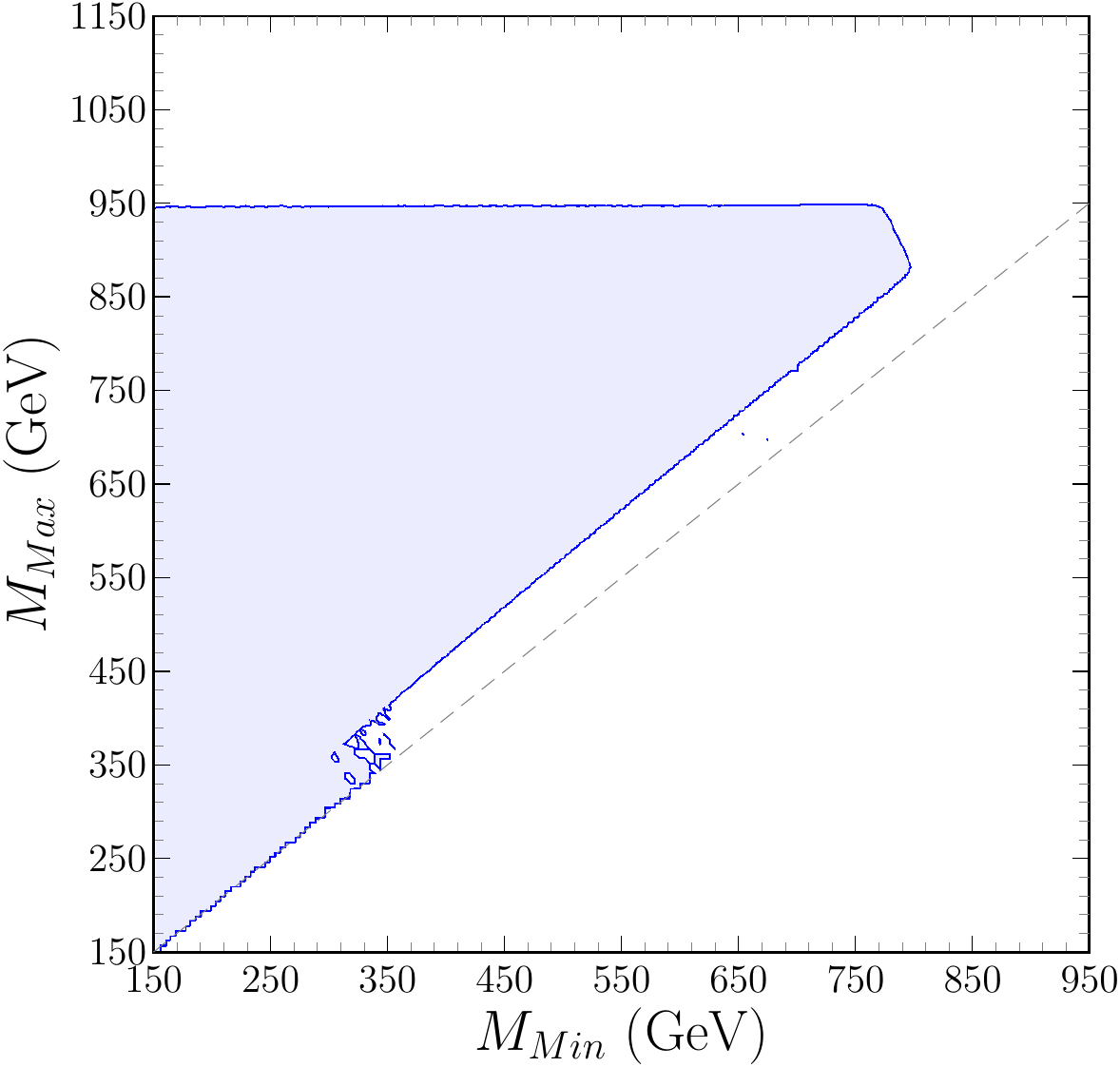}}\quad 
\subfigure[$\tan\beta$ vs. $\abs{\sin 2\theta}$.\label{FIG:logtanbetalogs2theta:goodScalar}]{\includegraphics[width=0.3\textwidth]{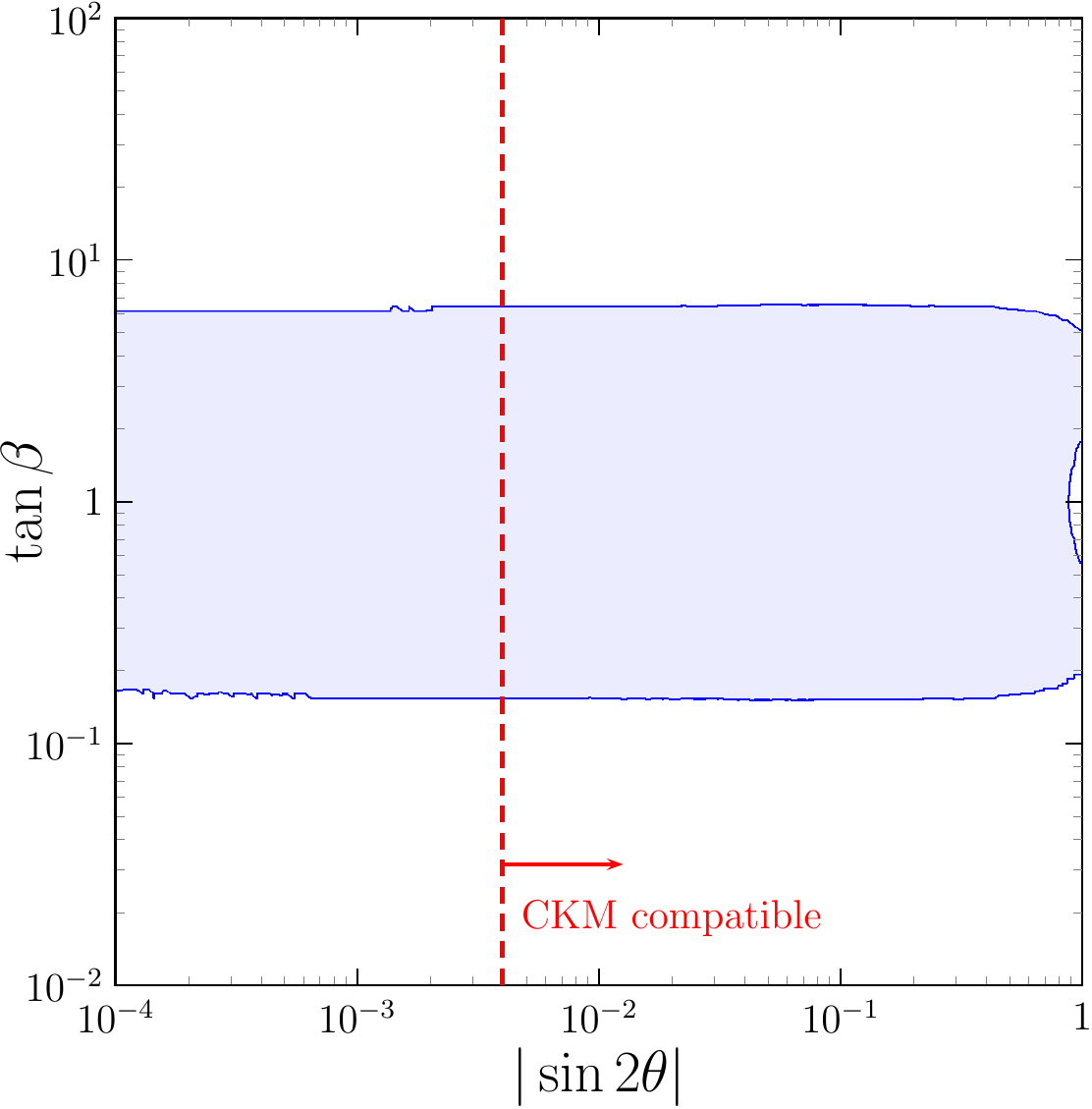}}
\caption{Regions allowed at 99\% C.L. by the requirements on the scalar sector.\label{FIG:goodScalar:1}}
\end{center}
\end{figure}

\begin{figure}[h!tb]
\begin{center}
\subfigure[$\mA$ vs. $\mcH$.\label{FIG:mA:mcH:goodScalar}]{\includegraphics[width=0.3\textwidth]{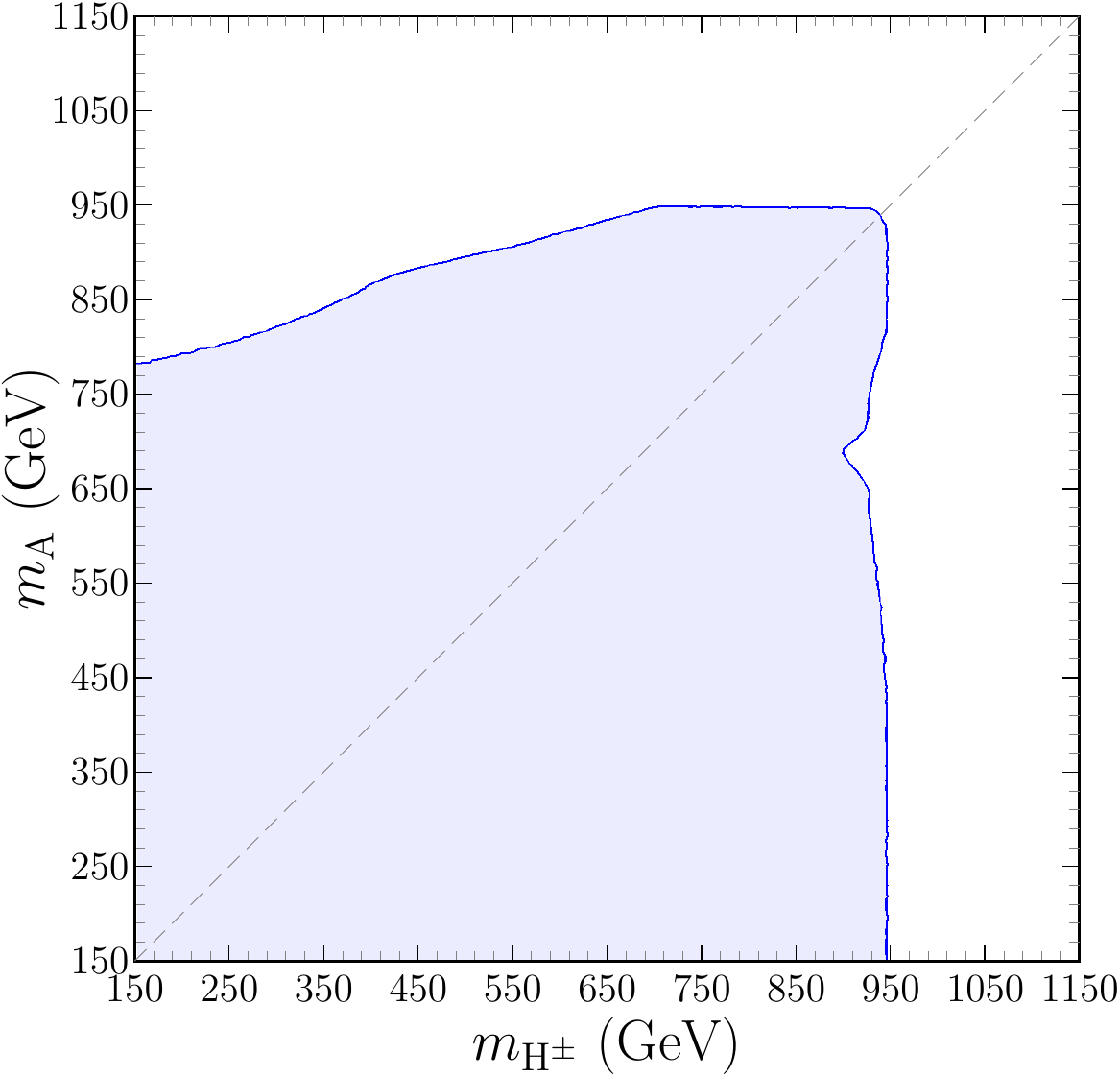}}\quad
\subfigure[$\mA$ vs. $\mH$.\label{FIG:mA:mH:goodScalar}]{\includegraphics[width=0.3\textwidth]{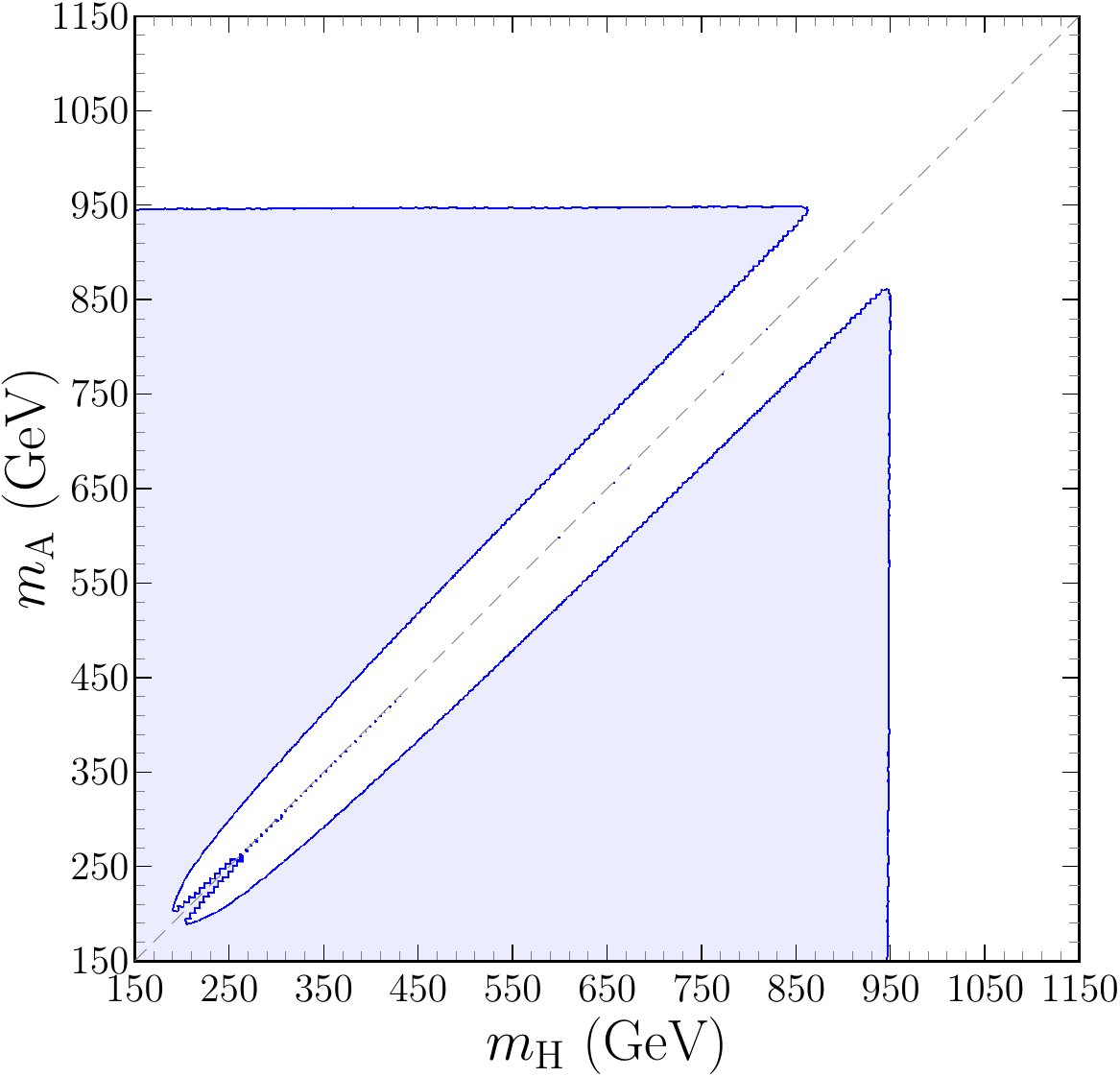}}\quad
\subfigure[$\mH$ vs. $\mcH$.\label{FIG:mH:mcH:goodScalar}]{\includegraphics[width=0.3\textwidth]{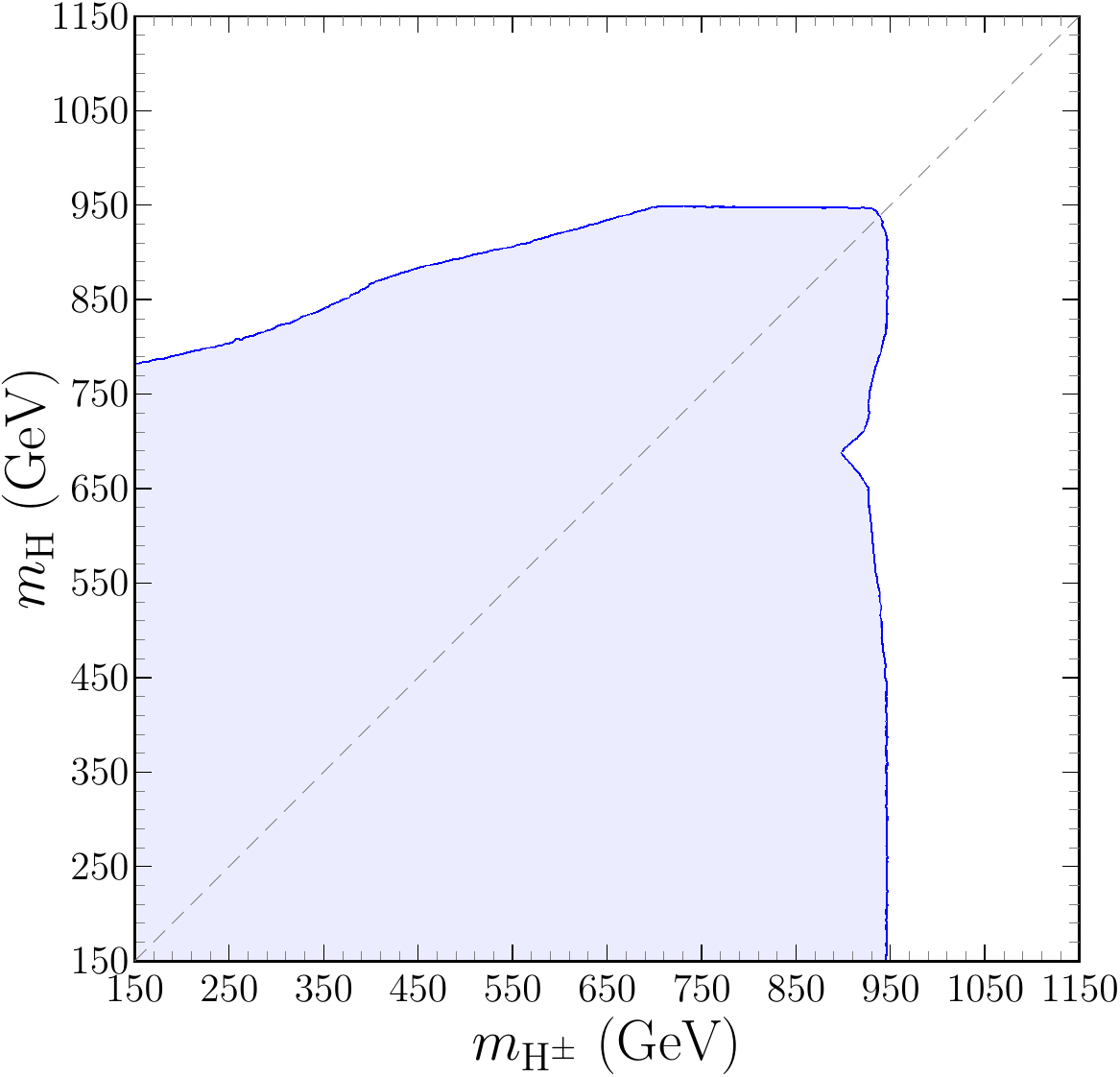}}
\caption{Regions allowed at 99\% C.L. by the requirements on the scalar sector.\label{FIG:goodScalar:2}}
\end{center}
\end{figure}

As a first step in the direction of the complete analysis of section \ref{SEC:Pheno}, in this subsection we analyse the available parameter space of the scalar sector of the model, considering the following constraints.
\begin{itemize}
\item Agreement with electroweak precision data, in particular the oblique parameters $S$ and $T$ \cite{Grimus:2008nb}.
\item Boundedness of the scalar potential and perturbative unitarity of the scattering processes, controlled by the scalar quartic couplings $\lambda_j$, as described, respectively, in appendices \ref{sAPP:minimum} and \ref{sAPP:perturbative}.
\item We only consider $\mcH$, $\mH$, $\mA \geq 150$ GeV; although masses of new scalars below $150$ GeV are not automatically excluded by existing constraints, they would require specific analyses, interesting on their own, which are out of the scope of the present work. Furthermore, attending to \refEQ{eq:M2:det} and the related discussion, imposing this requirement on $\mH$ and $\mA$ translates into a \emph{lower} bound on $\sbb^2$ and $\stCP^2$. For a simple estimate one can take $\lambda_5(\lambda_1\lambda_2-\lambda_{345}^2)< 10^2$, which gives (for $\mH$, $\mA \geq 150$ GeV) $\sbb^2$, $\stCP^2>10^{-4}$. In terms of $\tb$, this means $10^{-2}<\tb<10^{2}$. On the contrary, since the quantity relevant for the obtention of a realistic CKM matrix is $\sin 2\tCP$ rather than $\sin\tCP$, $\stCP^2>10^{-4}$ is only relevant for $\tCP\sim 0,\pi$, while $\abs{\sin 2\tCP}\ll 1$ with $\tCP\sim \frac{\pi}{2},\frac{3\pi}{2}$ is allowed.
\item The analyses of Higgs signal strengths from the ATLAS and CMS collaborations, e.g. \cite{Khachatryan:2016vau}, put constraints on different couplings of $\nh$. Overall, the resulting picture corresponds to an $\nh$ which is quite SM-like. For that reason, in order to discard from this simple analysis the regions of parameter space that these constraints will in any case eliminate in the complete analysis of section \ref{SEC:Pheno}, we require here $\abs{\ROT{11}}\geq 0.9$.
\end{itemize}
Although the analysis of section \ref{sSEC:CKMnumeric} already sets a lower bound $\abs{\sin 2\theta}\geq 4\times 10^{-3}$ in order to obtain the correct CKM matrix, we do not impose it here (it corresponds to the dashed vertical line in Figure \ref{FIG:logtanbetalogs2theta:goodScalar}).
A detailed discussion of one convenient parametrisation of all quantities related to the scalar sector is given in appendix \ref{sAPP:parameters}.\\ 
With these ingredients, the allowed regions in Figures \ref{FIG:goodScalar:1} and \ref{FIG:goodScalar:2} are obtained. We introduce
\begin{equation}
M_{Min}\equiv\text{min}(\mH,\mA,\mcH),\qquad M_{Max}\equiv\text{max}(\mH,\mA,\mcH)\,.
\end{equation}
Figure \ref{FIG:Mminlogtanbeta:goodScalar} shows that, with the simple requirements enumerated above, all new scalars cannot have, simultaneously, masses above $\sim 750$ GeV. These values are in rough agreement with the previous naive estimates. Figure \ref{FIG:Mmaxlogtanbeta:goodScalar} shows in addition that no new scalar can be heavier than $\sim 950$ GeV. It is also clear that the largest values of the scalar masses correspond to $\tb\simeq 1$, while only a reduced range of values of $\tb$ is allowed, $10^{-1}<\tb<10$.. Figure \ref{FIG:MmaxMmin:goodScalar} shows that the limitations on allowed $M_{Min}$ and $M_{Max}$ appear to be rather independent: for example, $M_{Max}\sim 850$ GeV is compatible with any value of $M_{Min}$ below $750$ GeV.\\
Figures \ref{FIG:mA:mcH:goodScalar}--\ref{FIG:mH:mcH:goodScalar} illustrate that any ordering of the masses $\mH$, $\mA$, $\mcH$ is allowed, and no particular restriction arises besides the impossibility of having $\mH\simeq\mA>250$ GeV.

\noindent Having introduced the physical scalars and analysed some relevant aspects of the scalar sector, we can now turn back to $\mathscr L_{\rm Y}$ in \refEQ{eq:Yukawa:01} and discuss the Yukawa couplings of the physical quarks and scalars.


\clearpage
\section{Physical Yukawa couplings\label{SEC:Yukawa}}
The Yukawa lagrangian in \refEQ{eq:Yukawa:01} 
\begin{equation}
\mathscr L_{\rm Y}=\mathscr L_{{\rm \bar qq}}+\mathscr L_{G{\rm \bar qq}}+\mathscr L_{\nh{\rm \bar qq}}+\mathscr L_{\nH{\rm \bar qq}}+\mathscr L_{\nA{\rm \bar qq}}+\mathscr L_{\cH{\rm \bar qq}},
\end{equation}
gives the mass terms for quarks $\mathscr L_{{\rm \bar qq}}=-\bar d_L\mMD d_R-\bar u_L\mMU u_R+\text{H.c}$, the couplings to the would-be Goldstone bosons $\mathscr L_{G{\rm \bar qq}}$,
\begin{multline}\label{eq:qqG:00}
\mathscr L_{G{\rm \bar qq}}=
-i\frac{G^0}{v}\left[\bar d\gamma_5 d-\bar u\gamma_5 u\right]\\
-\frac{\sqrt{2}G^+}{v}\left[\bar u_L\CKM\mMD d_R-\bar u_R\mMU\CKM d_L\right]-\frac{\sqrt{2}G^-}{v}\left[\bar d_R\mMD\CKMd u_L-\bar d_L\CKMd\mMU u_R\right],
\end{multline}
and the Yukawa couplings to the neutral and charged scalars $\mathscr L_{{\rm S\bar qq}}$, $S=\nh$, $\nH$, $\nA$, $\cH$.
Introducing the hermitian and antihermitian combinations
\begin{equation}\label{eq:HqAq:00}
\mHQ{q}\equiv\frac{\mNQ{q}+\mNQd{q}}{2},\quad \mAQ{q}\equiv\frac{\mNQ{q}-\mNQd{q}}{2},
\end{equation}
we have
\begin{align}
\mathscr L_{S{\rm \bar qq}}=
&-\frac{S}{v}
\left\{
\bar d\left[\ROT{1s}\mMD+\ROT{2s}\mHD+i\ROT{3s}\mAD\right]d+\bar d\left[\ROT{2s}\mAD+i\ROT{3s}\mHD\right]\gamma_5 d
\right\}\nonumber\\
&-\frac{S}{v}
\left\{
\bar u\left[\ROT{1s}\mMU+\ROT{2s}\mHU-i\ROT{3s}\mAU\right]u+\bar u\left[\ROT{2s}\mAU-i\ROT{3s}\mHU\right]\gamma_5 u
\right\},
\label{eq:qqS:00}
\end{align}
with $s=1,2,3$ for $S=\nh,\nH,\nA$, respectively, and
\begin{equation}
\mathscr L_{\cH{\rm \bar qq}} = -\frac{\sqrt{2}\cHp}{v}\left[\bar u_L\CKM\mND d_R-\bar u_R\mNUd\CKM d_L\right]-\frac{\sqrt{2}\cHm}{v}\left[\bar d_R \mNDd\CKMd u_L-\bar d_L \CKMd\mNU u_R\right].
\end{equation}
With \refEQS{eq:Nq:gen:00:1}--\eqref{eq:Nq:gen:00:2}, $[\mHQ{q}]_{ij}$ and $[\mAQ{q}]_{ij}$ in \refEQ{eq:HqAq:00} read
\begin{align}
[\mHQ{q}]_{ij}&=\tb\delta_{ij}m_{q_i}-(\tti)\unC{q}{i}\un{q}{j}\frac{m_{d_i}+m_{d_j}}{2},\label{eq:Hq:gen:00}\\
[\mAQ{q}]_{ij}&=(\tti)\unC{q}{i}\un{q}{j}\frac{m_{d_i}-m_{d_j}}{2}.\label{eq:Aq:gen:00}
\end{align}
We recall -- see for example \cite{Nebot:2015wsa} -- that, for flavour changing Yukawa couplings of quarks $q_j$, $q_k$ and a scalar $S$, of the form
\begin{equation}
\mathscr L_{{\rm Sq_jq_k}}=-S\bar q_j(a_{jk}+ib_{jk}\gamma_5)q_k+\text{H.c},\quad a_{jk},b_{jk}\in \mathbb{C},
\end{equation}
CP conservation requires $\re{a_{jk}^\ast b_{jk}}=0$. In this model
\begin{equation}
\re{a_{jk}^\ast b_{jk}}\propto 
\ROT{2s}\ROT{3s}(\tti)^2 m_{q_j}m_{q_k}\abs{\unC{q}{j}\un{q}{k}}^2\,,
\end{equation}
and thus, with $\ROTmat$ mixing all three neutral scalars, the flavour changing Yukawa couplings are CP violating.
For the charged scalar, we have
\begin{equation}
\re{a_{jk}^\ast b_{jk}}\propto \im{(\CKM\mND)_{jk}^\ast(\mNUd\CKM)_{jk}}\,,
\end{equation}
and thus in general, even for real $\mNQ{q}$, the Yukawa couplings of $\cH$ are also CP violating.\\ 
For flavour conserving Yukawa couplings
\begin{equation}\label{eq:CP:FCons:00}
\mathscr L_{{\rm Sqq}}=-S\bar q(a+ib\gamma_5)q,\quad a,b\in \mathbb{R},
\end{equation}
CP conservation requires $ab=0$. Then, for the coupling of the neutral scalar $S$, with \refEQS{eq:Hq:gen:00}--\eqref{eq:Aq:gen:00}, we have
\begin{equation}\label{eq:CP:FCons:01}
ab\propto \ROT{3s}m_{q_j}^2\left(\ROT{1s}+\ROT{2s}[\tb-(\tb+\tbinv)\abs{\un{q}{j}}^2]\right)\left(\tb-(\tb+\tbinv)\abs{\un{q}{j}}^2\right)\,,
\end{equation}
and thus the flavour conserving Yukawa couplings violate CP as long as the mixing in the scalar sector connects $\nA$ with $\nh$, $\nH$. Contributions to the electric dipole moment of the neutron arise from \refEQ{eq:CP:FCons:01}, but the suppression due to the $m_{q_j}^2$ factor for $q_j=u,d$, together with the need of different non-zero mixings in the scalar sector, keep them within experimental bounds \cite{Jung:2013hka}.

\section{Phenomenology\label{SEC:Pheno}}
\subsection{Analysis and constraints\label{sSEC:PhenoAnalysis}}
\begin{figure}[h!tb]
\begin{center}
\subfigure[\label{FIG:MesonMix}]{\raisebox{10pt}{\includegraphics[width=0.45\textwidth]{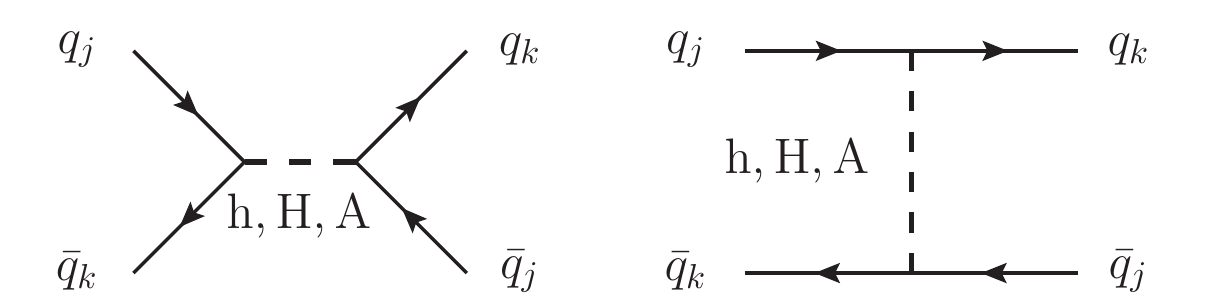}}}\quad
\subfigure[\label{FIG:bsg}]{\includegraphics[width=0.45\textwidth]{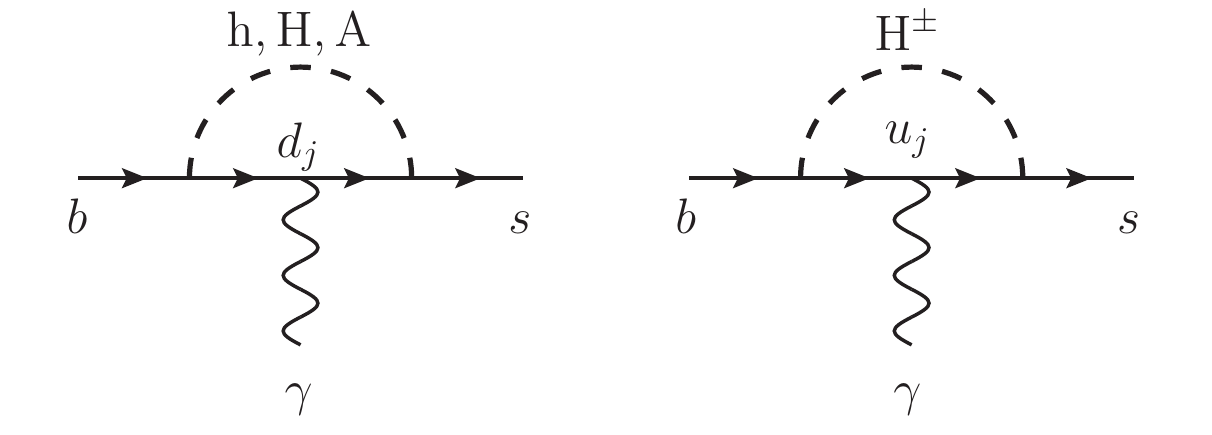}}
\caption{Diagrams contributing to: (a) meson mixing, (b) $b\to s\gamma$.\label{FIG:Diagrams}}
\end{center}
\end{figure}

In section \ref{SEC:CKM} we have shown that the model can give a CKM mixing matrix in agreement with data. We have also explored some aspects of the scalar sector in section \ref{SEC:scalar}. In this section we analyse the model considering simultaneously (i) obtention of an adequate CKM matrix (moduli $\abs{\V{ij}}$ in the first and second rows and phase $\gamma$ in agreement with data), (ii) a scalar sector verifying boundedness, perturbative unitarity, oblique parameter constraints and $\mH$, $\mA$, $\mcH>150$ GeV, and (iii) a number of constraints, to be discussed in the following, which involve both the quark Yukawa couplings and the scalar sector.
\begin{itemize}
\item \underline{Production $\times$ decay signal strengths of the 125 GeV Higgs-like scalar $\nh$}.\\ 
Agreement with the combined results of ATLAS and CMS from the LHC-Run I \cite{Khachatryan:2016vau}, together with additional data, involving in particular $\nh\to b\bar b$ \cite{Aaboud:2017xsd,Sirunyan:2017elk} from LHC-Run II, constrains the scalar mixings $\ROT{j1}$ and the diagonal entries of the $\mND$ and $\mNU$ matrices (see, for example, \cite{Botella:2018gzy}). Notice that the requirement $\abs{\ROT{11}}\geq 0.9$ used in section \ref{SEC:scalar} to mimic coarsely the effect of these results is, of course, not imposed here.
\item \underline{Neutral meson mixings}.\\
One of the most relevant characteristics of the model is the presence of tree level flavour changing couplings of the neutral scalars: they produce the contributions to neutral meson mixing represented in Figure \ref{FIG:MesonMix}. They affect mass differences and CP violating observables \cite{Patrignani:2016xqp,Amhis:2014hma}. For $B_d^0$--$\bar B_d^0$ and $B_s^0$--$\bar B_s^0$ we impose agreement with the mass differences $\Delta M_{B_d}$, $\Delta M_{B_s}$ and the mixing $\times$ decay CP asymmetries in $B_d\to J/\Psi K_S$ and $B_s\to J/\Psi\Phi$, respectively. For $K^0$--$\bar K^0$, we impose that the scalar mediated short distance contribution to $M_{12}^K$ does not yield sizable contributions to $\epsilon_K$ and $\Delta M_K$; in particular, for $\Delta M_K$, we require $2\abs{M_{12}^K}_{SFCNC}<\Delta M_K$. For $D^0$--$\bar D^0$, we impose, similarly, that the short distance contribution to $M_{12}^D$ verifies $\abs{M_{12}^D}<3\times 10^{-2}\,\text{ps}^{-1}$. In summary, neutral meson mixings constrain scalar mixings $\ROT{ij}$ and masses, together with off-diagonal entries of $\mND$ and the $12$, $21$, elements of $\mNU$.\\
Besides the SM one loop contribution, we only consider the scalar mediated tree level contributions to the Wilson coefficients of the different operators of interest; their QCD evolution from the electroweak scale to low energies follows \cite{Ciuchini:1997bw,Buras:2000if,Aebischer:2017gaw}. For the operator matrix elements and bag factors, we use \cite{Carrasco:2015pra} and \cite{Bazavov:2016nty} (see also \cite{Aoki:2016frl}).
\item \underline{$b\to s\gamma$}.\\
One loop diagrams like, for example, the ones in Figure \ref{FIG:bsg}, contribute to $\text{Br}(B\to X_s\gamma)$, and further constrain $\mNU$ and $\mND$, the neutral scalar mixings and masses, and $\mcH$ (in the previous constraints $\mcH$ does only appear in the one loop $\nh\to\gamma\gamma$ amplitude). Details of the calculation follow \cite{Misiak:2006zs,Crivellin:2013wna}.
\item \underline{Rare top decays $t\to\nh q$.}\\
Current bounds \cite{Aaboud:2017mfd,Khachatryan:2016atv,Sirunyan:2017uae} on $t\to\nh c,\nh u$ are at the $10^{-3}$ level, and have to be included in the analysis.
\end{itemize}

\noindent The analysis has two main goals:
\begin{enumerate}
\item to establish that the model is viable after a reasonable set of constraints is imposed;
\item to explore the prospects for the observation of some definite non-SM signal. We concentrate in particular on flavour changing decays $t\to\nh c,\nh u$ and $\nh\to bs,bd$, of interest, respectively, for the LHC and the ILC \cite{Barducci:2017ioq}. These are the most interesting tree level induced neutral flavour changing decays, since  $\nh\to uc,ds$ are more suppressed by the light fermion mass factors in $\mNU$ and $\mND$ (in addition, the experimental analysis is also more difficult having only light quarks in the final state).\\ 
We also consider a representative low energy observable, the time dependent CP violating asymmetry in $B_s\to J/\Psi\Phi$, $A^{CP}_{J/\Psi\Phi}$, for which the SM prediction is $A^{CP}_{J/\Psi\Phi}\simeq -0.04$, while current results \cite{Amhis:2016xyh} give $-0.030\pm 0.033$, leaving significant room for New Physics contributions.
\end{enumerate}
Further implications for the phenomenology of $\nH$, $\nA$ and $\cH$, vary significantly between allowed regions in the parameter space of the model (for example, the relevant decay modes change depending on which scalar is heavier and on the values of the $\mNQ{q}$ matrices), and would also involve, and be sensitive to, the couplings of the scalars with leptons, which we have not discussed in this work. As commented in precedence, we do not address such implications here. The fact that we have not included a description of the leptonic sector prevents (i) the use of constraints such as, for example, $\text{Br}(B_s\to\mu^+\mu^-)$ or $\text{Br}(K_L\to\mu^+\mu^-)$, and (ii) considerations on potential New Physics signals which involve leptons, as the so-called ``B anomalies'' \cite{Albrecht:2018vsa}.

\subsection{Results\label{sSEC:PhenoResults}}
\begin{figure}[h!tb]
\begin{center}
\subfigure[$s_{\theta_d} s_{\varphi_d}$ vs. $s_{\theta_d} c_{\varphi_d}$\label{FIG:hem:goodAll}]{\includegraphics[width=0.3\textwidth]{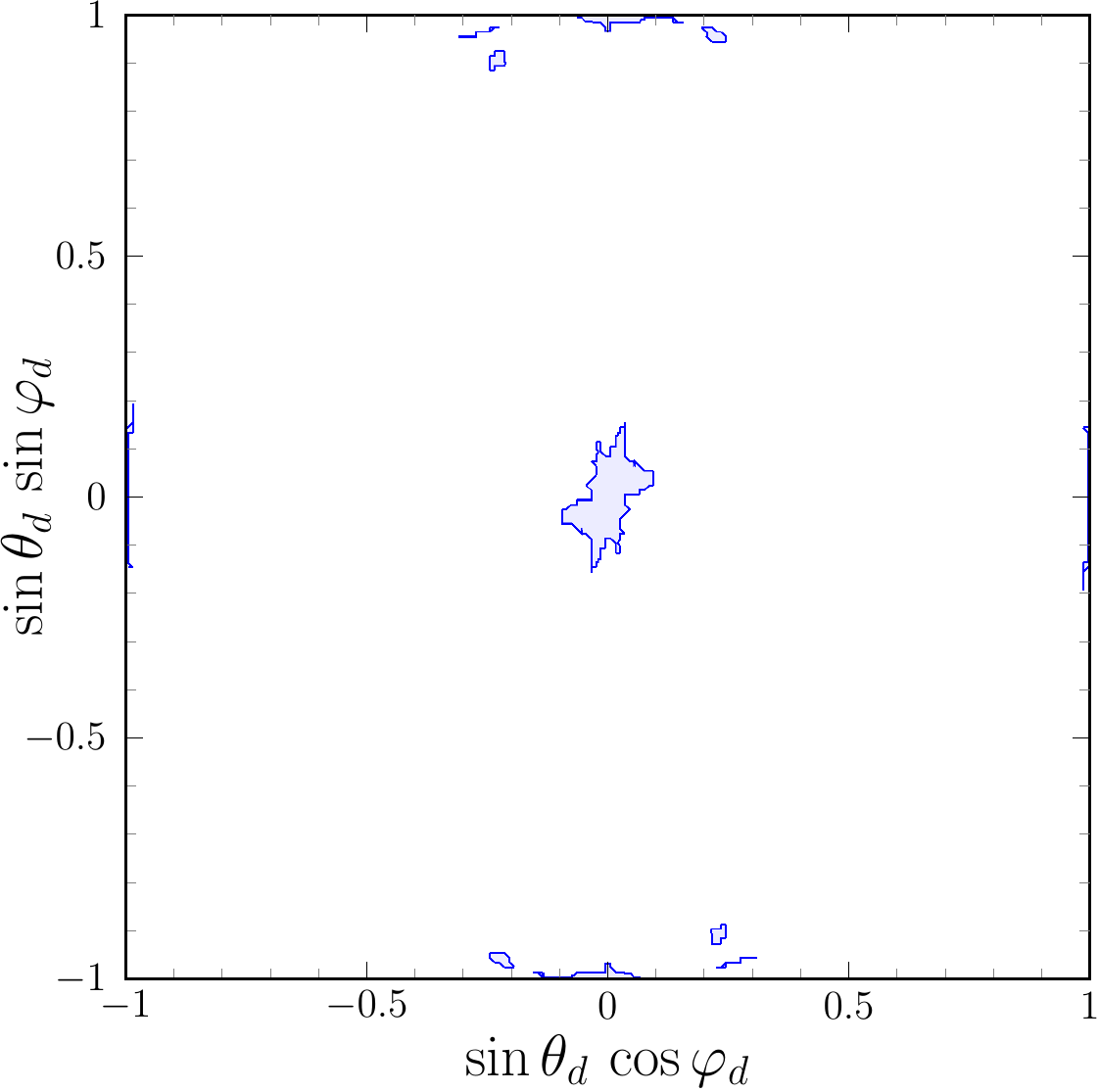}}\quad
\subfigure[$|\rnd{Min}|$ vs. $|\rnd{Mid}|$.\label{FIG:rd:Min:goodAll}]{\includegraphics[width=0.3\textwidth]{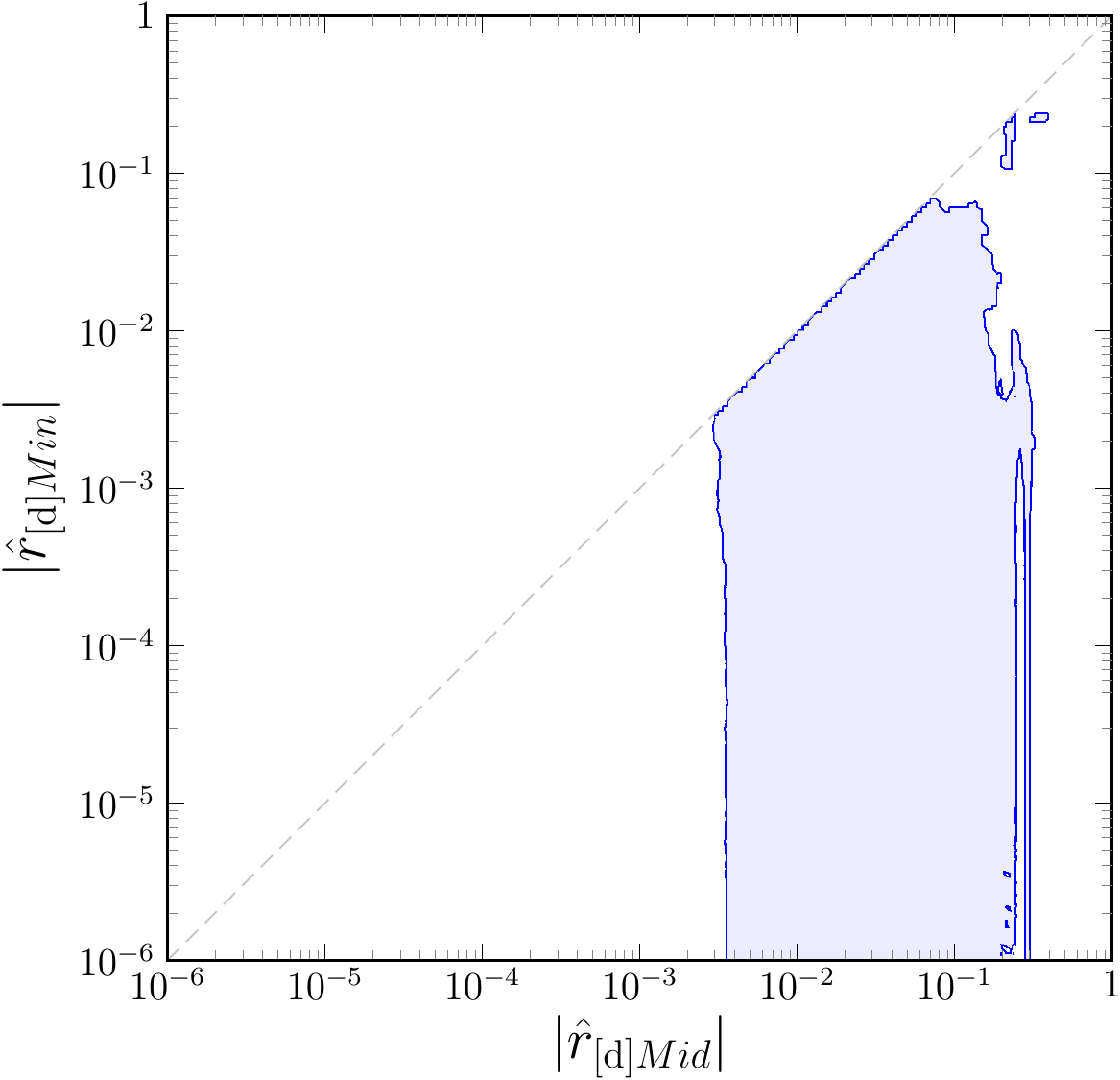}}\quad
\subfigure[$|\rnu{Min}|$ vs. $|\rnu{Mid}|$.\label{FIG:ru:Min:goodAll}]{\includegraphics[width=0.3\textwidth]{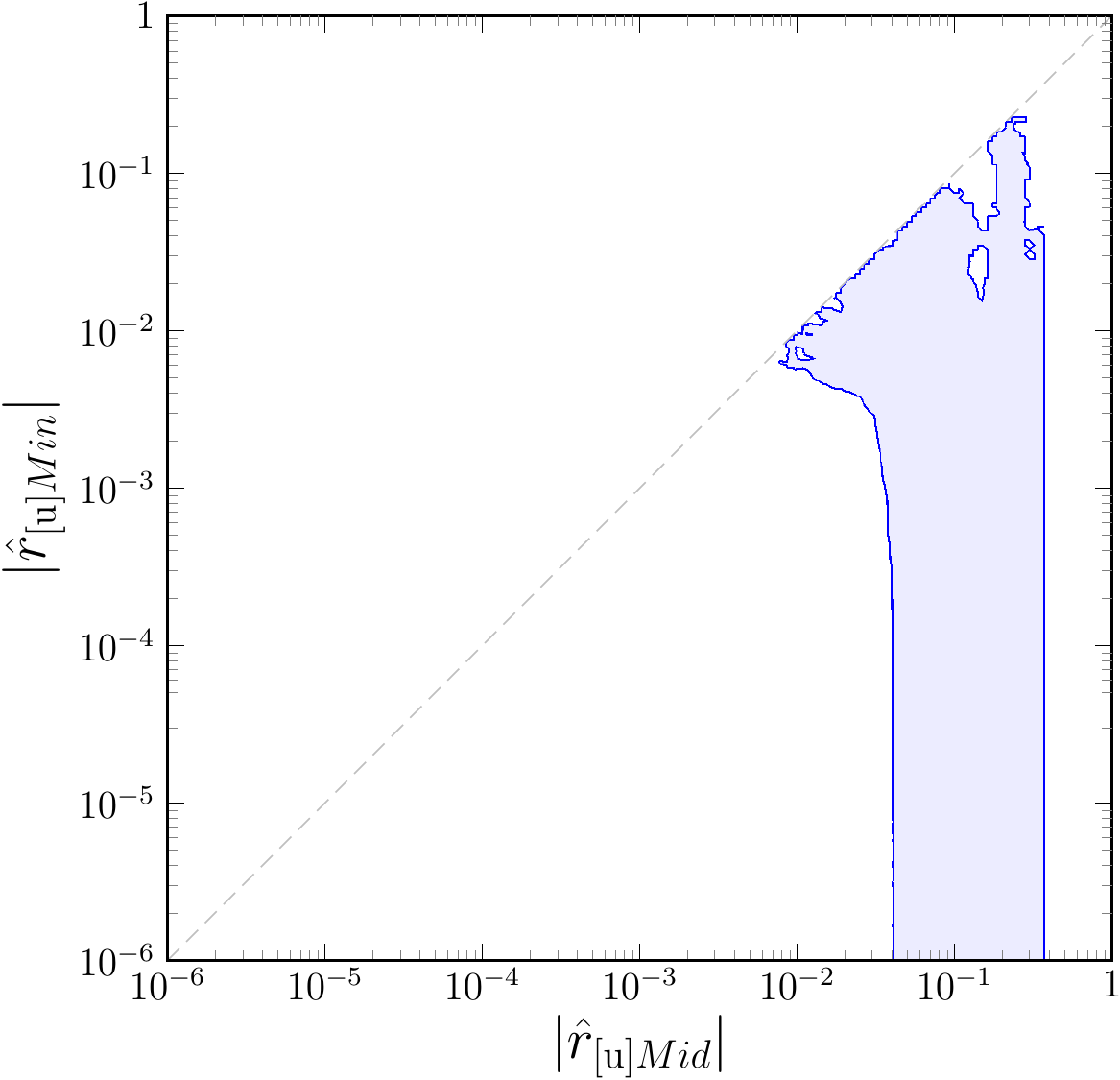}}\\
\subfigure[$\rnd{3}$ vs. $\rnd{2}$\label{FIG:hemlog:d100}]{\includegraphics[width=0.29\textwidth]{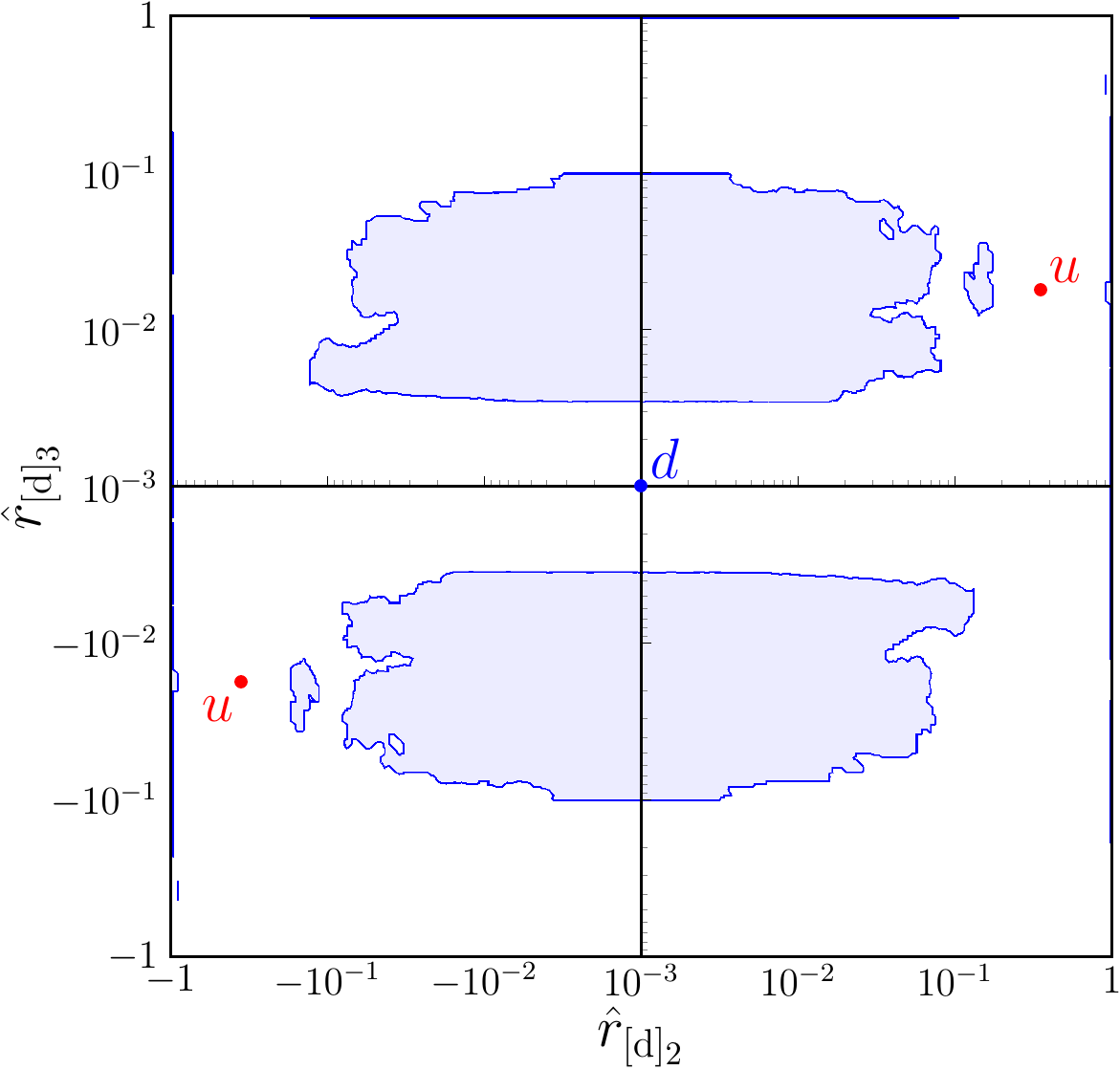}}\quad
\subfigure[$\rnd{3}$ vs. $\rnd{1}$\label{FIG:hemlog:d010}]{\includegraphics[width=0.29\textwidth]{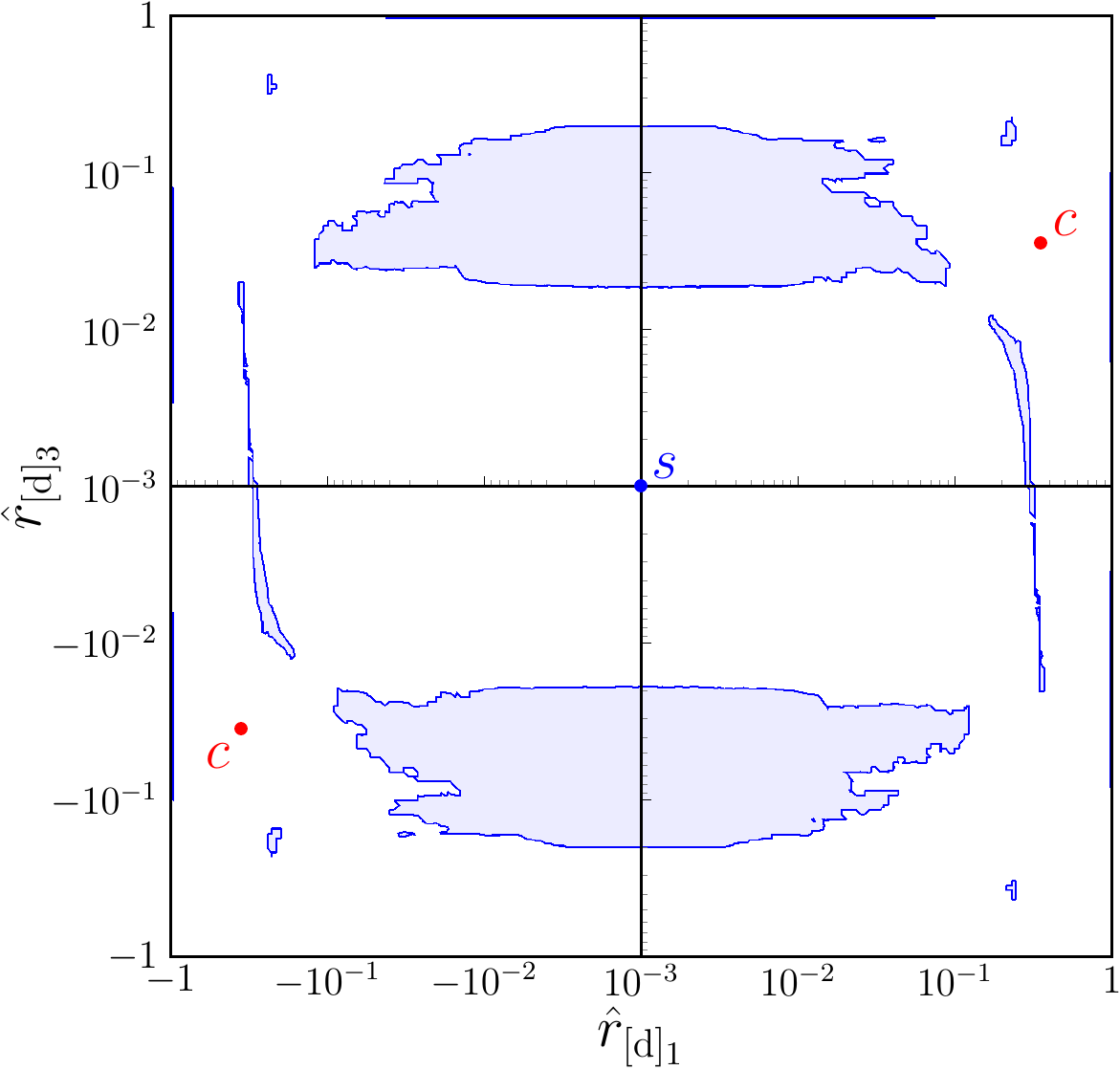}}\quad
\subfigure[$\rnd{2}$ vs. $\rnd{1}$\label{FIG:hemlog:d001}]{\includegraphics[width=0.29\textwidth]{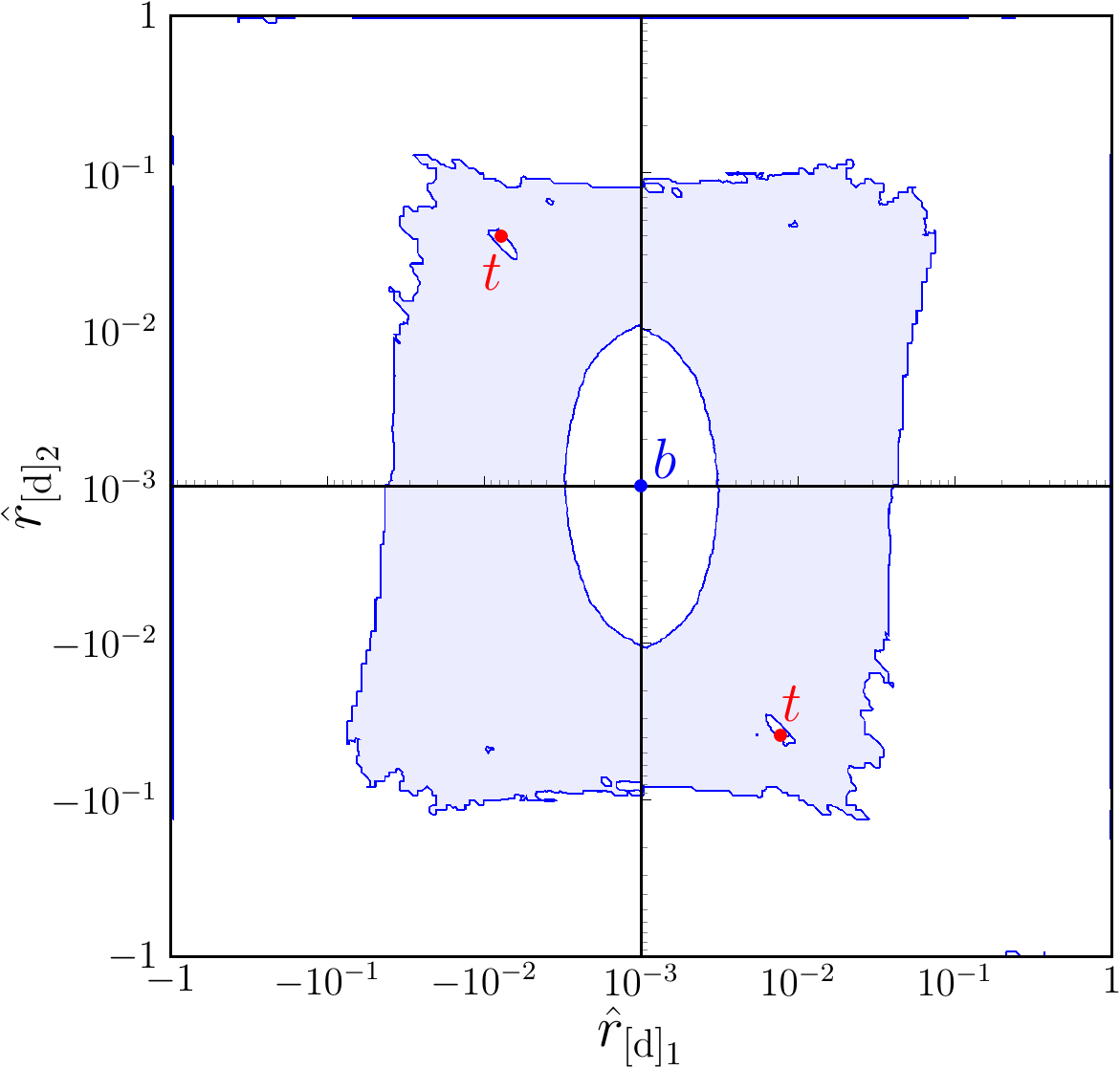}}\\
\caption{Regions allowed at 99\% C.L. by the requirements of the full analysis.\label{FIG:goodAll:1}}
\end{center}
\end{figure}

\begin{figure}[h!tb]
\begin{center}
\subfigure[$M_{Min}$ vs. $\tan\beta$.\label{FIG:Mminlogtanbeta:goodAll}]{\includegraphics[width=0.3\textwidth]{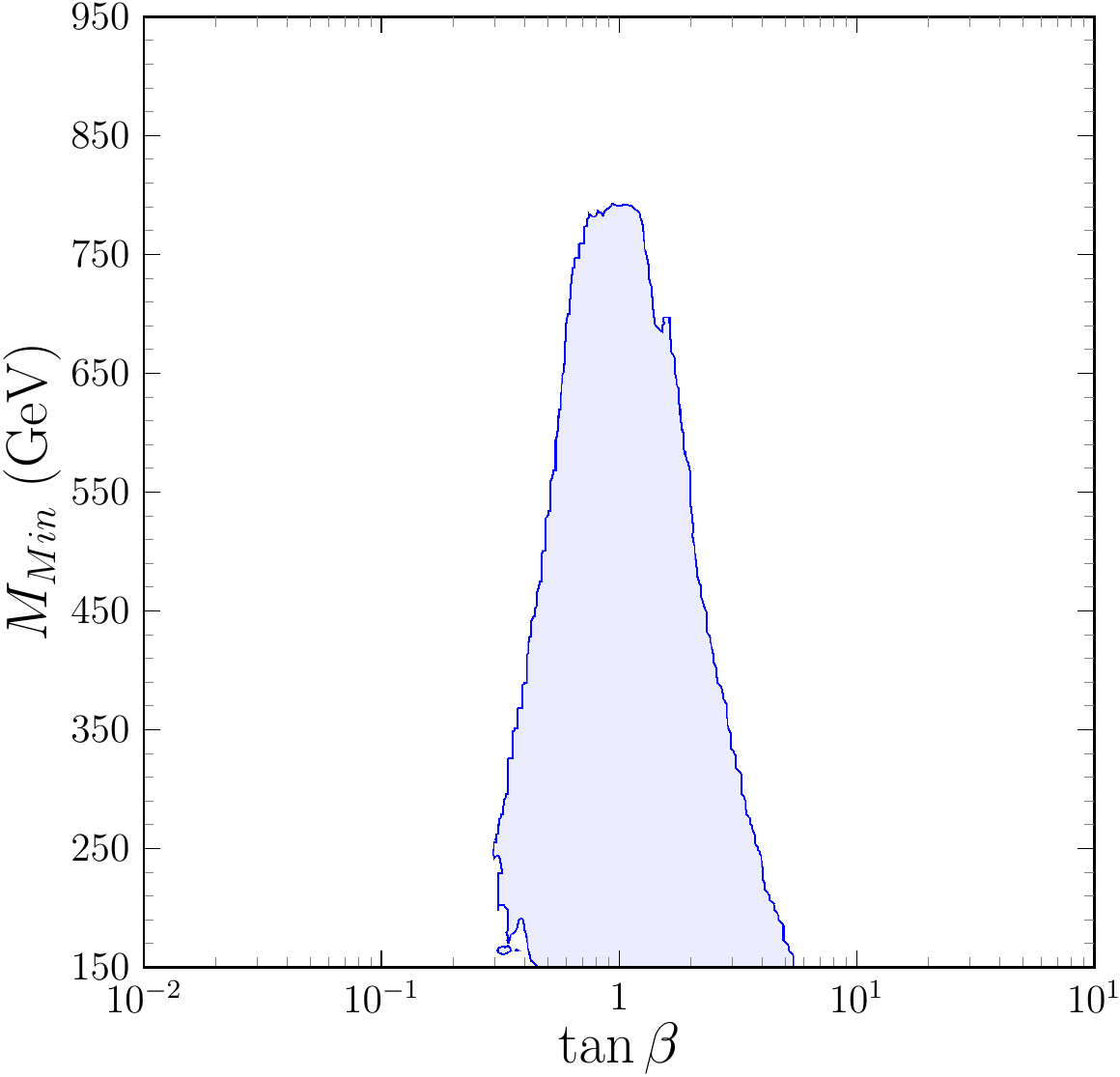}}\qquad
\subfigure[$M_{Max}$ vs. $\tan\beta$.\label{FIG:Mmaxlogtanbeta:goodAll}]{\includegraphics[width=0.3\textwidth]{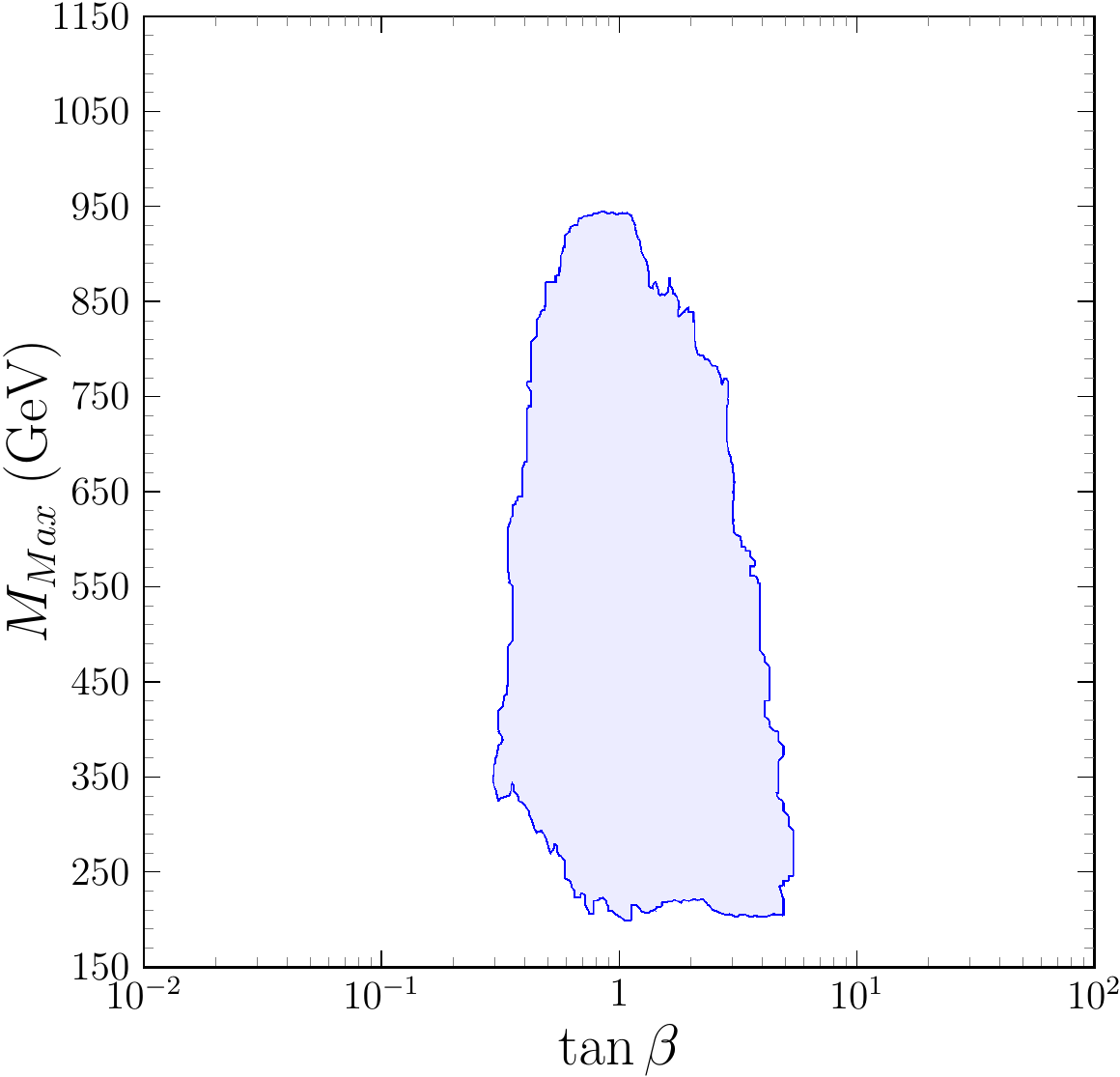}}\\
\subfigure[$M_{Max}$ vs. $M_{Min}$.\label{FIG:MmaxMmin:goodAll}]{\includegraphics[width=0.315\textwidth]{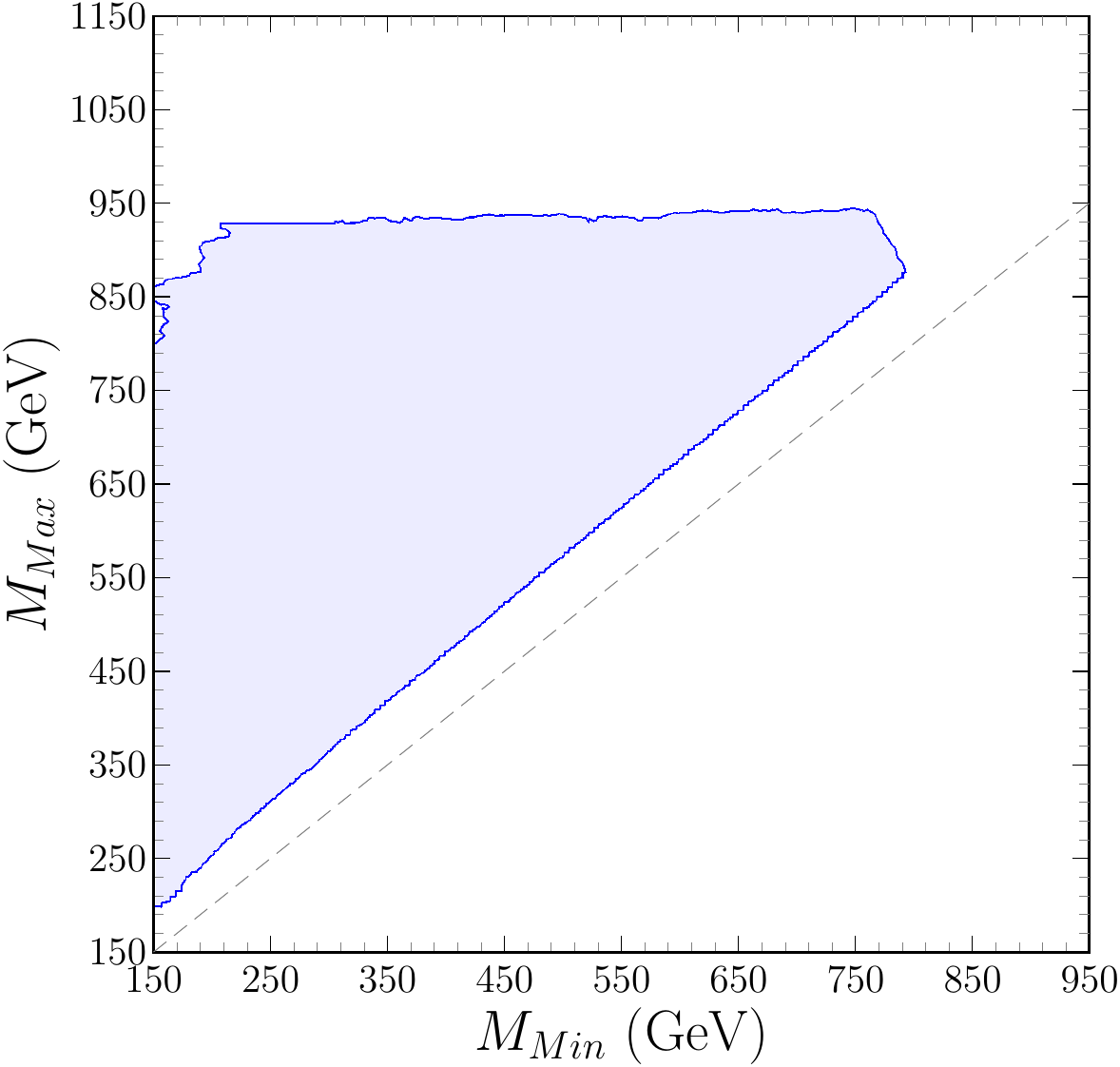}}\qquad 
\subfigure[$\tan\beta$ vs. $\abs{\sin 2\theta}$.\label{FIG:logtanbetalogs2theta:goodAll}]{\includegraphics[width=0.3\textwidth]{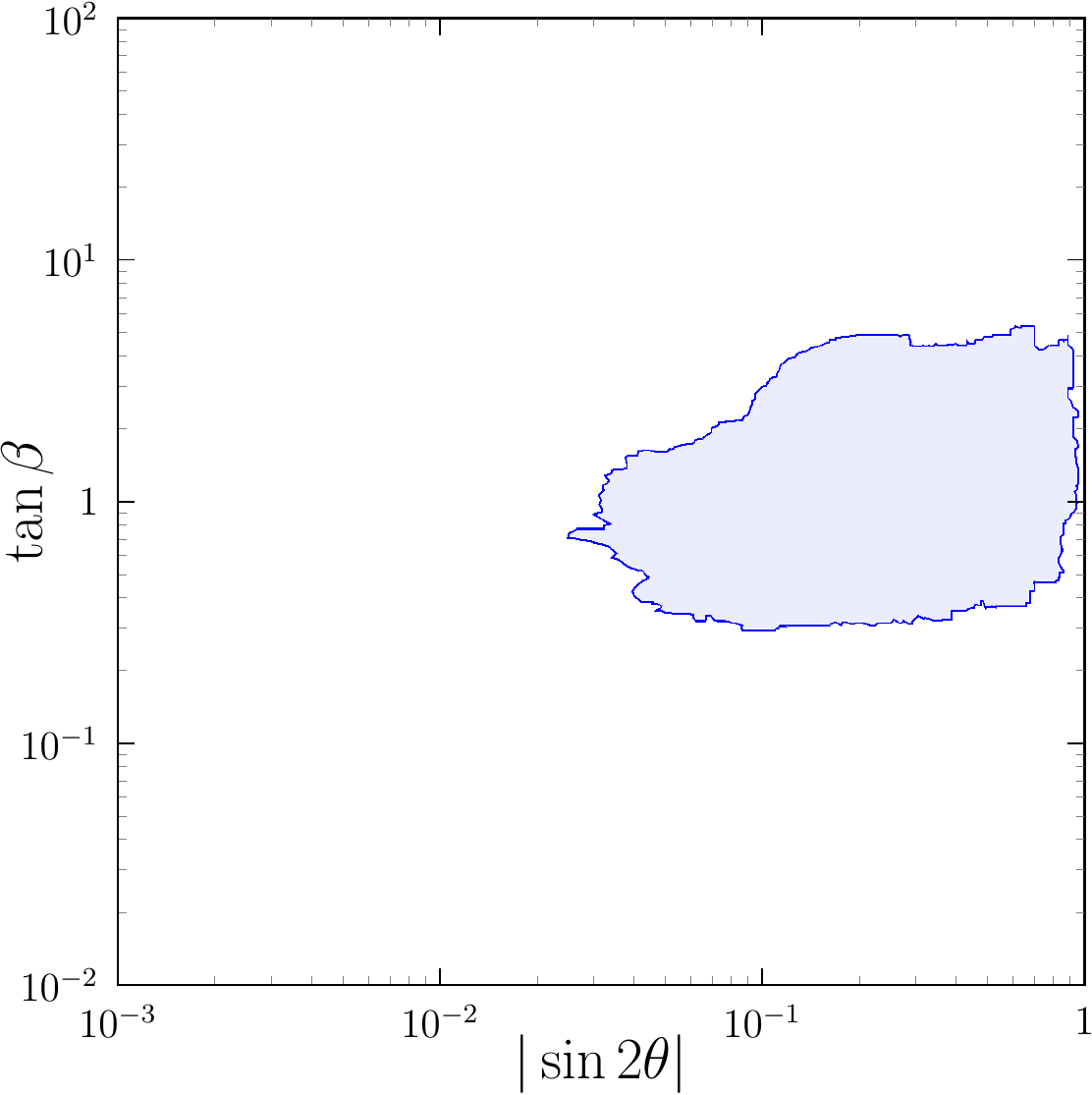}}
\caption{Regions allowed at 99\% C.L. by the requirements of the full analysis.\label{FIG:goodAll:2}}
\end{center}
\end{figure}

\begin{figure}[h!tb]
\begin{center}
\subfigure[$\mA$ vs. $\mcH$.\label{FIG:mA:mcH}]{\includegraphics[width=0.3\textwidth]{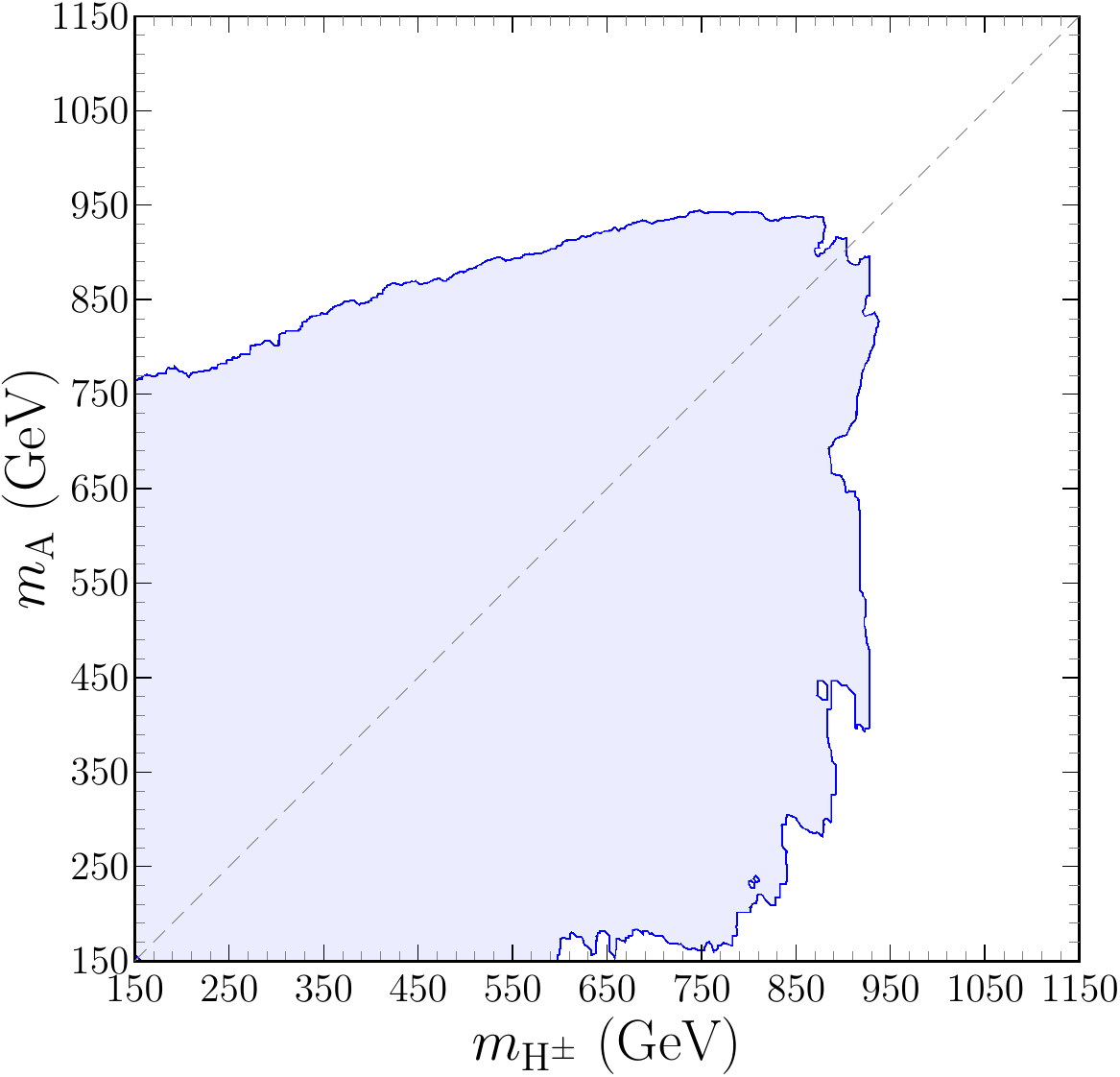}}\quad
\subfigure[$\mA$ vs. $\mH$.\label{FIG:mA:mH}]{\includegraphics[width=0.3\textwidth]{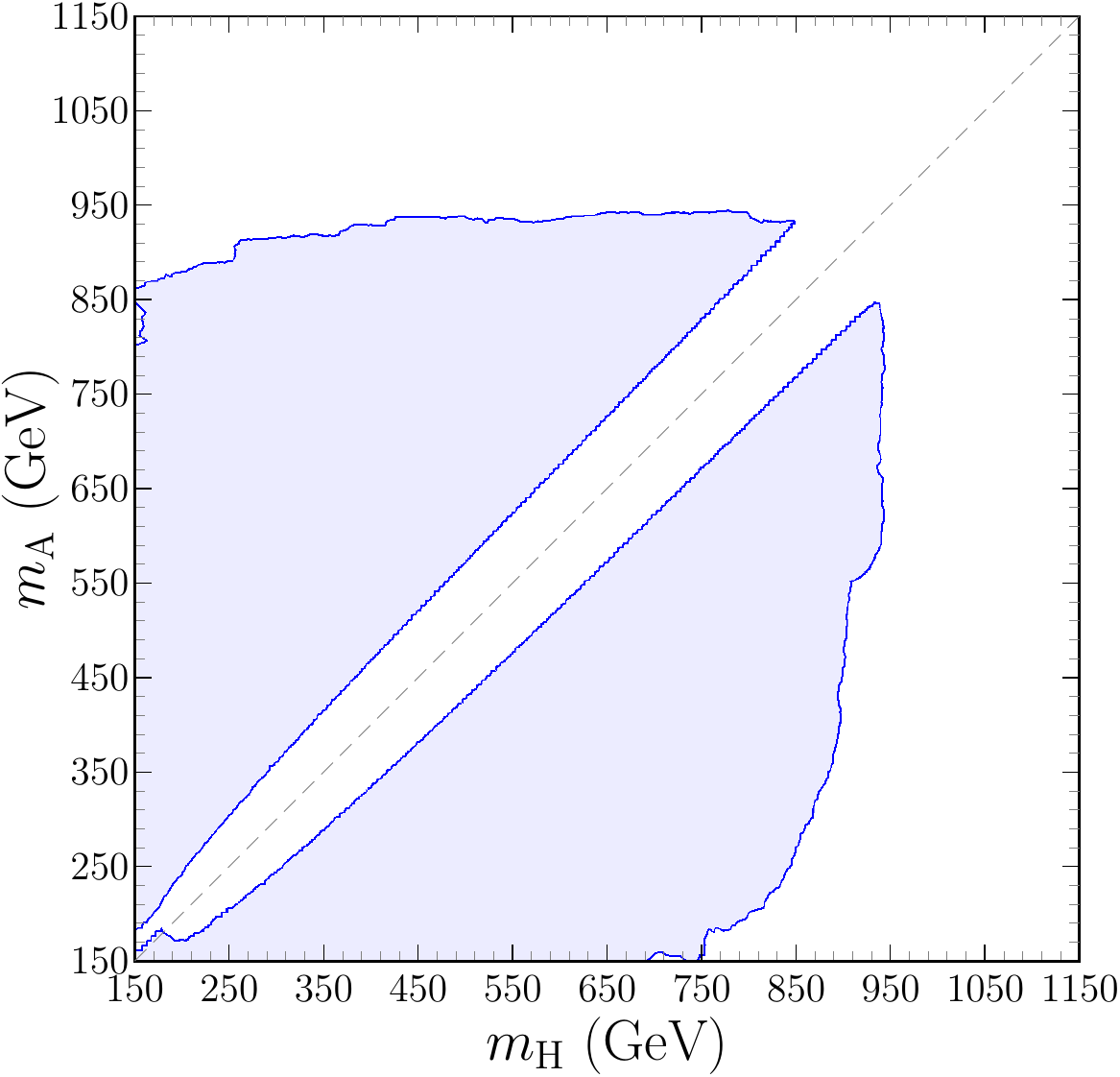}}\quad
\subfigure[$\mH$ vs. $\mcH$.\label{FIG:mH:mcH}]{\includegraphics[width=0.3\textwidth]{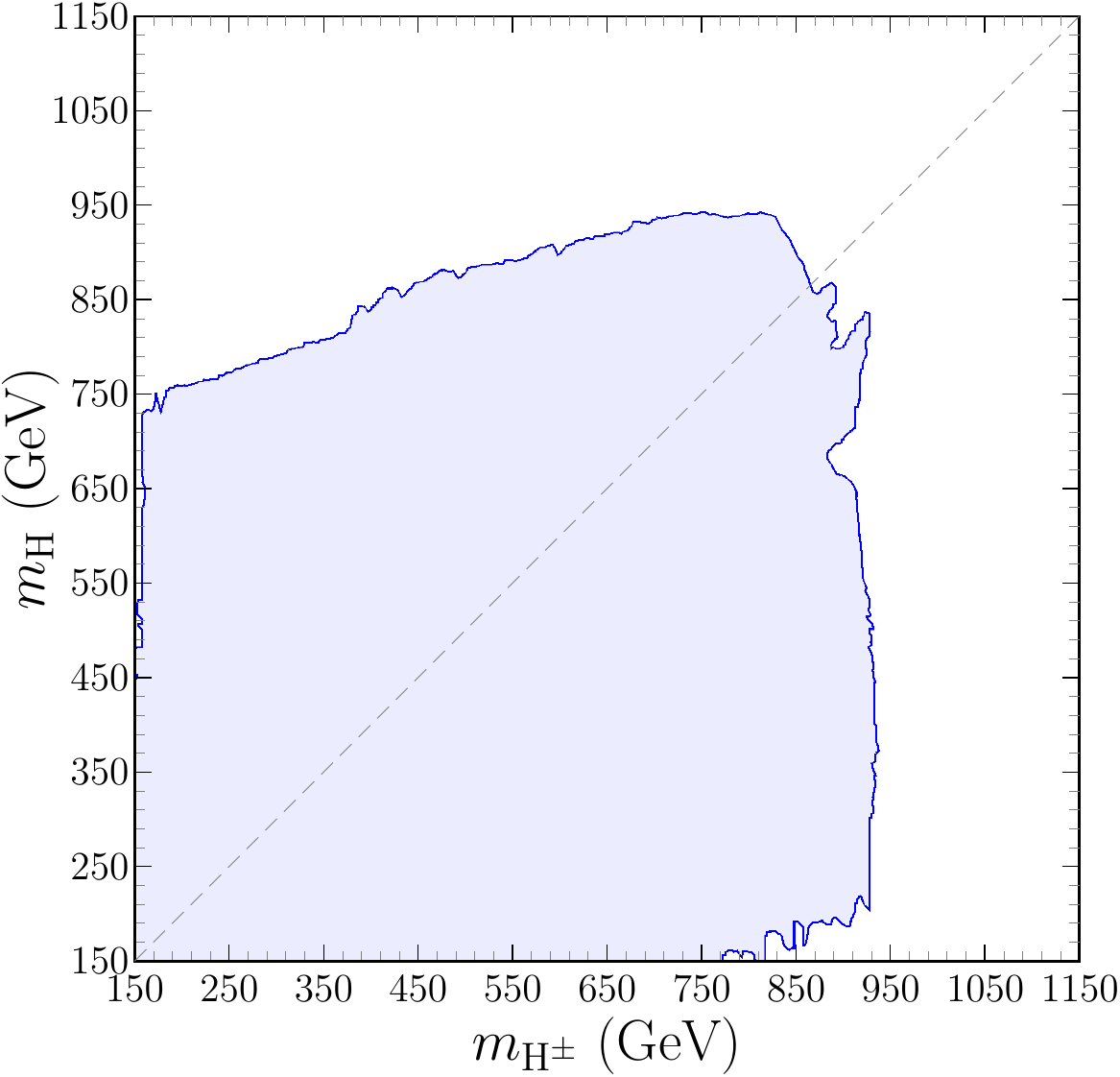}}
\caption{Regions allowed at 99\% C.L. by the requirements of the full analysis.\label{FIG:goodAll:3}}
\end{center}
\end{figure}

\begin{figure}[h!tb]
\begin{center}
\subfigure[$\tb$ vs. $\abs{\ROT{11}}$.\label{FIG:logtb:R11}]{\includegraphics[width=0.3\textwidth]{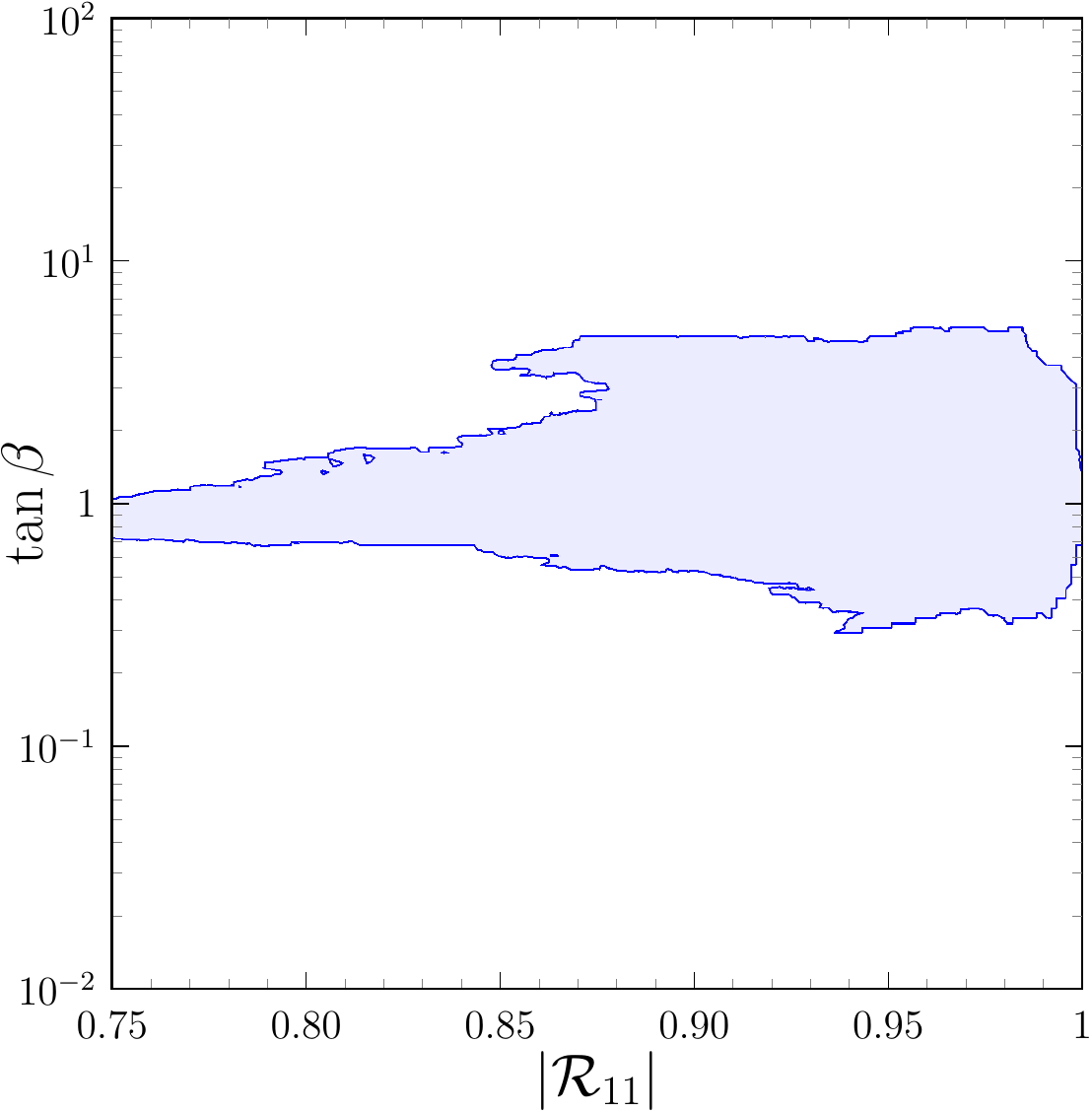}}\quad
\subfigure[$\abs{\sin 2\theta}$ vs. $\abs{\ROT{11}}$.\label{FIG:logs2t:R11}]{\includegraphics[width=0.3\textwidth]{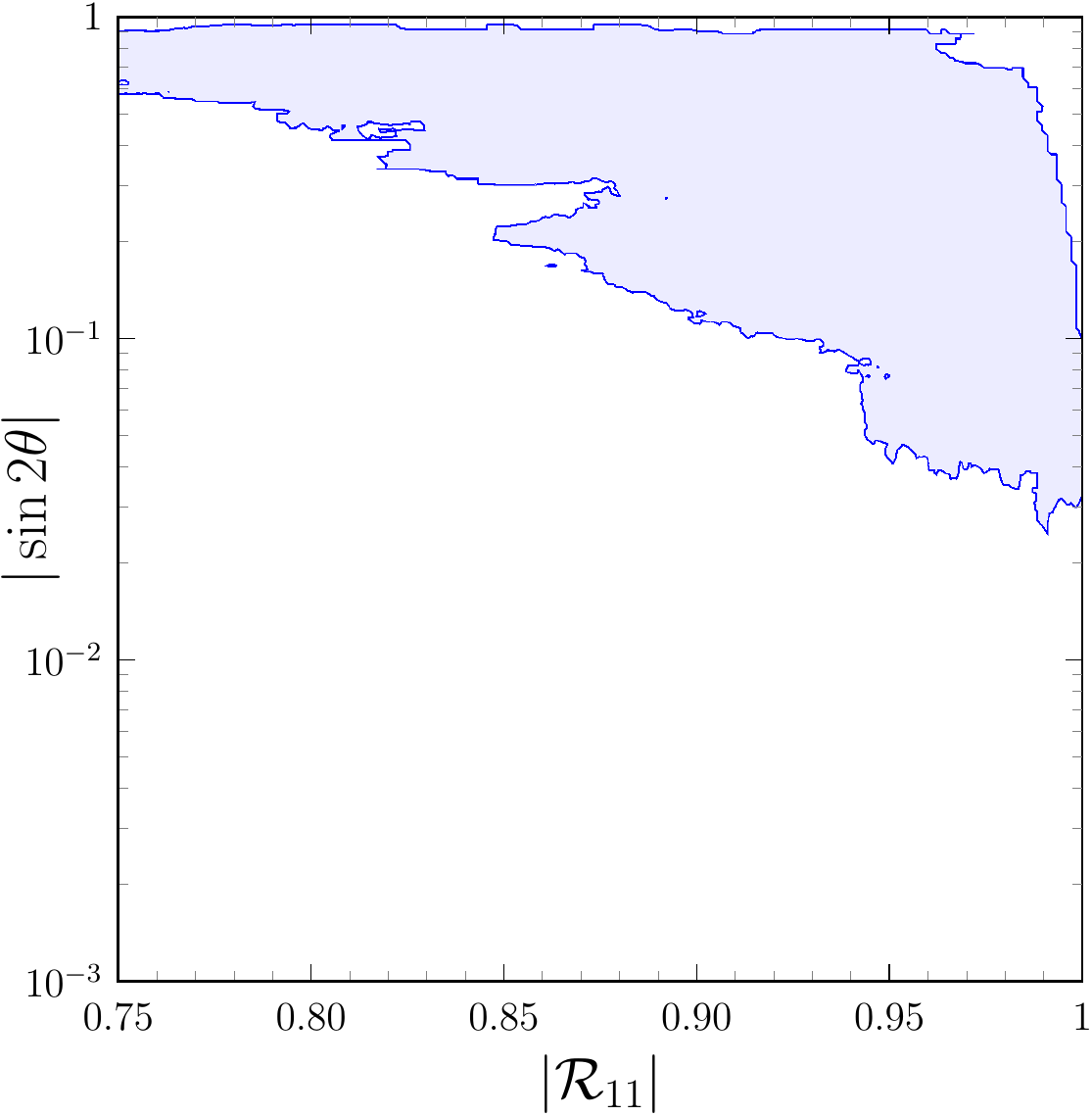}}\quad
\subfigure[$\abs{\sin 2\theta}$ vs. $\ROT{31}$.\label{FIG:logs2t:R31}]{\includegraphics[width=0.3\textwidth]{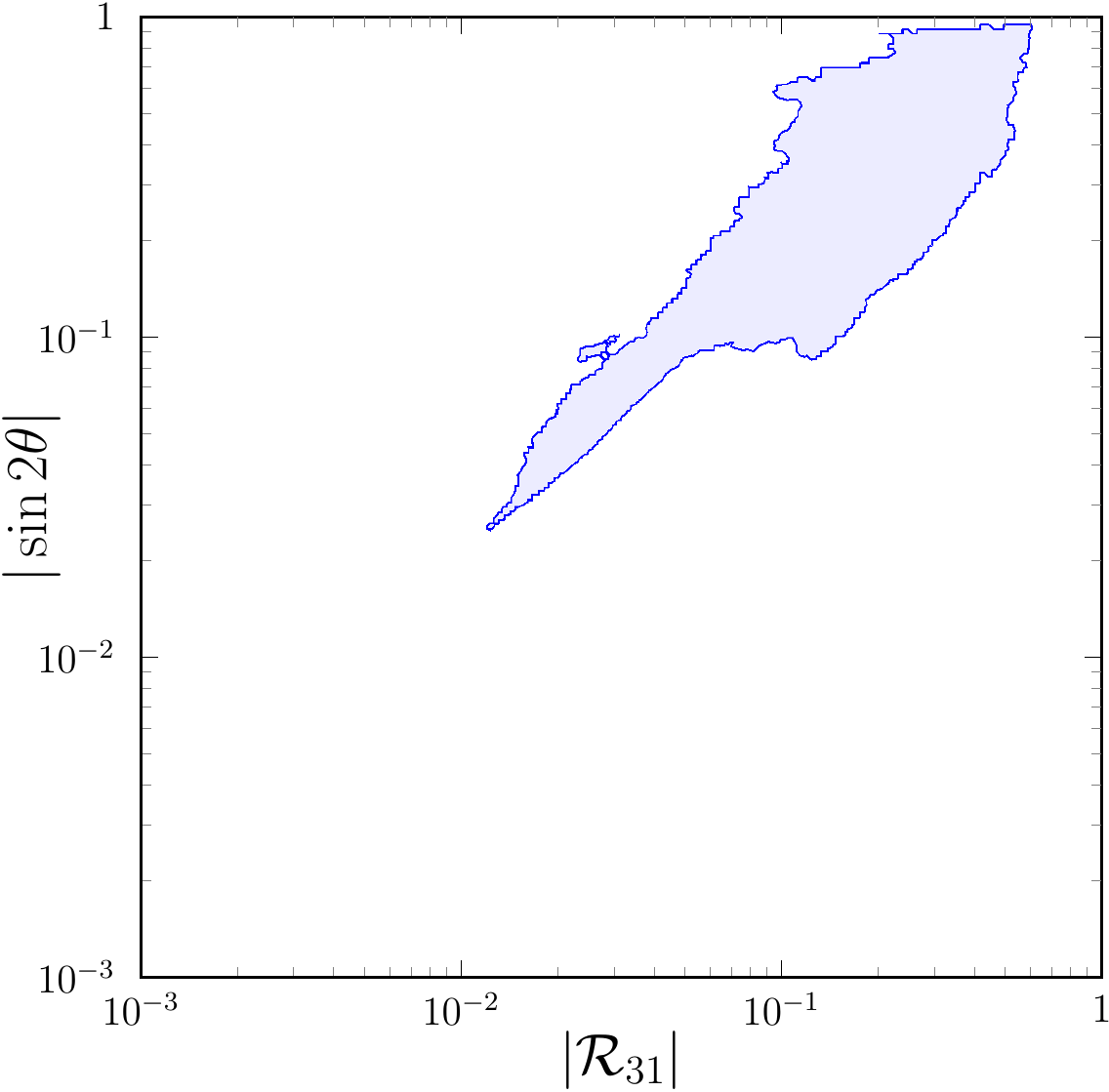}}
\caption{Regions allowed at 99\% C.L. by the requirements of the full analysis.\label{FIG:goodAll:4}}
\end{center}
\end{figure}

\begin{figure}[h!tb]
\begin{center}
\subfigure[$\text{Br}(\nh\to bs)$ vs. $A^{CP}_{J/\Psi\Phi}$.\label{FIG:hbs:CPBs}]{\includegraphics[width=0.345\textwidth]{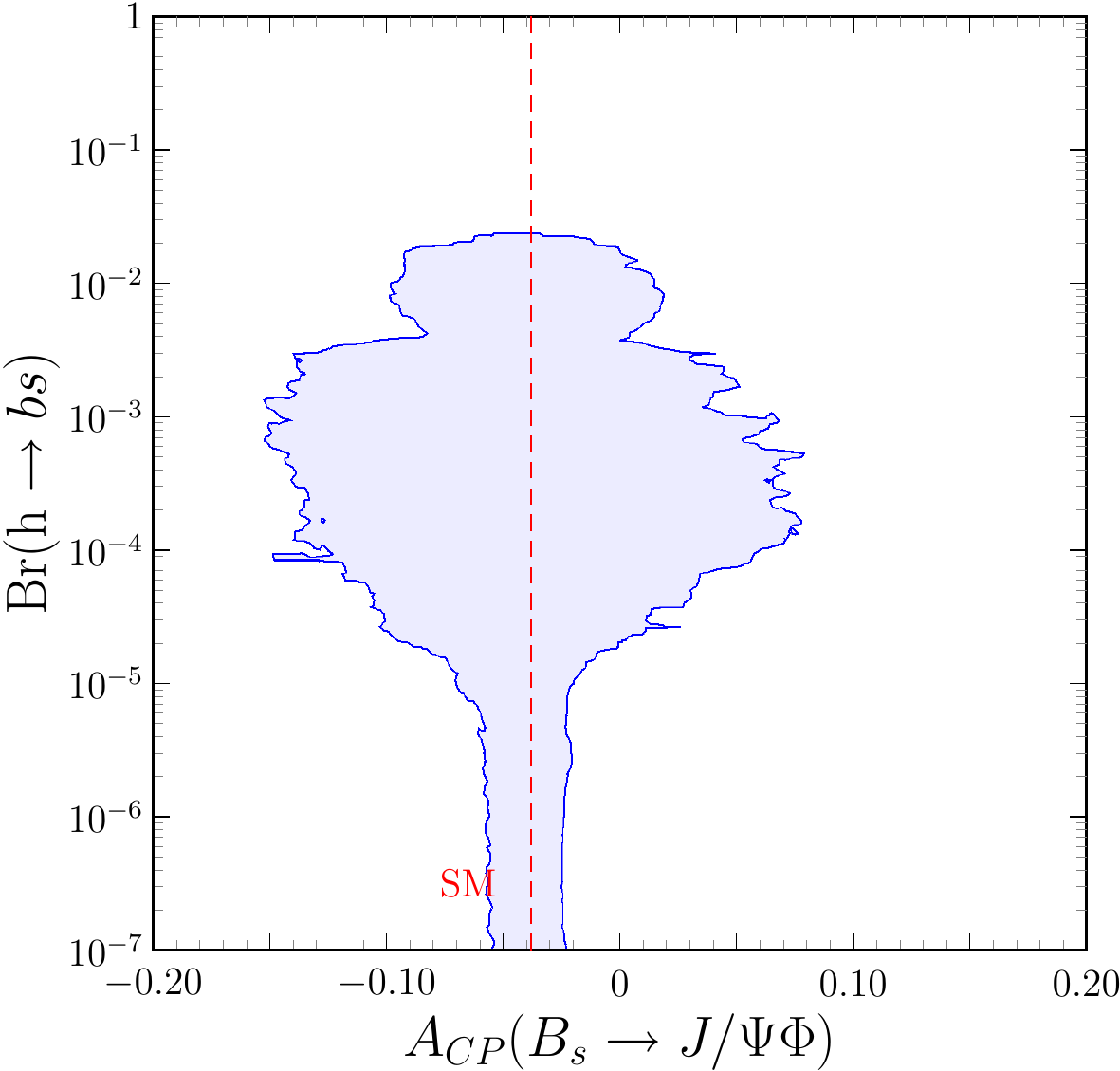}}\qquad
\subfigure[$\text{Br}(\nh\to q_1q_2)$ vs. $\tb$.\label{FIG:hqq:tb}]{\includegraphics[width=0.345\textwidth]{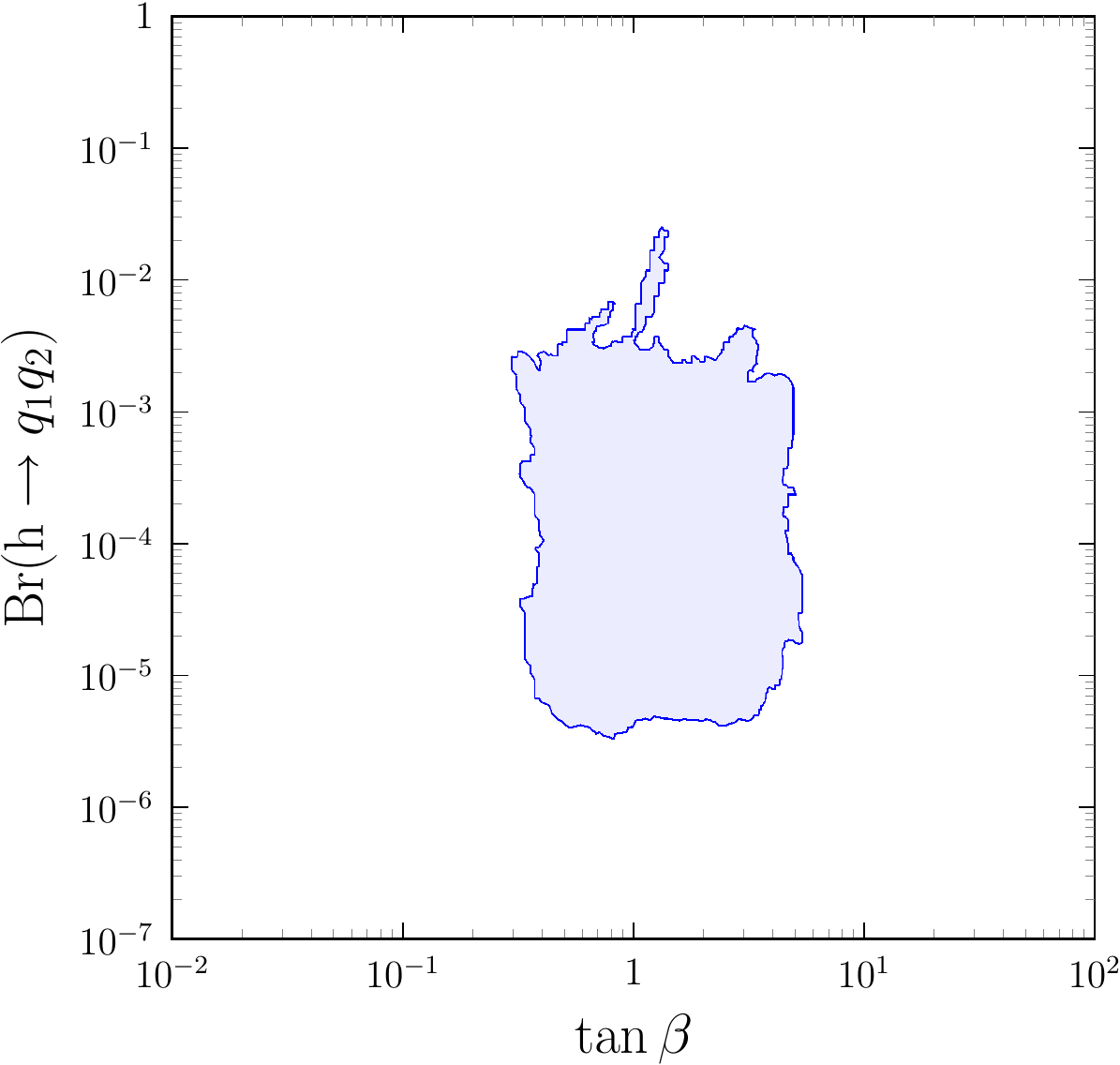}}
\caption{Region allowed at 99\% C.L. by the requirements of the full analysis.\label{FIG:goodAll:5}}
\end{center}
\end{figure}

\begin{figure}[h!tb]
\begin{center}
\subfigure[$t\to\nh q$ vs. $\nh\to q_1q_2$.\label{FIG:thq:hqq}]{\includegraphics[width=0.3\textwidth]{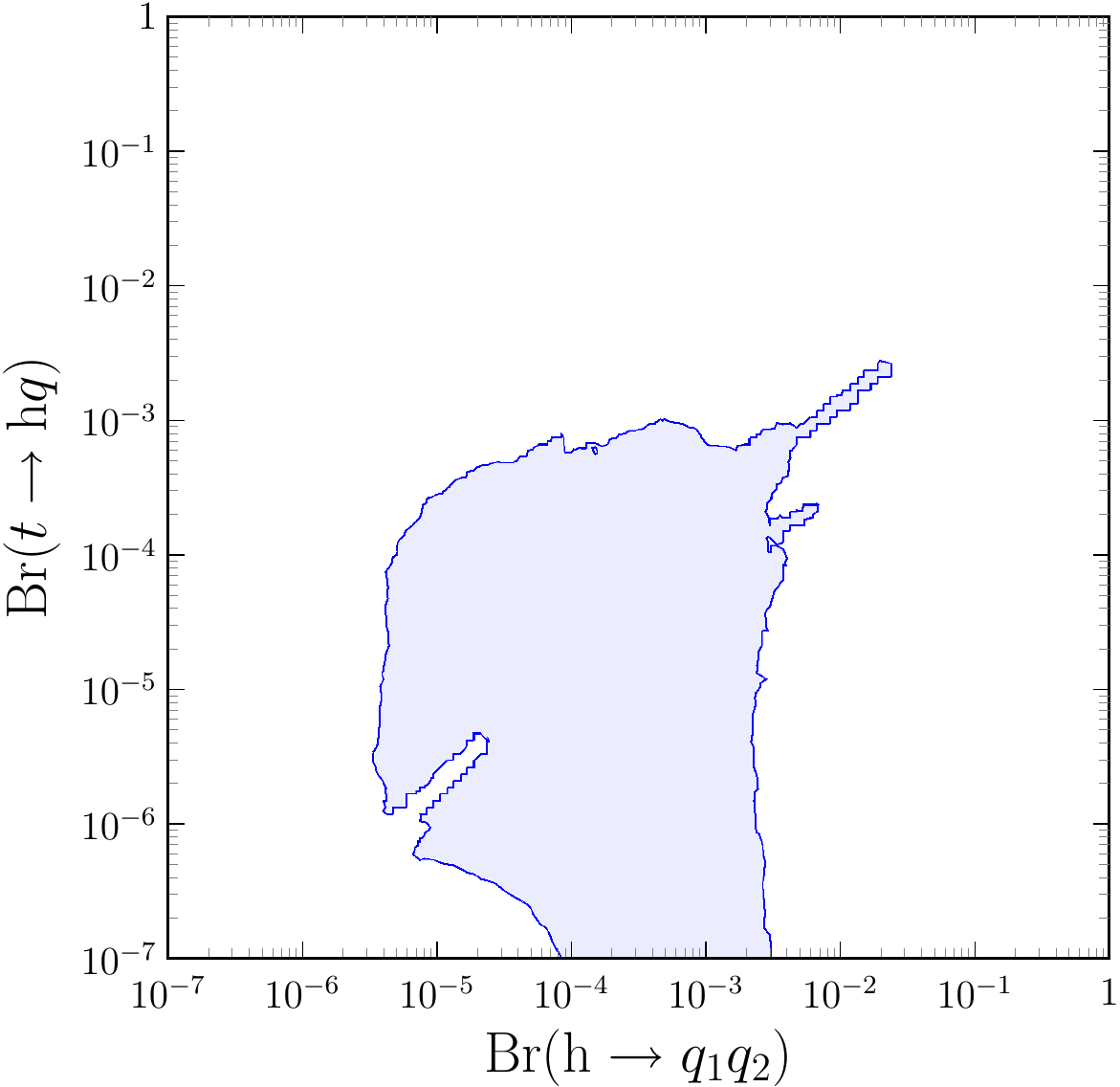}}\quad
\subfigure[$t\to\nh q$ vs. $\nh\to bq$.\label{FIG:thq:hbq}]{\includegraphics[width=0.3\textwidth]{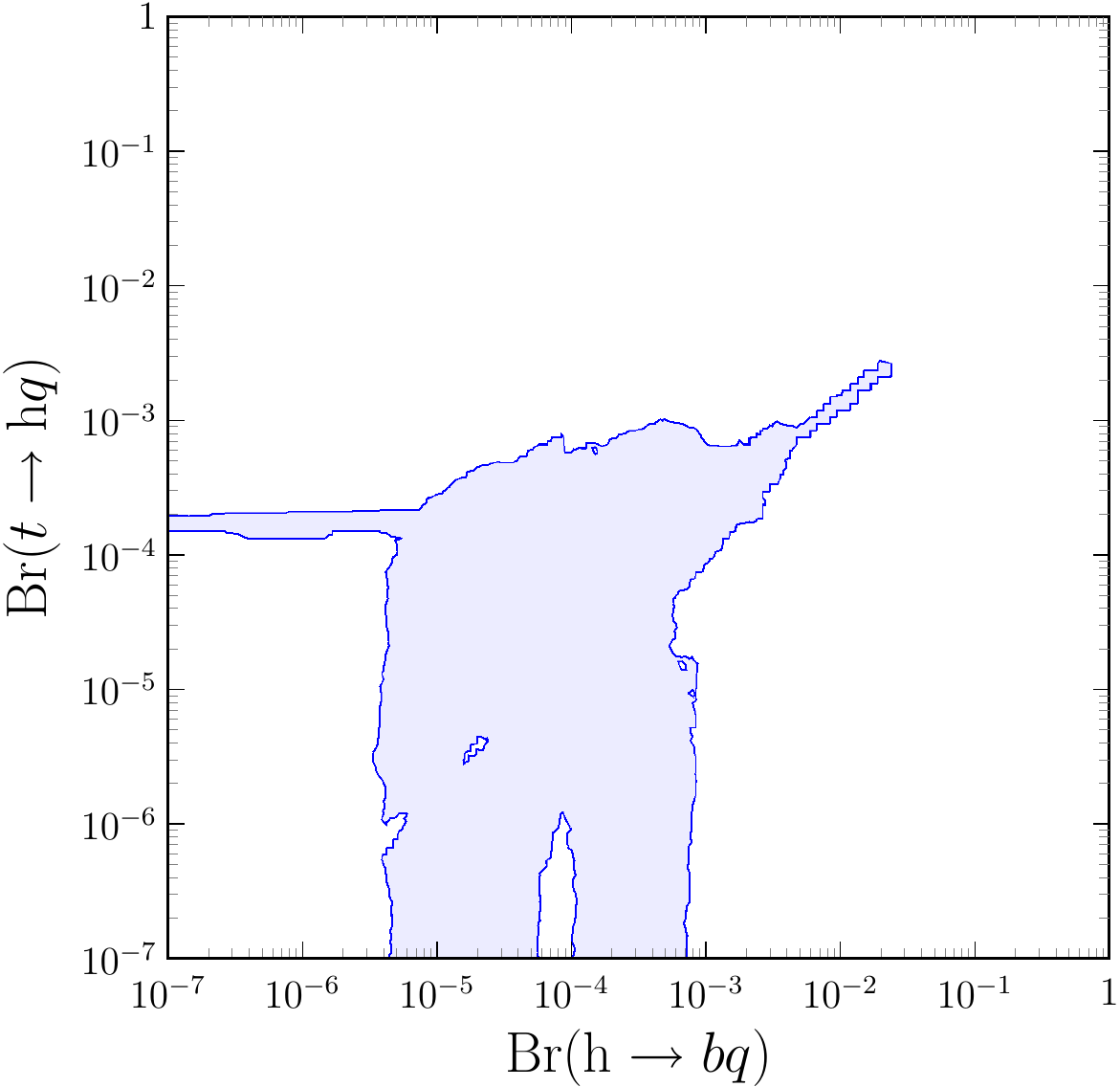}}\quad
\subfigure[$\nh\to q_1q_2$ vs. $\nh\to bq$.\label{FIG:hqq:hbq}]{\includegraphics[width=0.3\textwidth]{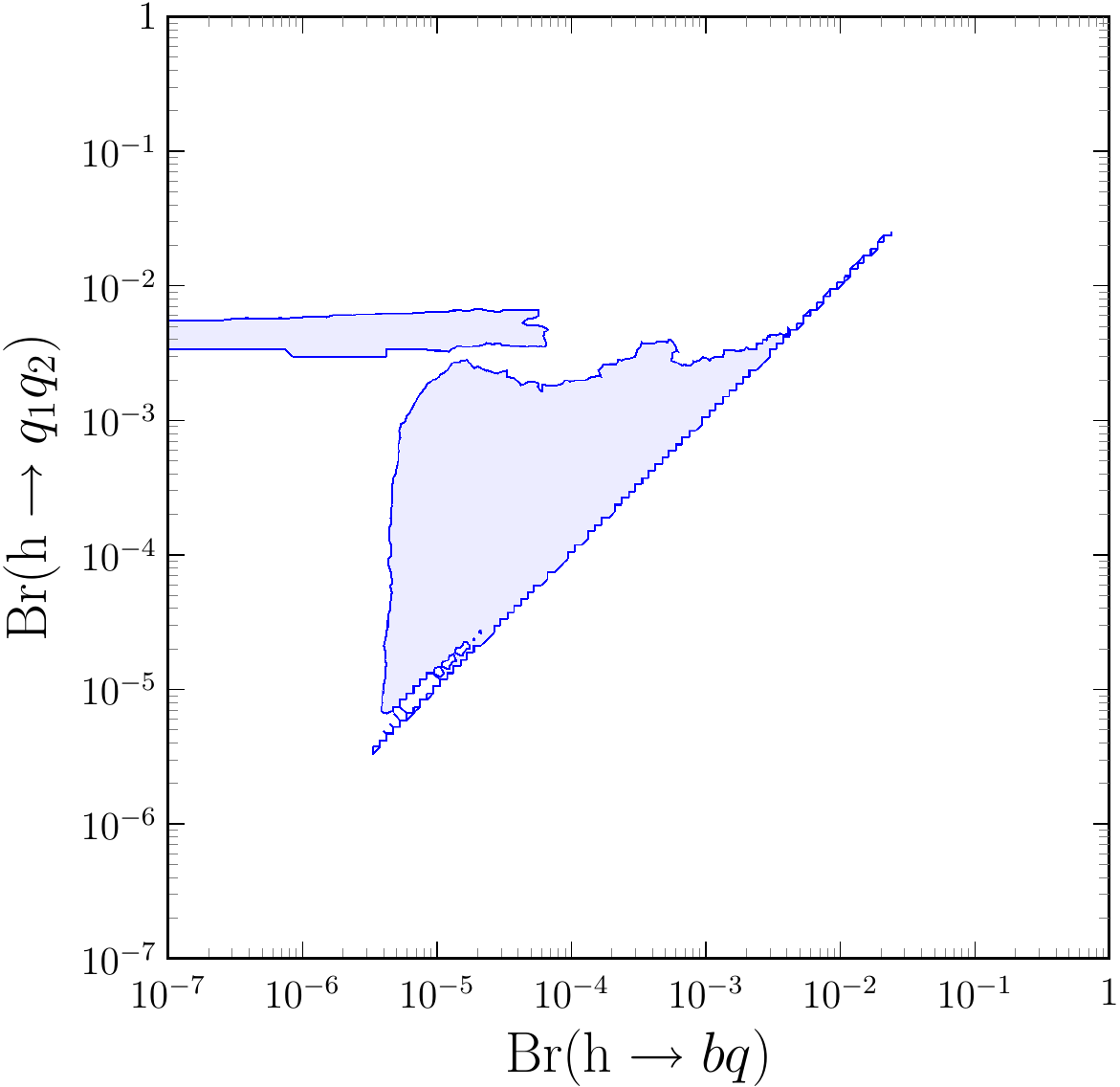}}\\
\subfigure[$t\to\nh u$ vs. $t\to\nh c$.\label{FIG:thu:thc}]{\includegraphics[width=0.3\textwidth]{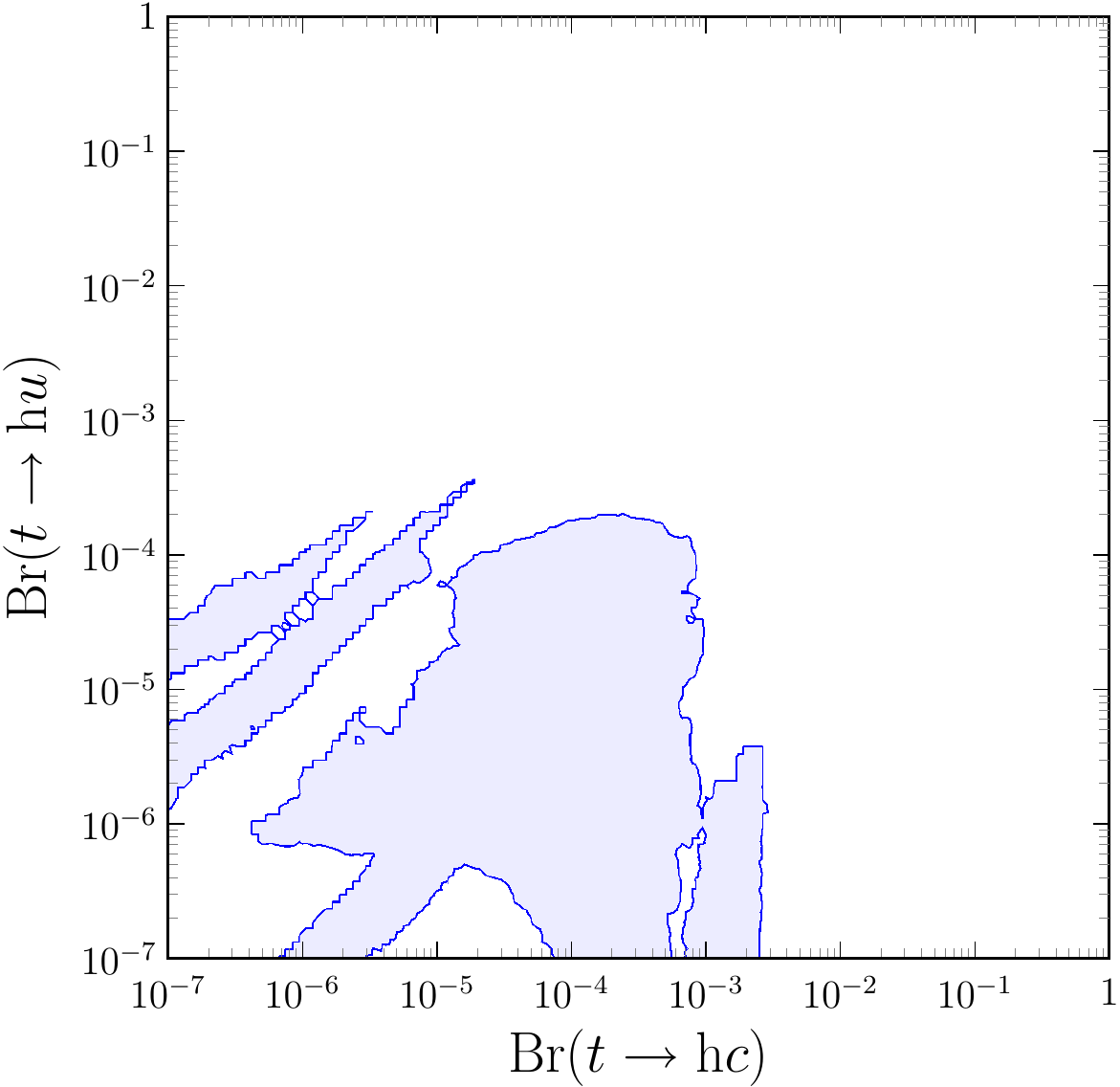}}\quad
\subfigure[$\nh\to bd$ vs. $\nh\to bs$.\label{FIG:hbs:hbd}]{\includegraphics[width=0.3\textwidth]{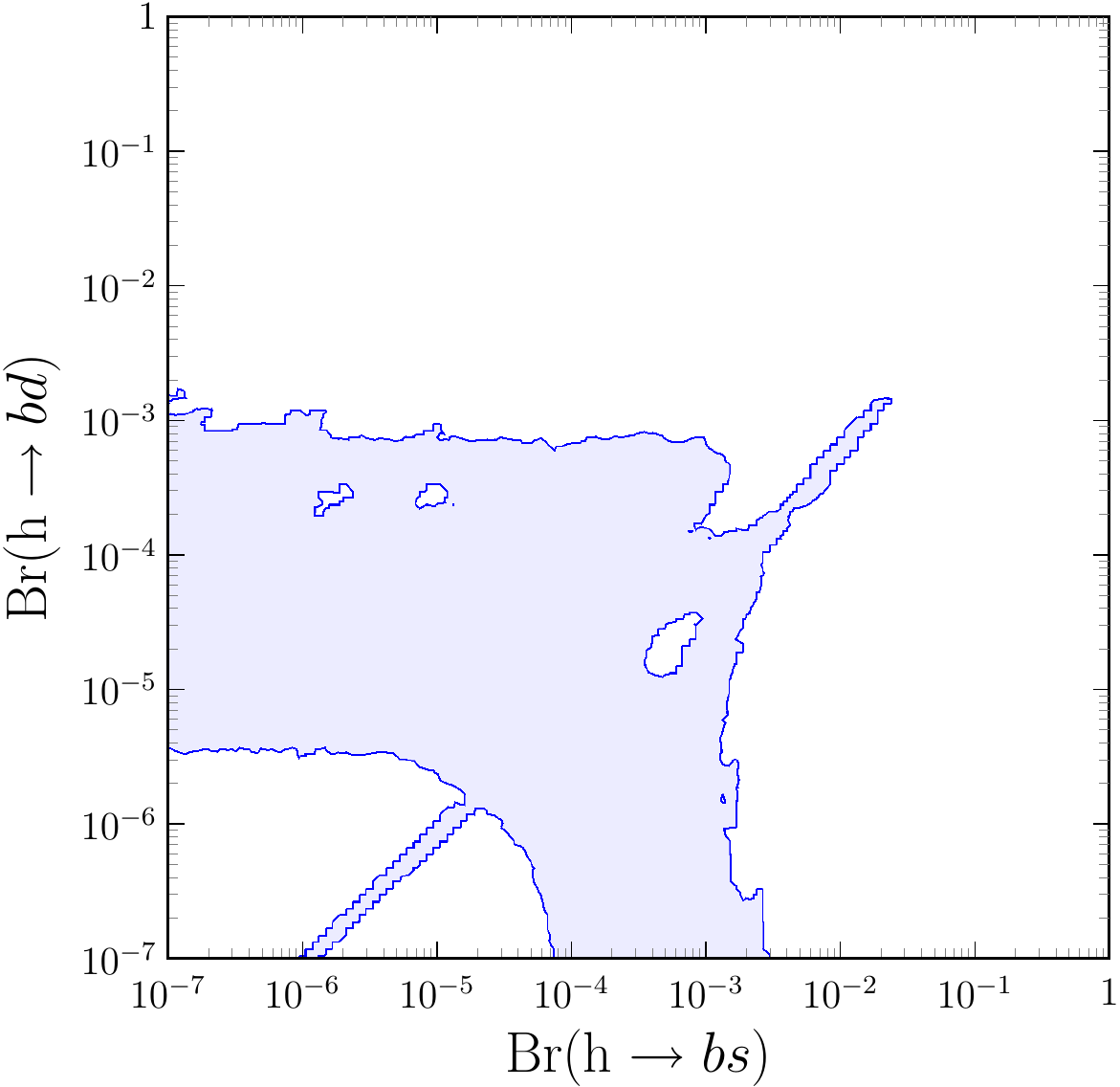}}\quad
\subfigure[$t\to\nh c$ vs. $\nh\to bs$.\label{FIG:thc:hbs}]{\includegraphics[width=0.3\textwidth]{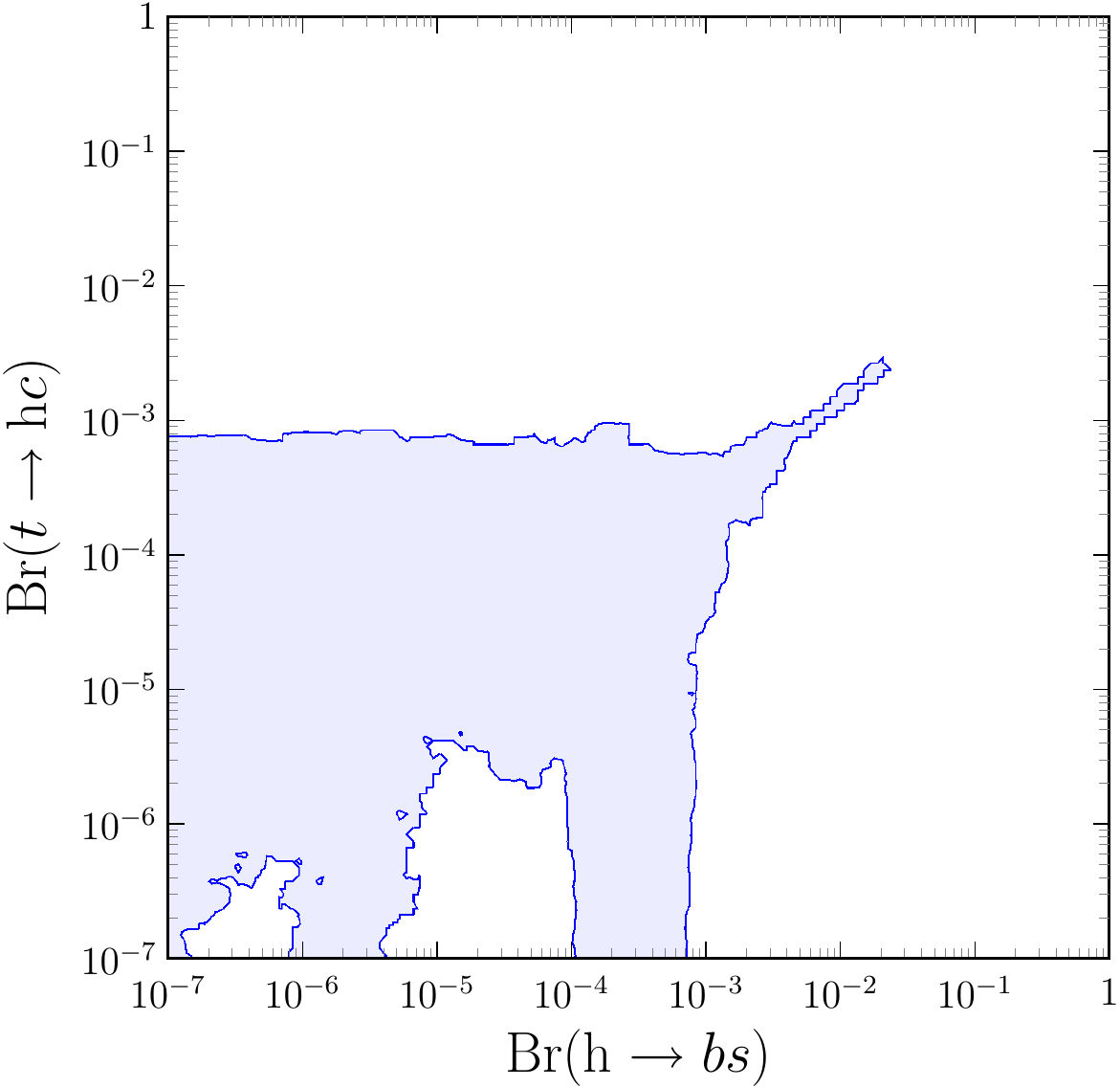}}
\caption{Regions allowed at 99\% C.L. by the requirements of the full analysis.\label{FIG:goodAll:6}}
\end{center}
\end{figure}

The main results of the full analysis are presented in Figures \ref{FIG:goodAll:1} to \ref{FIG:goodAll:6}.\\ 
Figure \ref{FIG:hem:goodAll} corresponds to Figures \ref{FIG:goodCKM:1}--\ref{FIG:goodCKM:3} of the analysis in section \ref{SEC:CKM}: as one could anticipate, it is to be noticed that the allowed regions, where the model satisfies all the constraints, are much reduced with respect to the simple requirement of section \ref{SEC:CKM}, i.e. just reproducing a realistic CKM matrix. 
In particular, the only allowed regions for $\theta_d$ and $\varphi_d$ correspond to having one component of $\rndvec$ close to $\pm 1$ (that is close to the points $(0,\pm 1)$, $(\pm 1,0)$, $(0,0)$ in Figure \ref{FIG:hem:goodAll}), and the remaining two components much smaller: this naturally suppresses neutral flavour changing couplings, since they depend on the products of different components.
As discussed in subsection \ref{sSEC:FCNC:CKM} and in appendix \ref{APP:FCNC:CKM}, \emph{without actually reaching that exact point}, at which the CKM matrix becomes CP conserving. From this point of vue, those regions are ``close to'' (but not exactly) the different types of BGL models (as discussed in \cite{Alves:2017xmk}), in which (i) tree SFCNC are absent in one of the quark sectors and (ii) the scalar potential does not permit spontaneous CP violation. This is clearly illustrated by Figures \ref{FIG:hemlog:d100}, \ref{FIG:hemlog:d010} and \ref{FIG:hemlog:d001}, that are enlargements of Figure \ref{FIG:hem:goodAll} with peculiar logarithmic scales where values below $10^{-3}$ have been collapsed to the central point. These central points correspond, respectively, to the flavour structures of the BGL models of types $d$, $s$ and $b$. For example, the $b$ BGL model corresponds to $\rndvec=(0,0,\pm 1)$. The figures show that the allowed regions exclude the BGL models, but remain close. Correspondingly, the regions close to the flavour structures of up type BGL models (which are also indicated in the Figures), remain empty, i.e. they are not allowed. For example, the $t$ BGL model corresponds to the small holes in the allowed region in Figure \ref{FIG:hemlog:d001}.\\ 
Figures \ref{FIG:rd:Min:goodAll} and \ref{FIG:ru:Min:goodAll} correspond to Figures \ref{FIG:rd:Min} and  \ref{FIG:ru:Min} in subsection \ref{sSEC:CKMnumeric}: the range of allowed values for $|\rnd{Mid}|$, $|\rnu{Mid}|$ is reduced in the full analysis, in particular the largest allowed values are now smaller than $0.3$.\\ 
Figure \ref{FIG:goodAll:2} corresponds to Figure \ref{FIG:goodScalar:1} of the analysis of the scalar sector in section \ref{SEC:scalar}. It is clear that the constraints of the full analysis reduce the available room for $\tb$, leaving only $1/4<\tb<4$. Furthermore, the allowed region in $M_{Max}$ vs. $M_{Min}$ in Fig. \ref{FIG:MmaxMmin:goodAll} is slightly reduced with respect to Fig. \ref{FIG:MmaxMmin:goodScalar}. Notice in particular that the region with all new scalars light, i.e. $M_{Max}<250$ GeV, is now almost excluded. On the contrary, the largest values of $M_{Max}$ and $M_{Min}$ coincide with those in section \ref{SEC:scalar}, that is, they are still limited by the requirements on the scalar sector itself. The same comments apply to Figure \ref{FIG:goodAll:3}, which corresponds to Figure \ref{FIG:goodScalar:2} of the analysis of section \ref{SEC:scalar}. Notice that $\abs{\sin 2\theta}$ is now required to be in the range $[0.03;1.0]$.\\ 
Figure \ref{FIG:logtb:R11} shows that deviations from $\abs{\ROT{11}}=1$ can be achieved for almost all values of $\tb$ within the allowed range, while Figures \ref{FIG:logs2t:R11} and \ref{FIG:logs2t:R31} illustrate, as expected, that for large values of $\abs{\sin 2\tCP}$, $\ROT{31}$ (which controls the amount of pseudoscalar $\nHI$ entering $\nh$), reaches the larger allowed values, while $\abs{\ROT{11}}$ is reduced. Notice in particular that, overall, $\abs{\ROT{31}}\geq 10^{-2}$ and that values as large as $\abs{\ROT{31}}\sim 0.4$ are allowed. In any case, even if $\abs{\ROT{11}}$ can reach values very close to $1$, $\abs{\ROT{11}}<1$.


\noindent Finally, Figures \ref{FIG:goodAll:5} and \ref{FIG:goodAll:6} illustrate some New Physics prospects in different flavour changing neutral transitions.\\ 
Figure \ref{FIG:hbs:CPBs} shows\footnote{The notation is $\text{Br}(\nh\to bs)\equiv \text{Br}(\nh\to \bar bs+b\bar s)$, $\text{Br}(\nh\to bd)\equiv \text{Br}(\nh\to \bar bd+b\bar d)$, etc.} $\text{Br}(\nh\to bs)$ vs. $A^{CP}_{J/\Psi\Phi}$; it is interesting to notice that: (i) $\text{Br}(\nh\to bs)$ can reach values as large as $10^{-2}$, relevant for searches at the ILC, and (ii) significant deviations of the SM expectation $A^{CP}_{J/\Psi\Phi}\simeq -0.036$ can arise. An interesting correlation among New Physics effects follows: $A^{CP}_{J/\Psi\Phi}$ values neatly different from SM expectations (the dashed vertical line in Figure \ref{FIG:hbs:CPBs}) would necessarily require values of $\text{Br}(\nh\to bs)$ in the range $10^{-4}$-$10^{-2}$. The origin of such a correlation is clear: the tree level couplings that induce $\nh\to bs$ at that level also contribute significantly to the dispersive amplitude $M_{12}^{B_s}$ in $B^0_s$--$\bar B^0_s$ mixing, changing its phase while maintaining $\abs{M_{12}^{B_s}}$ (i.e. $\Delta M_{B_s}$). According to the discussion on the connection of SFCNC and CP violation in subsection \ref{sSEC:FCNC:CKM}, tree level SFCNC should give
\begin{align}
&\text{Br}(t\to\nh c)+\text{Br}(t\to\nh u)+\text{Br}(\nh\to cu)\neq 0\qquad \text{and}\nonumber\\
&\text{Br}(\nh\to bs)+\text{Br}(\nh\to bd)+\text{Br}(\nh\to sd)\neq 0\,.\label{EQ:hFCdecays:00}
\end{align}
We introduce
\begin{align}
&\text{Br}(t\to\nh q)\equiv \text{Br}(t\to\nh c)+\text{Br}(t\to\nh u),\nonumber\\
&\text{Br}(\nh\to bq)\equiv \text{Br}(\nh\to bs)+\text{Br}(\nh\to bd),\nonumber\\
&\text{Br}(\nh\to q_1q_2)\equiv \text{Br}(\nh\to bq)+\text{Br}(\nh\to sd)+\text{Br}(\nh\to cu).
\end{align}
Concentrating on the decays of $\nh$, \refEQ{EQ:hFCdecays:00} implies, for the total rate of flavour changing decays of $\nh$ $\text{Br}(\nh\to q_1q_2)$, $\text{Br}(\nh\to q_1q_2)\neq 0$. Figure \ref{FIG:hqq:tb} clearly shows that in any case $5\times 10^{-6}\leq\text{Br}(\nh\to q_1q_2)\leq 2\times 10^{-2}$. \\ 
Figure \ref{FIG:goodAll:6} shows different correlations among the branching ratios of flavour changing transitions involving $\nh$. It is important to notice that these New Physics signals are not confined to one particular sector (up or down type quarks) and that the largest allowed rates can be achieved for the transitions with second and third generation quarks, $t\to\nh c$ and $\nh\to bs$. Notice in particular that for $t\to\nh c$, the LHC bounds at the level of $10^{-3}$ do play a role in limiting the allowed regions. Of course, the remaining transitions, $t\to\nh u$ and $\nh\to bd,sd,cu$ are also interesting: even if the largest values of their rates are smaller than the largest values allowed for $t\to\nh c$ and $\nh\to bs$, in some regions of the parameter space they have larger rates than $t\to\nh c$ and $\nh\to bs$, and they can also be within experimental reach. 


\clearpage
\section*{Conclusions\label{SEC:conclusion}}
In this paper we have addressed the question: is it  possible to construct a realistic model with spontaneous CP violation, in the framework of a minimal two Higgs doublet extension of the Standard Model? We show that this is indeed possible. In order to accomplish this task, one has to surmount enormous obstacles, like having a natural suppression of SFCNC and generating a complex CKM matrix from the vacuum phase, with the correct strength of the invariant measure of the amount of CP violation in the quark mixing matrix.

We have shown that a minimal scenario is phenomenological viable, through the introduction of a flavoured $\mathbb{Z}_2$ symmetry, where one of the three quark families is odd under $\mathbb{Z}_2$ while the other two are even. A remarkable feature of the model is its prediction of New Physics which can be discovered at the LHC. More precisely, the model predicts that all the new scalars have a mass below 950 GeV with at least one of the masses below 750 GeV. This prediction is
obtained through a thorough study of the constraints arising from the 125 GeV Higgs signals, the size of neutral meson mixings, the size of $b\to s \gamma$, and reproducing a correct CKM matrix, including the size of CP violation. Constraints from the electroweak oblique parameters, and perturbative unitarity and boundedness of the scalar potential are also included.\\
We encounter a deep connection between the generation of a complex CKM matrix from a vacuum phase and the necessary appearance of SFCNC. In the New Physics predictions, we give special emphasis to processes like $t\to \nh c,\nh u$, $h\to bs,bd$, which are relevant for the LHC and the ILC. Interestingly, there is still room for important New Physics contributions to the phase of $B_s^0$--$\bar B_s^0$ mixing.

In the present model of SCPV, none of the new scalars can be heavier than 1 TeV, and the presence of SFCNC cannot be avoided. The experimental constraints select regions in parameter space where the SFCNC are kept under control, as happens in BGL models. It is indeed remarkable that these allowed regions are located close to BGL models: for example, in the neighbourhood of a down-type BGL flavour structure, we almost do not have SFCNC in the down sector while the SFCNC in the up sector are of the Minimal Flavour Violating type \cite{Botella:2009pq}. Apparently these are the only regions within this model, where one can have
an effective suppression of the dangerous SFCNC.

\section*{Acknowledgments}
This work is partially supported by Spanish MINECO under grant FPA2017-85140-C3-3-P and by the Severo Ochoa Excellence Center Project SEV-2014-0398, by Generalitat Valenciana under grant GVPROMETEOII 2014-049 and by Funda\c{c}\~ao para a Ci\^encia e a Tecnologia (FCT, Portugal) through the projects CERN/FIS-NUC/0010/2015 CFTP-FCT Unit 777 (UID/FIS/00777/2013) and PTDC/FIS-PAR/29436/2017 which are partially funded through POCTI (FEDER), COMPETE, QREN and EU. 
MN acknowledges support from FCT through postdoctoral grant SFRH/BPD/112999/2015.


\appendix

\section{SFCNC and CP Violating CKM\label{APP:FCNC:CKM}}
In subsection \ref{sSEC:FCNC:CKM} we have addressed the incompatibility between a CP violating CKM matrix and the absence of tree level SFCNC in one quark sector, in this model. In this appendix we provide a simple proof that completes the discussion.\\
Let us consider the case of the down quark sector. 
According to \refEQS{eq:N:diag:00} and \eqref{eq:PR3:01},
\begin{equation}\label{eq:Nd:FCCP:00}
\mND=[\tb\id-(\tti)\OdLt\PR{3}\OdL]\mMD\,,\quad [\OdLt\PR{3}\OdL]_{ij}=\rnd{i}\rnd{j}\,.
\end{equation}
Tree level SFCNC in the down sector are absent when $\OdLt\PR{3}\OdL$ is diagonal, that is $\rnd{i}=\delta_{ik}$ for some $k$ (1 or 2 or 3). In that case,
\begin{equation}
[\OdLt\PR{3}\OdL]_{ij}=\delta_{ik}\delta_{jk}=\delta_{ik}\delta_{ij}=[\PR{k}]_{ij}
\end{equation}
with the projectors
\begin{equation}
\PR{1}=\begin{pmatrix}1&0&0\\ 0&0&0\\ 0&0&0\end{pmatrix},\quad
\PR{2}=\begin{pmatrix}0&0&0\\ 0&1&0\\ 0&0&0\end{pmatrix},\quad
\PR{3}=\begin{pmatrix}0&0&0\\ 0&0&0\\ 0&0&1\end{pmatrix}.
\end{equation}
On the other hand, the CKM matrix in \refEQ{EQ:CKM:ROT:rd:ru} reads in that case
\begin{equation}
\CKM=\OuLt[\id+(e^{i2\theta}-1)\PR{3}]\OdL=\OROTmat\,[\id+(e^{i2\theta}-1)\OdLt\PR{3}\OdL]=\OROTmat\,[\id+(e^{i2\theta}-1)\PR{k}],
\end{equation}
it is the product of a real orthogonal matrix $\OROTmat$ and a diagonal matrix of phases, and hence CP conserving.

For the up quark sector, the reasoning is analogous: the absence of tree level SFCNC requires $\OuLt\PR{3}\OuL$ to be diagonal in 
\begin{equation}\label{eq:Nu:FCCP:00}
\mNU=[\tb\id-(\tti)\OuLt\PR{3}\OuL]\mMU\,,\quad [\OuLt\PR{3}\OuL]_{ij}=\rnu{i}\rnu{j}\,,
\end{equation}
that is $\rnu{i}=\delta_{ik}$ for some $k$, in which case $[\OuLt\PR{3}\OuL]_{ij}=[\PR{k}]_{ij}$ and 
\begin{equation}
\CKM=[\id+(e^{i2\theta}-1)\OuLt\PR{3}\OuL]\OROTmat=[\id+(e^{i2\theta}-1)\PR{k}]\OROTmat,
\end{equation}
with the CKM matrix the product of a diagonal matrix of phases and a real orthogonal matrix $\OROTmat$, hence CP conserving again.

\section{Scalar potential\label{APP:scalar}}
In this appendix we discuss different aspects concerning the scalar potential of section \ref{SEC:scalar}: in \ref{sAPP:parameters} the election of a convenient set of basic parameters, in \ref{sAPP:minimum} boundedness (from below) of the potential, then perturbativity requirements in \ref{sAPP:perturbative}, and finally, in \ref{sAPP:lambda5}, a simple proof that $\lambda_5>0$ is a necessary condition in the present scenario.

\subsection{Independent parameters\label{sAPP:parameters}}
It is important to discuss the number and nature of the independent parameters of interest in the scalar sector. The goal is to adopt the most convenient choice for them. Through the minimization conditions of section \ref{sSEC:min}, it is already clear that one can trade the three quadratic coefficients $\mu_{ij}^2$ for $v^2$, $\beta$ and $\theta$, and set $v=246\text{ GeV}$. At this stage one could already consider a set of values for $\{\lambda_j\}$, $j=1,\ldots,5$, compute $\mcH$, the mass matrix $\mNSc$, and from $\mNSc$, obtain, at least numerically, $\mh^2$, $\mH^2$ ,$\mA^2$ and the mixings $\ROTmat$. Of course, one would then need to impose appropriate conditions: for example $\mh^2>0$, $\mH^2>0$, $\mA^2>0$. This is hardly the most convenient strategy, since, besides the computational toll, one would like for example to impose $\mh=125$ GeV. The phenomenological conditions in section \ref{sSEC:PhenoAnalysis} can be imposed afterwards. With this in mind, one would prefer to have (beside $\beta$ and $\theta$), $\mh^2$, $\mH^2$, $\mA^2$, $\mcH^2$ and three angles $\alpha_j$ describing $\ROTmat$ as parameters, and the different $\lambda_j$ expressed in terms of them.\\ 
We have already noticed that $\lambda_4$ can be traded for $\mcH^2$ and $\lambda_5$ using \refEQ{eq:ChargedMass:00}; this leaves \emph{four} quantities, $\lambda_1$, $\lambda_2$, $\lambda_{345}$ and $\lambda_5$, that, together with $\beta$ and $\theta$, determine $\mNSc$ (i.e. six different matrix elements). On the other hand, in $\mNSc$, we have three masses, $\mh^2$, $\mH^2$, $\mA^2$, while $\ROTmat$ requires three parameters: \emph{six} quantities. It is to be expected that they cannot be chosen independently. One simple procedure that can be adopted is the following:
\begin{enumerate}
\item equating elements $[\mNSc]_{13}$, $[\mNSc]_{23}$ and $[\mNSc]_{33}$ in \refEQS{eq:M2:33} with their expressions in terms of $\mh^2$, $\mH^2$, $\mA^2$, and $\ROTmat$, they can be read as a linear system in $\lambda_5$, $\mH^2$, $\mA^2$ which can be solved, giving them in terms of $\mh^2$, $\ROTmat$ and of course $\beta$, $v^2$ and $\theta$;
\item then, equating elements $[\mNSc]_{11}$, $[\mNSc]_{22}$ and $[\mNSc]_{12}$ in \refEQS{eq:M2:33} with their expressions in terms of $\mh^2$, $\mH^2$, $\mA^2$, and $\ROTmat$, they can be read as a linear system in $\lambda_1$, $\lambda_2$ and $\lambda_{345}$, which can also be solved, giving them in terms of $\mh^2$, $\ROTmat$, $\beta$, $v^2$ and $\theta$;
\item $\lambda_4$ is simply given by $\lambda_4=\lambda_5-\mcH^2/v^2$ with $\lambda_5$ already known;
\item with $\lambda_{345}$, $\lambda_4$ and $\lambda_5$ known, $\lambda_3$ is trivially $\lambda_3=\lambda_{345}+\lambda_5-\lambda_4$.
\end{enumerate}
Summarising: with these simple steps, for given values of $\beta$, $v^2$, $\theta$, $\mh^2$, $\mcH^2$ and $\ROTmat$ (three $\alpha_j$), one can compute $\mH^2$, $\mA^2$ and all $\lambda_j$, $j=1$ to $5$. For that set of values to be acceptable, one should then require
\begin{itemize}
\item positive values of all masses ($\mh^2$ and $\mcH^2$ can be chosen and thus only $\mH^2>0$ and $\mA^2>0$ have to be checked),
\item boundedness from below of $\mathscr V(\Hd{1},\Hd{2})$ and absolute minimum for $\{v^2,\beta,\theta\}$,
\item perturbative unitarity requirements on $\lambda$'s.
\end{itemize}
In order to illustrate the procedure to express $\mA^2$, $\mH^2$ and all $\lambda_j$ in terms of the basic set of parameters $\{v^2$, $\beta$, $\theta$, $\mh^2$, $\mcH^2$, $\alpha_j\}$, we start with the use of $[\mNSc]_{13}$, $[\mNSc]_{23}$ and $[\mNSc]_{33}$ to obtain $\mA^2$, $\mH^2$ and $\lambda_5$. One needs to equate those elements in \refEQS{eq:M2:33} to
\begin{align}
& [\mNSc]_{13}=\mh^2\,\ROT{11}\ROT{31}+\mH^2\,\ROT{12}\ROT{32}+\mA^2\,\ROT{13}\ROT{33} ,\label{eq:M2:m:13}\\
& [\mNSc]_{23}=\mh^2\,\ROT{21}\ROT{31}+\mH^2\,\ROT{22}\ROT{32}+\mA^2\,\ROT{23}\ROT{33},\label{eq:M2:m:23}\\
& [\mNSc]_{33}=\mh^2\,\ROT{31}^2+\mH^2\,\ROT{32}^2+\mA^2\,\ROT{33}^2.\label{eq:M2:m:33}
\end{align}
From the orthonormality relations $(\ROTmat^T\ROTmat)_{ij}=\delta_{ij}$ we have
\begin{align}
& \mh^2\,\ROT{31}=\ROT{11}[\mNSc]_{13}+\ROT{21}[\mNSc]_{23}+\ROT{31}[\mNSc]_{33},\label{eq:M2:mh2:0}\\
& \mH^2\,\ROT{32}=\ROT{12}[\mNSc]_{13}+\ROT{22}[\mNSc]_{23}+\ROT{32}[\mNSc]_{33},\label{eq:M2:mH2:0}\\
& \mA^2\,\ROT{33}=\ROT{13}[\mNSc]_{13}+\ROT{23}[\mNSc]_{23}+\ROT{33}[\mNSc]_{33},\label{eq:M2:mA2:0}
\end{align}
and thus
\begin{align}
& \mh^2\,\ROT{31}=v^2\lambda_5\left[-\sbb\sttCP\ROT{11}-\cbb\sttCP\ROT{21}+2\stCP^2\ROT{31}\right],\label{eq:M2:mh2}\\
& \mH^2=\mh^2\frac{\ROT{31}}{\ROT{32}}\frac{\ROT{12}[\mNSc]_{13}+\ROT{22}[\mNSc]_{23}+\ROT{32}[\mNSc]_{33}}{\ROT{11}[\mNSc]_{13}+\ROT{21}[\mNSc]_{23}+\ROT{31}[\mNSc]_{33}},\label{eq:M2:mH2}\\
& \mA^2=\mh^2\frac{\ROT{31}}{\ROT{33}}\frac{\ROT{13}[\mNSc]_{13}+\ROT{23}[\mNSc]_{23}+\ROT{33}[\mNSc]_{33}}{\ROT{11}[\mNSc]_{13}+\ROT{21}[\mNSc]_{23}+\ROT{31}[\mNSc]_{33}}.\label{eq:M2:mA2}
\end{align}
The solution reads
\begin{align}
& \lambda_5=\frac{\mh^2}{2v^2}\frac{\ROT{31}}{\stCP}\frac{1}{\stCP\ROT{31}-\ctCP\cbb\ROT{21}-\ctCP\sbb\ROT{11}},\label{eq:sol:l5}\\
& \mH^2=\mh^2\frac{\ROT{31}}{\ROT{32}}\left[\frac{-\ctCP\sbb\ROT{12}-\ctCP\cbb\ROT{22}+\stCP\ROT{32}}{-\ctCP\sbb\ROT{11}-\ctCP\cbb\ROT{21}+\stCP\ROT{31}}\right],\label{eq:sol:mH2}\\
& \mA^2=\mh^2\frac{\ROT{31}}{\ROT{33}}\left[\frac{-\ctCP\sbb\ROT{13}-\ctCP\cbb\ROT{23}+\stCP\ROT{33}}{-\ctCP\sbb\ROT{11}-\ctCP\cbb\ROT{21}+\stCP\ROT{31}}\right].\label{eq:sol:mA2}
\end{align}
%
Next, equating elements $[\mNSc]_{11}$, $[\mNSc]_{22}$ and $[\mNSc]_{12}$ in \refEQS{eq:M2:33} to
\begin{align}
& [\mNSc]_{11}=\mh^2\,\ROT{11}^2+\mH^2\,\ROT{12}^2+\mA^2\,\ROT{13}^2 ,\label{eq:M2:m:11}\\
& [\mNSc]_{22}=\mh^2\,\ROT{21}^2+\mH^2\,\ROT{22}^2+\mA^2\,\ROT{23}^2,\label{eq:M2:m:22}\\
& [\mNSc]_{12}=\mh^2\,\ROT{11}\ROT{21}+\mH^2\,\ROT{12}\ROT{22}+\mA^2\,\ROT{13}\ROT{23},\label{eq:M2:m:12}
\end{align}
one can solve for $\lambda_1$, $\lambda_2$ and $\lambda_{345}$:
\begin{align}
\lambda_1&=\frac{1}{2v^2}\left[[\mNSc]_{11}+\tb^2[\mNSc]_{22}+2\tb[\mNSc]_{12}\right]-\lambda_5\ctCP^2\tb^2\,,\label{eq:sol0:l1}\\
\lambda_2&=\frac{1}{2v^2}\left[[\mNSc]_{11}+\tb^{-2}[\mNSc]_{22}-2\tb^{-1}[\mNSc]_{12}\right]-\lambda_5\ctCP^2\tb^{-2}\,,\label{eq:sol0:l2}\\
\lambda_{345}&=\frac{1}{2v^2}\left[[\mNSc]_{11}-[\mNSc]_{22}+(\tbinv-\tb)[\mNSc]_{12}\right]-\lambda_5\ctCP^2\,,\label{eq:sol0:l345}
\end{align}
that is
\begin{align}
\lambda_1&=\frac{\mh^2}{2v^2}(\ROT{11}-\tb\ROT{21})^2+\frac{\mH^2}{2v^2}(\ROT{12}-\tb\ROT{22})^2+\frac{\mA^2}{2v^2}(\ROT{13}-\tb\ROT{23})^2-\lambda_5\ctCP^2\tb^2\,,\label{eq:sol:l1}\\
\lambda_2&=\frac{\mh^2}{2v^2}(\ROT{11}+\tbinv\ROT{21})^2+\frac{\mH^2}{2v^2}(\ROT{12}+\tbinv\ROT{22})^2+\frac{\mA^2}{2v^2}(\ROT{13}+\tbinv\ROT{23})^2-\lambda_5\ctCP^2\tb^{-2}\,,\label{eq:sol:l2}\\
\lambda_{345}&=\frac{\mh^2}{2v^2}(\ROT{11}^2-\ROT{21}^2+(\tbinv-\tb)\ROT{11}\ROT{21})+\frac{\mH^2}{2v^2}(\ROT{12}^2-\ROT{22}^2+(\tbinv-\tb)\ROT{12}\ROT{22})\nonumber\\
&\ +\frac{\mA^2}{2v^2}(\ROT{13}^2-\ROT{23}^2+(\tbinv-\tb)\ROT{13}\ROT{23})-\lambda_5\ctCP^2\,,\label{eq:sol:l345}
\end{align}
with $\mH^2$, $\mA^2$ and $\lambda_5$ in \refEQS{eq:sol:l5}--\eqref{eq:sol:mA2}. To complete the procedure, we just need to recall
\begin{equation}\label{eq:finsol:l345}
\lambda_4=\lambda_5-\mcH^2/v^2,\qquad \lambda_3=\lambda_{345}-\lambda_4+\lambda_5\,.
\end{equation}

\subsection{Boundedness and absolute minimum\label{sAPP:minimum}}
The conditions to be imposed on the resulting $\lambda_j$'s for a scalar potential bounded from below are
\begin{equation}\label{eq:Vbounded}
\lambda_1>0,\quad \lambda_2>0,\quad \sqrt{\lambda_1\lambda_2}>-\lambda_3,\quad \lambda_{345}>-\sqrt{\lambda_1\lambda_2}\,.
\end{equation}
Notice that with the expression of $\det\mNSc$ in \refEQ{eq:M2:det}, with $\lambda_5>0$ (see subsection \ref{sAPP:lambda5} below) it follows from $\det\mNSc>0$ that
\begin{equation}
\sqrt{\lambda_1\lambda_2}>\lambda_{345}>-\sqrt{\lambda_1\lambda_2}\,.
\end{equation}
One last concern on the scalar potential is the possibility that the local minimum for $\{v^2,$ $\beta,$ $\theta\}$ is not the absolute minimum of the potential, but instead a metastable minimum which can decay to the ``true'' absolute minimum (such a situation is sometimes dubbed the \emph{panic vacuum} \cite{Ivanov:2015nea}). From general studies of the minimization problem in 2HDM \cite{Ivanov:2007de,Barroso:2012mj,Barroso:2013awa,Ivanov:2015nea}, it follows that $\{v^2,$ $\beta,$ $\theta\}$ and $\{v^2,$ $\beta,$ $-\theta\}$ (this discrete ambiguity arised already in \refEQ{eq:thetaCP}) give indeed the absolute minima of the potential.

\subsection{Perturbative unitarity\label{sAPP:perturbative}}
Requiring perturbative unitarity of tree level scattering processes translates into the following bounds \cite{Kanemura:1993hm,Ginzburg:2005dt} (one loop corrections in a CP conserving 2HDM scenario have been addressed in \cite{Grinstein:2015rtl})
\begin{align}
&\lambda_1+\lambda_2\pm\sqrt{(\lambda_1-\lambda_2)^2+4\lambda_5^2}<\Lambda,\nonumber\\
&2(\lambda_3+\lambda_4)<\Lambda,\nonumber\\
&2(\lambda_3-\lambda_4)<\Lambda,\nonumber\\
&\lambda_1+\lambda_2\pm\sqrt{(\lambda_1-\lambda_2)^2+4\lambda_4^2}<\Lambda,\nonumber\\
&2(\lambda_3\pm\lambda_5)<\Lambda,\nonumber\\
&3(\lambda_1+\lambda_2)\pm\sqrt{9(\lambda_1-\lambda_2)^2+4(2\lambda_3+\lambda_4)^2}<\Lambda,\nonumber\\
&2(\lambda_3+2\lambda_4\pm 3\lambda_5)<\Lambda,
\end{align}
with $\Lambda=16\pi$.

\subsection{$\lambda_5>0$\label{sAPP:lambda5}}
As anticipated in section \ref{SEC:scalar}, the necessary condition $\lambda_5>0$ follows from a simple requirement on the scalar potential. If $V(\vev{1},\vev{2},\theta)$ is the absolute minimum of the potential, it is obviously necessary that $V(\vev{1},\vev{2},\theta)<V(\vev{1},\vev{2},0)$ and $V(\vev{1},\vev{2},\theta)<V(\vev{1},\vev{2},\pi)$. Notice that, although $\{\vev{1},\vev{2},0\}$ and $\{\vev{1},\vev{2},\pi\}$ fulfill \refEQ{eq:V:min:theta}, they do not fulfill \refEQS{eq:V:min:v1}--\eqref{eq:V:min:v2}, that is, $V$ cannot have a minimum for $\{\vev{1},\vev{2},0\}$, $\{\vev{1},\vev{2},\pi\}$. Since the $\theta$-independent terms of $V$ are common to all three cases, we only need to analyse the $\theta$-dependent part, $V_\theta$.
\begin{equation}
V_\theta(\vev{1},\vev{2},\begin{smallmatrix}0\\ \pi\end{smallmatrix})=\vev{1}\vev{2}\left[\pm\mu_{12}^2+\frac{1}{2}\vev{1}\vev{2}\lambda_5 \right]=\vev{1}^2\vev{2}^2\lambda_5\left[\frac{1}{2}\mp 2\cos\theta \right]\,,
\end{equation}
while
\begin{equation}
V_\theta(\vev{1},\vev{2},\theta)=\vev{1}\vev{2}\left[\mu_{12}^2\cos\theta+\frac{1}{2} \vev{1}\vev{2}\lambda_5(2\cos^2\theta-1)\right]\\
=-\vev{1}^2\vev{2}^2\lambda_5\left[\cos^2\theta+\frac{1}{2}\right]\,.
\end{equation}
Then
\begin{multline}
V_\theta(\vev{1},\vev{2},\theta)<V_\theta(\vev{1},\vev{2},\begin{smallmatrix}0\\ \pi\end{smallmatrix})\Leftrightarrow\\
-\vev{1}^2\vev{2}^2\lambda_5\left[\cos^2\theta+\frac{1}{2}\right]<\vev{1}^2\vev{2}^2\lambda_5\left[\frac{1}{2}\mp 2\cos\theta \right]
\Leftrightarrow
-\lambda_5(1\mp\cos\theta)^2<0\,,
\end{multline}
that is $\lambda_5>0$.

\section{Rephasings\label{APP:rephasing}}
The diagonalisation of the mass matrices $\wMD$ and $\wMU$ is only defined up to rephasings of the quark mass eigenstates. With
\begin{equation}
\mMD=\text{diag}(m_{d_j})=\UdLd\wMD\UdR\,,\quad \mMU=\text{diag}(m_{u_j})=\UuLd\wMU\UuR\,,
\end{equation}
and the rephasings
\begin{equation}
R_d=\text{diag}(e^{i\varphi^{d}_j})\,,R_u=\text{diag}(e^{i\varphi^{u}_j})\,,\quad R_q^\dagger=R_q^{-1}=R_q^{\ast}\,,
\end{equation}
it is clear that
\begin{equation}
R_d^\dagger\mMD R_d=\mMD=\UdLd\wMD\UdR\,,\quad R_u^\dagger\mMU R_u=\mMU=\UuLd\wMU\UuR\,.
\end{equation}
Consequently the diagonalising unitary matrices $\UdL$, $\UuL$, $\UdR$ and $\UuR$ are only given up to common redefinitions
\begin{equation}
\UdL\mapsto\UdL\,R_d\,,\ \UdR\mapsto\UdR\,R_d\,,\qquad \UuL\mapsto\UuL\,R_u\,,\ \UuR\mapsto\UuR\,R_u\,.
\end{equation}
Under such rephasings, the CKM matrix is transformed into
\begin{equation}
\CKM\mapsto R_u^\dagger\,\CKM\,R_d\,,\qquad \V{jk}\mapsto e^{i(\varphi^{d}_k-\varphi^{u}_j)}\V{jk}\,.
\end{equation}
The off-diagonal elements of the matrices $\mND$ and $\mNU$ are also transformed under rephasings,
\begin{equation}
\mND\mapsto R_d^\dagger\mND R_d,\quad \mNU\mapsto R_u^\dagger\mNU R_u,
\end{equation}
with
\begin{equation}
\und{j}=[\UdL]_{3j}\mapsto [\UdL R_d]_{3j}=e^{i\varphi^{d}_j}\und{j}\,,\quad
\undC{j}\und{k}\mapsto e^{i(\varphi^{d}_k-\varphi^{d}_j)}\undC{j}\und{k}\,,
\end{equation}
and
\begin{equation}
\unu{j}=[\UuL]_{3j}\mapsto [\UuL R_u]_{3j}=e^{i\varphi^{u}_j}\unu{j}\,,\quad
\unuC{j}\unu{k}\mapsto e^{i(\varphi^{u}_k-\varphi^{u}_j)}\unuC{j}\unu{k}\,.
\end{equation}


\newpage

\providecommand{\href}[2]{#2}\begingroup\raggedright\endgroup

\end{document}